\documentclass[sigconf]{acmart}

\usepackage{enumitem}
\usepackage{bbm}
\usepackage{algorithm}
\usepackage{algpseudocode}
\usepackage{graphicx} 
\usepackage{subcaption} 

\newtheorem{problem}{Problem}

\AtBeginDocument{%
  }

\setcopyright{acmlicensed}
\copyrightyear{2025}
\acmYear{2025}
\acmDOI{XXXXXXX.XXXXXXX}

\acmConference[Conference acronym 'XX]{Make sure to enter the correct
  conference title from your rights confirmation email}{June 03--05,
  2018}{Woodstock, NY}
\acmISBN{978-1-4503-XXXX-X/2018/06}

\acmSubmissionID{472}

\begin{document}

\title{Poisoning Attacks to Local Differential Privacy for Ranking Estimation}


\author{Pei Zhan}
\orcid{0009-0002-7690-309X}
\affiliation{%
  \institution{School of Cyber Science and Technology, Shandong University}
  \city{Qingdao}
  \country{China}
}
\additionalaffiliation{%
    \institution{Quan Cheng Laboratory, Jinan, China, and State Key Laboratory of Cryptography and Digital Economy Security, Shandong University, Qingdao, China}}
\email{peizhan@mail.sdu.edu.cn}

\author{Peng Tang}
\affiliation{%
  \institution{School of Cyber Science and Technology, Shandong University}
  \city{Qingdao}
  \country{China}
}
\authornotemark[1]
\authornote{Corresponding author.}
\email{tangpeng@sdu.edu.cn}

\author{Yangzhuo Li}
\affiliation{%
  \institution{School of Cyber Science and Technology, Shandong University}
  \city{Qingdao}
  \country{China}
}
\additionalaffiliation{%
    \institution{State Key Laboratory of Cryptography and Digital Economy Security, Shandong University, Qingdao, China}}
\email{liyangzhuo@mail.sdu.edu.cn}

\author{Puwen Wei}
\affiliation{%
  \institution{School of Cyber Science and Technology, Shandong University}
  \city{Qingdao}
  \country{China}
}
\authornotemark[3]
\email{pwei@sdu.edu.cn}

\author{Shanqing Guo}
\affiliation{%
  \institution{School of Cyber Science and Technology, Shandong University}
  \city{Qingdao}
  \country{China}
}
\authornotemark[3]
\email{guoshanging@sdu.edu.cn}










\begin{abstract}
Local differential privacy (LDP) involves users perturbing their inputs to provide plausible deniability of their data. However, this also makes LDP vulnerable to poisoning attacks. In this paper, we first introduce novel poisoning attacks for ranking estimation. These attacks are intricate, as fake attackers do not merely adjust the frequency of target items. Instead, they leverage a limited number of fake users to precisely modify frequencies, effectively altering item rankings to maximize gains.
To tackle this challenge, we introduce the concepts of \textit{attack cost} and \textit{optimal attack item (set)}, and propose corresponding strategies for kRR, OUE, and OLH protocols. For kRR, we iteratively select optimal attack items and allocate suitable fake users. For OUE, we iteratively determine optimal attack item sets and consider the incremental changes in item frequencies across different sets. Regarding OLH, we develop a \textit{harmonic cost function} based on the pre-image of a hash to select that supporting a larger number of effective attack items. Lastly, we present an attack strategy based on confidence levels to quantify the probability of a successful attack and the number of attack iterations more precisely. We demonstrate the effectiveness of our attacks through theoretical and empirical evidence, highlighting the necessity for defenses against these attacks.
The source code and data have been made available at https://github.com/LDP-user/LDP-Ranking.git.
\end{abstract}





\maketitle

\section{Introduction}

Differential privacy (DP)~\cite{dwork2006calibrating, dwork2006differential} has increasingly become the standard for data privacy in the research community. Recently, techniques for satisfying differential privacy in the local setting, referred to as LDP, have been developed and deployed.
In the local setting of DP, there are multiple users and one aggregator. Unlike the centralized setting, the aggregator does not have access to any user's private data. Instead, each user sends randomized information to the aggregator, who then attempts to infer the data distribution.
LDP facilitates the collection of statistics from decentralized users while safeguarding privacy without necessitating a trusted third party. LDP has wide-ranging applications, including data aggregation~\cite{DBLP:journals/sensors/RenLLHDZ16}, federated learning (FL)~\cite{DBLP:journals/inffus/BarrosoSJRMGLVH20}, blockchain technology~\cite{DBLP:journals/jpdc/HassanRC20}, context-free privacy~\cite{DBLP:conf/icml/AcharyaBKRS20}, and facial recognition~\cite{DBLP:journals/jpc/JosephRUW20}, and has been adopted by major companies  such as Google~\cite{DBLP:conf/ccs/ErlingssonPK14} in Chrome, Apple~\cite{abadi2017learning} in iOS, and Microsoft~\cite{ding2017collecting} in Windows 10.

LDP is a privacy-preserving technique in which users introduce perturbations to their data inputs, ensuring plausible deniability without the need for a trusted intermediary. This decentralized approach empowers individuals to share their information while upholding a certain level of privacy assurance. However, the fundamental nature of LDP, relying on user-controlled perturbations, presents potential vulnerabilities, particularly against poisoning attacks. In these attacks, fake users manipulate the server's analytics results by sending meticulously crafted data.

The existing attacks primarily focus on frequency estimation, which involves increasing the frequency of target items~\cite{DBLP:conf/uss/CaoJG21} and altering the mean or variance of user data~\cite{DBLP:conf/uss/LiLSG023}. However, in reality, LDP usually faces another common type of attack, namely ranking estimation attacks, which include elevating or lowering the ranking of target items.
For instance, within the healthcare sector, deceitful patients may attempt to manipulate the ranking of a particular disease diagnosis by providing inaccurate medical history details or adjusting symptom descriptions in the medical system, possibly resulting in misdiagnoses or treatment errors. In the financial realm, nefarious individuals or institutions could tamper with their financial data or transaction records to influence their risk assessment rankings, potentially causing financial system instability and the spread of risk. In public health, malevolent departments or organizations may endeavor to manipulate epidemic data or disseminate false information to alter the epidemic risk rankings of specific regions or populations, consequently posing a threat of a public health crisis.

Compared to attacks targeting frequency estimation, attacks targeting ranking estimation are more complex, primarily due to the allocation of fake data. Specifically, for attacks on frequency estimation, the attack's effectiveness is measured by the total change in the frequency of target items. Therefore, fake data can be randomly and uniformly allocated to target items. Even when target items have different weights, fake data can simply be distributed in proportions corresponding to those weights. However, for ranking attacks, one must consider the position of target items and the frequency differences between target and non-target items, and allocate fake data based on this information. Especially when it comes to lowering the ranking of target items, we need to allocate fake data to non-target items. Moreover, the number of non-target items is often much larger than that of target items, which further complicates the allocation of fake data. These challenges are discussed in detail in Section~\ref{sec:problem}.

To comprehensively tackle these challenges, we first formalize the problem by defining it as an optimization problem, and introduce the concepts of \textit{attack cost} and \textit{optimal attack item (set)} within the framework of LDP. These concepts are pivotal in formulating strategic approaches to effectively mitigate potential threats.
Based on these concepts, tailored attack strategies have been carefully crafted for the kRR, OUE, and OLH protocols. These strategies are finely tuned to exploit vulnerabilities within each protocol while considering the unique characteristics and constraints of the respective mechanisms.

In particular, in the context of the kRR protocol, our approach involves the iterative selection of optimal attack items and the precise allocation of fake users to maximize the impact of the attack. This strategic allocation aims to disrupt the integrity of the data while minimizing the attack cost.
For the OUE protocol, as each output supports multiple inputs, our strategy revolves around the iterative selection of optimal attack item sets. We need to carefully analyze the incremental changes in the frequencies of identical items across various combinations to identify strategic points of exploitation, enhancing the efficacy of the attack.
In the case of the OLH protocol, due to the randomness of hash mapping, determining the optimal attack item sets becomes more challenging.
To solve this problem, we devise a \textit{harmonic cost function} based on the pre-image of a hash function. This function aids in the selection of a hash function that can accommodate a larger number of effective attack items, thereby enhancing the effectiveness of the attack strategy.

Furthermore, recognizing the randomness in LDP, we have developed a sophisticated attack strategy formulation scheme that leverages confidence levels to quantify the probability of a successful attack and determine the optimal number of attack iterations. This approach enhances the precision and efficiency of our attack strategies.
Through a combination of theoretical analysis and empirical validation, we have demonstrated the effectiveness of our attack strategies, underscoring the critical need for innovative defense mechanisms to counteract and mitigate the impact of these sophisticated attacks. Our findings emphasize the urgency of developing robust defenses to safeguard LDP systems against evolving threats and vulnerabilities.

Our key contributions are summarized as follows:
\begin{itemize}[leftmargin=*]
  \item
  We first conduct a systematic study on poisoning attacks to LDP protocols for ranking estimation, introducing the concepts of \textit{attack cost} and \textit{optimal attack item (set)} within the LDP framework.

  \item
  We intricately design customized attack strategies for the kRR, OUE, and OLH protocols, and create an advanced attack strategy formulation scheme that utilizes confidence levels to account for the stochastic nature of LDP.
  \item
  We demonstrate the effectiveness of our attacks through theoretical and empirical evidence, highlighting the necessity for defenses against these attacks.
\end{itemize}

\noindent \textbf{Roadmap.} The remainder of this paper is organized as follows.
In Section~\ref{relatedwork}, we discuss the related work.
In Section~\ref{preliminaries}, we review the primitives that underlie our
proposed schemes.
In Section~\ref{problem}, we present the problem definition and threat model, and propose two baseline attacks including random input attack (RIA) and random output attack (ROA).
In Sections~\ref{ranking} and~\ref{heavy}, we discuss the attacks on kRR, OUE, OLH, and PEM protocols.
We show our experimental results in Section~\ref{experiments}.
We discuss defenses against ranking estimation attacks in Section~\ref{sec:defense}, and conclude in Section~\ref{conclusions}.

\section{Related Work} \label{relatedwork}
\textbf{Data Poisoning Attacks to LDP Protocols. }
Recent research highlights that although LDP protocols offer significant advantages in protecting user privacy, their distributed nature also makes it vulnerable to data poisoning attacks. Specifically, Cheu et al.~\cite{DBLP:journals/corr/abs-1909-09630} demonstrated that because the LDP aggregator is highly sensitive to changes in the distribution of perturbed data, attackers can degrade the overall performance of LDP by injecting fake data through the manipulation of fake users. Cao et al.~\cite{DBLP:conf/uss/CaoJG21} showed that poisoning attacks can be formulated as a single-objective optimization problem, where the attacker's goal is to maximize the estimated frequency of the items of interest. Wu et al.~\cite{DBLP:conf/uss/WuCJG22} explored LDP poisoning attacks against key-value data and proposed a dual-objective optimization scheme aimed at simultaneously increasing the estimated frequency and mean for certain attacker-selected target keys. Li et al.~\cite{DBLP:conf/uss/LiLSG023} proposed a fine-grained data poisoning attack method that allows the attacker to manipulate the post-attack mean and variance estimates to match pr e-determined values, enabling precise result manipulation across a wider range of scenarios.

In this paper, we extend attack strategies from frequency estimation to ranking estimation, empowering attackers to manipulate the rankings of specific items to either elevate or diminish their positions. Contrasted with prior research, poisoning attacks on LDP protocols for ranking estimation demonstrate a heightened level of intricacy. This complexity stems from attackers not merely modifying the frequency of target items but employing a controlled group of fake users to meticulously fine-tune frequencies, ultimately reshaping the rankings of these targets with precision for maximal influence.

\noindent\textbf{Defenses against Data Poisoning Attacks to LDP. }
To address the threats posed by poisoning attacks on LDP protocols, several countermeasures have been proposed in prior research. Cheu et al.~\cite{DBLP:journals/corr/abs-1909-09630} enhanced the robustness of LDP protocols by selecting appropriate parameters, such as the privacy budget. Cao et al.~\cite{DBLP:conf/uss/CaoJG21} introduced defense strategies against poisoning attacks on frequency estimation protocols, including normalization and fake user detection, which can mitigate the impact of such attacks to some extent. Subsequently, Song et al.~\cite{DBLP:journals/tifs/SongXZ23} designed verifiable LDP protocols that leverage zero-knowledge set membership proofs to open multiple commitments during the verification phase, preventing fake users from forging commitment vectors. Li et al.~\cite{DBLP:journals/corr/abs-2403-19510} systematically evaluated the robustness of LDP protocols for numerical data against data poisoning attacks. Leveraging the distribution properties of the reported data, they proposed a zero-shot attack detection method to effectively identify poisoning attacks. More recently, Sun et al.~\cite{DBLP:conf/icde/Sun0HDW0Y24} proposed LDPRecover, a method capable of recovering relatively accurate aggregated frequencies even when the server is unaware of the attack details. However, when applying the latest defense methods to counter our proposed attacks, experimental results reveal that although these methods are highly effective against frequency-based attacks, their performance against ranking attacks remains insufficient.

\noindent\textbf{Data Poisoning Attacks to ML. }
A series of poisoning attacks against machine learning systems~\cite{DBLP:conf/icml/BiggioNL12, DBLP:conf/uss/Carlini21, DBLP:journals/corr/abs-1712-05526, DBLP:conf/uss/FangCJG20, DBLP:conf/www/FangG020, DBLP:conf/acsac/FangYGL18, DBLP:journals/corr/abs-1708-06733, DBLP:journals/popets/HidanoMKKH20, DBLP:conf/ccs/JiZJLW18, DBLP:conf/cns/JiZW17, DBLP:conf/nips/LiWSV16, DBLP:conf/ndss/LiuMALZW018, DBLP:conf/ccs/Munoz-GonzalezB17, DBLP:conf/nsdi/NelsonBCJRSSTX08, DBLP:conf/ndss/YangGC17,DBLP:conf/aaai/AlfeldZB16, DBLP:conf/sp/JagielskiOBLNL18, DBLP:conf/aaai/JiaCG21, DBLP:conf/aaai/MeiZ15, DBLP:journals/titb/KermaniSRJ15, DBLP:conf/ccs/NewellPXN14, DBLP:conf/imc/RubinsteinNHJLRTT09, DBLP:conf/nips/ShafahiHNSSDG18, DBLP:conf/uss/WangWZZ14} has been developed recently. In these attacks, attackers poison carefully selected training samples or tamper with the training process, resulting in a manipulated model. For example, poisoning attacks on training data have been applied to support vector machines~\cite{DBLP:conf/icml/BiggioNL12} and regression models~\cite{DBLP:conf/sp/JagielskiOBLNL18}. Additionally, Shafahi et al.~\cite{DBLP:conf/nips/ShafahiHNSSDG18} proposed a novel training data poisoning attack against neural networks, where the attacker can manipulate the trained model to make predictions aligned with their intent. Gu et al.~\cite{DBLP:journals/corr/abs-1708-06733} and Liu et al.~\cite{DBLP:conf/ndss/LiuMALZW018} proposed a new type of poisoning attack against neural networks, known as backdoor attacks, where the attacker can select a label of interest, causing the trained model to predict that label for certain specific test samples. Moreover, training data poisoning attacks are also applicable to recommender systems~\cite{DBLP:conf/www/FangG020, DBLP:conf/acsac/FangYGL18, DBLP:conf/ndss/HuangMGL0X21, DBLP:conf/nips/LiWSV16,DBLP:conf/ndss/YangGC17}. On the other hand, poisoning attacks that tamper with the training process have been studied in the context of federated learning~\cite{DBLP:conf/aistats/BagdasaryanVHES20, DBLP:conf/icml/BhagojiCMC19,DBLP:conf/uss/FangCJG20}.
Our data poisoning attacks differ from these because the computational process of ranking estimation protocols or heavy hitter identification is significantly different from the methods machine learning uses to derive predictive models from training data.

\section{Background} \label{preliminaries}


\subsection{Frequency Estimation}
In frequency estimation, LDP protocols consist of three key steps: \textit{encode}, \textit{perturb}, and \textit{aggregate}. Encoding is to transform each user's raw data into a numerical form suitable for privacy-preserving processing. We denote the space of encoded values as \( D \). The perturb step involves randomly perturbing the values within the space \( D \), after which the perturbed values are then sent to the aggregator. In the aggregate step, the aggregator uses the perturbed values from all users to estimate the frequency of items. We denote the perturbed encoded value of item \( v \) as \( PE(v) \). Informally, if the probability of obtaining the same value after perturbation is similar for any two items, the protocol satisfies LDP, which is formally defined as follows:

\begin{definition}[\textnormal{Local Differential Privacy}]
If for any pair of items \( v_1, v_2 \in [d] \) and any perturbed value \( y \in D \), it holds that \( \Pr(PE(v_1) = y) \leq e^\epsilon \Pr(PE(v_2) = y) \), then the protocol satisfies \( \epsilon \)-Local Differential Privacy (\( \epsilon \)-LDP), where \( \epsilon > 0 \) is the privacy budget and \( PE(v) \) is the randomly perturbed encoded value of item \( v \).
\end{definition}

Furthermore, an LDP protocol is termed Pure LDP if it satisfies the following definition:

\begin{definition}[\textnormal{Pure LDP~\cite{DBLP:conf/uss/WangBLJ17}}]
An LDP protocol is termed Pure LDP if there exist two probability parameters \( 0 < q < p < 1 \) such that the following equations hold for any pair of items \( v_1, v_2 \in [d] \), \( v_1 \neq v_2 \):
\begin{equation}
\Pr(PE(v_1) \in \{ y \mid v_1 \in S(y) \}) = p
\end{equation}
\begin{equation}
\Pr(PE(v_2) \in \{ y \mid v_1 \in S(y) \}) = q,
\end{equation}
where \( S(y) \) is the support set of \( y \).
\end{definition}

The definition of the support \( S(y) \) depends on the specific LDP protocol. For example, in some LDP protocols ~\cite{DBLP:conf/allerton/DuchiJW13, DBLP:conf/uss/WangBLJ17}, the support \( S(y) \) of a perturbed value \( y \) consists of items whose encoded value could be \( y \). In a pure LDP protocol, the aggregate step is as follows:
\begin{equation}
\tilde{n}_v = \frac{\sum_{i=1}^{n} \mathbbm{1}_{S(y_i)}(v) - nq}{p - q},
\label{eq:freq_estimation}
\end{equation}
where \( \tilde{n}_v \) is the estimated frequency of item \( v \in [d] \), \( y_i \) is the perturbed value of the \( i \)-th user, and \( \mathbbm{1}_{S(y_i)}(v) \) is an \textit{indicator function} that outputs 1 if and only if \( y_i \) supports item \( v \), i.e., if \( v \in S(y_i) \), \( \mathbbm{1}_{S(y_i)}(v) = 1 \), otherwise outputs 0.

Next, we describe three state-of-the-art pure LDP protocols: kRR~\cite{DBLP:conf/allerton/DuchiJW13}, OUE~\cite{DBLP:conf/uss/WangBLJ17}, and OLH~\cite{DBLP:conf/uss/WangBLJ17}. These protocols are suitable for different scenarios. When the number of items is small, i.e., \( d < 3e^\epsilon + 2 \), kRR achieves the minimal estimation errors. When the number of items is large, both OUE and OLH can achieve minimal estimation errors. OUE has a higher communication cost, while OLH incurs higher computation costs for the aggregator. Therefore, OLH is recommended when communication cost is constrained; otherwise, OUE is preferred.

\subsubsection{kRR}\
\label{sec:kRR}

\noindent\textbf{Encode: } kRR encodes the item \( v \) as itself. Therefore, the encoded space \( D \) for kRR is the same as the domain of items, i.e., \( D = [d] \).

\noindent\textbf{Perturb: } kRR keeps the encoded value unchanged with probability \( p \) and perturbs it to another random item \( a \in D \) with probability \( q \). Thus, we have:
\begin{equation}
\Pr(y = a) = 
\begin{cases} 
\frac{e^\epsilon}{e^\epsilon + d - 1} \triangleq p, & \text{if } a = v, \\
\frac{1}{e^\epsilon + d - 1} \triangleq q, & \text{otherwise},
\end{cases}
\end{equation}
where \( y \) is the random perturbed value sent to the aggregator when a user's item is \( v \).

\noindent\textbf{Aggregate: } The key for aggregation is deriving the support set. In kRR, the perturbed value \( y \) only supports itself, i.e., \( S(y) = \{y\} \). Given the support set, the frequency of items can be estimated using Eq.~\eqref{eq:freq_estimation}.

\subsubsection{OUE}\
\label{sec:OUE}

\noindent\textbf{Encode: } OUE encodes an item \( v \) as a \( d \)-bit binary vector \( \textbf{e}_v \), where the \( v \)-th bit is 1 and all other bits are 0. The encoded space for OUE is \( D = \{0,1\}^d \), where \( d \) is the domain size of items.

\noindent\textbf{Perturb: } OUE independently perturbs each bit of the encoded binary vector. Specifically, for each bit of the encoded binary vector, if it is 1, it remains 1 with probability \( p \). If it is 0, it changes to 1 with probability \( q \). Thus, we have:
\begin{equation}
\Pr[y_i = 1] = 
\begin{cases} 
\frac{1}{2} \triangleq p, & \text{if } i = v, \\
\frac{1}{e^\epsilon + 1} \triangleq q, & \text{otherwise},
\end{cases}
\end{equation}
where the vector \( \textbf{y} = [y_1, y_2, \ldots, y_d] \) is the perturbed value of item \( v \).

\noindent\textbf{Aggregate: } The perturbed value \( \textbf{y} \) supports item \( v \) if and only if the \( v \)-th bit \( y_v = 1 \). Therefore, \( S(\textbf{y}) = \{v \mid v \in [d] \text{ and } y_v = 1\} \).

\subsubsection{OLH}\
\label{sec:OLH}

\noindent\textbf{Encode: } OLH uses a family of hash functions $\mathbf{H}$, where each hash function maps item \( v \in [d] \) to a value \( h \in [d'] \), with \( d' < d \). For optimal performance~\cite{DBLP:conf/uss/WangBLJ17}, in OLH, we set \( d' = e^\epsilon + 1 \). The family of hash functions $\mathbf{H}$ can be implemented using xxhash with different seeds, where each seed is a non-negative integer representing a different xxhash hash function. In the encode step, OLH randomly selects a hash function \( H \) from $\mathbf{H}$. When using xxhash, randomly selecting a hash function is equivalent to randomly selecting a non-negative integer as a seed. Then, OLH computes the hash value of item \( v \), \( h = H(v) \), and the tuple \( (H, h) \) is the encoded value of item \( v \). The encoded space for OLH is \( D = \{(H, h) \mid H \in \mathbf{H} \text{ and } h \in [d']\} \).

\noindent\textbf{Perturb: } OLH only perturbs the hash value \( h \), without changing the hash function \( H \). During perturbing, the hash value remains unchanged with probability \( p' \) and changes to another value in \( [d'] \) with probability \( q' \). Thus, we have:
\begin{equation}
\Pr[y = (H, a)] = 
\begin{cases} 
\frac{e^\epsilon}{e^\epsilon + d' - 1} \triangleq p', & \text{if } a = H(v), \\
\frac{1}{e^\epsilon + d' - 1} \triangleq q', & \text{if } a \neq H(v),
\end{cases}
\end{equation}
where \( y \) is the perturbed value sent to the aggregator by the user with item \( v \). Thus, the total probability parameters \( p \) and \( q \) are \( p = p' = \frac{e^\epsilon}{e^\epsilon + d' - 1} \) and \( q = \frac{1}{d'} \cdot p' + (1 - \frac{1}{d'}) \cdot q' = \frac{1}{d'} \).

\noindent\textbf{Aggregate: } If \( v \) maps to hash value \( h \) through hash function \( H \), then the perturbed value \( y = (H, h) \) supports item \( v \in [d] \), i.e., \( S(y) = \{v \mid v \in [d] \text{ and } H(v) = h\} \).


\section{Problem Definition and Threat Model} \label{problem}

\subsection{Problem Definition} \label{sec:problem}
\subsubsection{Challenges Analysis}\

In this work, we focus on ranking estimation attacks, which include both elevating and lowering the rankings of target items. In our problem, each user, whether genuine or fake, holds an item from a certain domain. To obtain rankings of items, we first need to estimate their frequencies. Consequently, changing the frequency of items allows us to alter their rankings. For instance, increasing the frequency of a target item elevates its ranking, while increasing the frequency of non-target items can lower the ranking of the target item. Note that, in poisoning attacks to LDP, we can generally only increase the frequency of an item by injecting fake data.

However, compared to existing frequency estimation attacks, our problem poses new challenges.
The complexity of the problem has shifted from linear to exponential.
Specifically, for attacks on frequency estimation, the attack's effectiveness is measured by the total change in the frequency of target items. Therefore, fake data can be randomly and evenly allocated to target items. Even when target items have different weights, fake data can be simply distributed in proportions corresponding to those weights, whose complexity is linear. In contrast, for ranking estimation attacks, we need to consider the position of target items and the frequency differences between target and non-target items. The smaller the difference, the lower the attack cost, meaning less fake data needs to be allocated. This way, the overall gain is maximized. Additionally, once a target item reaches a certain position, continuing to allocate fake data to it may not yield any further gains. Therefore, we need to consider how to allocate fake data reasonably to maximize gains, whose complexity is exponential. In particular, to lower the ranking of a target item, we need to allocate fake user data to non-target items. The number of non-target items is often much larger than that of target items, further complicating the allocation of fake data.

To better address this problem, we formally define it. Considering that lowering the ranking of target items is more complex, we will focus on discussing this issue subsequently.
The proposed approaches are equally effective for elevating the ranking of target items.

\subsubsection{Problem Definition}\

The problem of lowering the ranking of target items can be defined as follows.
\begin{problem}
Given $n$ genuine users and $m$ fake users, each possessing an item from a specific domain $[d]$, the aggregator employs frequency estimation protocols under LDP to collect item frequencies to determine item rankings. Attackers choose a set of target items and aim to lower the rankings of target items.
Let $T = \{t_1, t_2, \dots, t_r\}$ denote the set of target items, and $A = \{a_1, a_2, \dots, a_s\}$ denote the set of non-target items.
And let $\tilde{r}_{a,1}$ and $\tilde{r}_{t,1}$ denote the estimated rankings of non-target and target items before the attack, and $\tilde{r}_{a,2}$ and $\tilde{r}_{t,2}$ denote their estimated rankings after the attack. Given a set of perturbed values $\mathbf{Y}$ provided by fake users, the overall gain $G$ can be defined as:
\begin{equation}
G(Y) = \sum_{a \in A} \sum_{t \in T} \mathbbm{1}(\tilde{r}_{a,1} > \tilde{r}_{t,1} \text{ and } \tilde{r}_{a,2} < \tilde{r}_{t,2}).
\end{equation}
The attacker attempts to devise an attack strategy $ Y = \left\{ y_{fake_1 } , \ldots\right.$ $\left. y_{fake_i } , \ldots y_{fake_m }  \right\}$ that maximizes the gain, where $y_{fake_i }$ denotes the perturbed value of the $i^{th}$ fake user. This is,

\begin{equation} \label{Eq.optimal}
\begin{split}
&\mathop {\max }\limits_Y G\left( Y \right)\\
\text{subject to }
&\sum\limits_{j = 1}^s {\left| {Alloc_{a_j } } \right| \le } \sum\limits_{i = 1}^m {\left| {S\left( {y_{fake_i } } \right)} \right|},
\end{split}
\end{equation}
where $ S\left( {y_{fake_i } } \right)$ is the support set of $ y_{fake_i }$, $ Alloc_{a_j }$ denotes the set of perturbed values in $Y$ that support $ a_j $, i.e., $ Alloc_{a_j }  = \left\{ y_{fake_i } |a_j  \in\right.$ $\left. S\left( {y_{fake_i } } \right),1 \le i \le m\right\}
$, and $\left|\cdot \right|$ denotes the size of a set.
To facilitate the understanding of Eq.~\eqref{Eq.optimal}, a concrete numerical example is provided in Appendix~\ref{app:example}.
\end{problem}


Note that, in practice, different target items may have varying levels of importance, so different weights can be assigned to the ranking gains of each target item. However, to simplify the problem description, we assume that all target items have the same priority.
The notations adopted in this paper are listed in Table~\ref{tab:notation}.

\begin{table}[t]
\centering
\caption{Notations.} 
\label{tab:notation}
\begin{tabular}{cl}
\toprule
\textbf{Notation} & \textbf{Description} \\
\midrule
$r,s$ & Number of target and non-target items \\
$T,A$ & Sets of target and non-target items \\
$\tilde{r}_{v,1}$,$\tilde{r}_{v,2}$ & Estimated rankings before and after the attack \\
$A_{\text{eff}},\ A_{\text{ineff}}$ & Sets of effective and ineffective attack items  \\
$T_{a_i} $ & Set  that contains all \(k_i\) target items of ${a_i}$ \\
$ Y $ & Perturbed value attack strategy \\
$G$ & Overall gain of attack \\
$ S\left( \cdot \right)$ & Support set  \\
$ Alloc_{a_j }$ & Set of perturbed values that support $a_j$\\
$n,m$ & Number of genuine and fake users \\
$s'$ & Number of effective attack items \\
$d,D$ & Domain size and encoding space\\
$\delta,\delta_*$ & Attack cost and minimal cost among $\Delta$\\
$n_{v},n'_{v}$ & Genuine frequency and expected perturbed frequency\\
\(\Delta \) & Cost array of all effective attack items \\
\(a_{\text{opt}}\) & Optimal attack item \\
\( B , B_{\text{opt}}\) & Effective and optimal attack item set \\
$l,\ d'$ & Number and range size of hash functions \\
$C_{\left(i,j\right)}$ & Scoring function of $(H_i, h_j)$ \\
$P$ &  Probability  of success in attack\\
$x$ &  Number of attack rounds\\
$\epsilon$ &  Privacy budget\\
\bottomrule
\end{tabular}
\end{table}

\subsection{Threat Model}
We assume that attackers can inject fake users into an LDP protocol at a low cost. These fake users can send arbitrary data from the encoded space to the aggregator. The existing work~\cite{DBLP:conf/uss/ThomasMGKP13} has shown that attackers can acquire a large number of fake users from services like Twitter and Facebook. Specifically, our assumption involves the presence of $n$ genuine users, with the attacker introducing $m$ fake users into the system, culminating in a combined total of $m+n$ users.

Secondly, we assume that attackers can estimate the frequencies $n_v$ of each item $v$ using publicly available historical data, or by compromising a small number of genuine users.
This assumption is based on the fact that, in practice, the aggregator may periodically collect user information to make informed decisions based on statistical results. Additionally, for various reasons such as transparency, research and regulatory compliance~\cite{DBLP:conf/ccs/ErlingssonPK14}, these results are made public.
This allows the attacker to access the corresponding information. If data collection is conducted multiple times within a short period, the estimation results tend to ``stabilize'', indicating that the obtained values are very close. This estimation is denoted as $n_v^*$ where $n_v^* \approx n_v$. For simplicity, we use $n_v$ instead of $n_v^*$ in the following text.

Thirdly, given that the LDP protocol performs encode and perturb steps locally on the user's device, attackers possess knowledge of several parameters of the LDP protocol. In essence, attackers are aware of the domain size $d$, the encoding space $D$, the perturbation function, and the support set $S(y)$ for each perturbed value $y \in D$.


\subsection{Baseline Attack} \label{baseline}
We first propose two baseline attacks: Random Input Attack (RIA) and Random Output Attack (ROA).

\subsubsection{Random Input Attack (RIA)}\

In RIA, each fake user uniformly and randomly picks a non-target item from the set $A$, encodes and perturbs this value using the LDP protocol, and then reports the perturbed value to the server.

\subsubsection{Random Output Attack (ROA)}\

In the ROA, when targeting the kRR protocol, each fake user uniformly and randomly selects a non-target item and uses this item itself as the perturbed value. When attacking the OUE protocol, each fake user generates a random binary vector with $E_1$ (where $E_1 = p + (d-1)q$) non-target item bits set to 1 and the remaining bits set to 0, which is then considered the perturbed value. In the case of attacking the OLH protocol, each fake user uniformly and randomly chooses a non-target item and subsequently selects a hash function at random to map this non-target item to a specific hash value. The pair of hash function and hash value is then adopted as the perturbed value.

\section{Our Work: Attacking Ranking Estimation} \label{ranking}
To maximize the overall gain $G$ in the attack, the allocation of fake users to different non-target items is crucial. This involves assessing gains for all potential allocation schemes to determine the most profitable one. However, even in the kRR protocol, where $m$ fake users are distributed among 
$s$ non-target items, there exist $s^m$ possible schemes, rendering the computation highly complex and infeasible. To address this, a greedy strategy is proposed, focusing on incremental gain, which is referred to as \textit{Optimal Item Attack (OIA)}.

\begin{figure}
\begin{center}
\includegraphics[width=\linewidth]{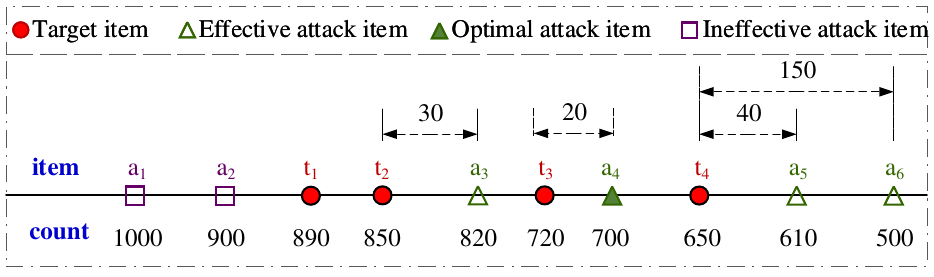}
\end{center}
\caption{\label{fig:data}Illustration of effective attack items and ineffective attack items.}
\end{figure}

\subsection{Optimal Item Attack (OIA)}
\label{sec:OIA}

OIA is a form of output poisoning attack that seeks to maximize gains by strategically selecting the optimal attack items or sets with the lowest attack cost. In the following, we will present formal definitions for both \textit{attack cost} and the \textit{optimal attack item}.

\begin{algorithm}[t]
  \caption{Attacking kRR}
  \label{alg:kRR}
  \begin{algorithmic}[1]
    \Require
      $A_{\text{eff}}$: effective attack item set; $T_{a_i}$: target item subset; $\Delta $: cost array; $m$: number of fake users
    \Ensure
      $M$: fake user allocation vector
    \State $M = \left[0,0,\dots,0\right]$
    \label{alg:kRR:1} 
    \label{code:fram:extract}
    \While {$m > 0$} \label{alg:kRR:while} 
      \State $\delta_* = min\left( \Delta\right)$
      \Comment{find the minimum attack cost}
      \State select $a_{opt}$ from $A_{\text{eff}}$ according to $\delta_*$;
      \If {$m>\delta_*$}
        \State $m_i = \delta_*$
        \label{alg:kRR:6} 
        \Comment{determine the number of fake users}
        \State $M[a_{\text{opt}}] = M[a_{\text{opt}}] + m_i$
        \label{alg:kRR:7} 
        \State \text{Update} $T_{a_{\text{opt}}}$, and $\Delta[a_{\text{opt}}]$
        \label{alg:kRR:8} 
        \State $m = m - m_i$
        \Comment{update the number of fake users}
      \Else
        \State $M\left[a_{opt}\right] = M\left[a_{opt}\right] + m$
        \State $m = 0$
      \EndIf
    \EndWhile
    \State\Return $M$; \label{alg:kRR:return} 
  \end{algorithmic}
\end{algorithm}

\begin{definition}[\textnormal{Effective Attack Items}]
For a non-target item $a\in A$, if there exists a target item $t \in T$ such that $r_{a,1} > r_{t,1}$, then we refer to $a$ as an effective attack item.
\end{definition}

In addition, the set of effective attack items can be defined as \(A_{\text{eff}} = \{a_j, a_{j+1}, \dots, a_s\} \subseteq A\).
The items in $A_{\text{eff}}$ are sorted in descending order based on their frequency, e.g., $\{a_3, a_4, a_5, a_6\}$ in Figure~\ref{fig:data}. Besides, for each effective attack item \(a_i\), we construct a set that contains all \(k_i\) target items whose frequencies are higher than \(a_i\), which is referred to as
\begin{equation}
T_{a_i} = \{t_1, t_2, \dots, t_{k_i}\} \subseteq T.
\end{equation}
Similarly, for a non-target item \(a \in A\), if \(r_{a,1} < r_{t,1}\) for all target items \(t \in T\), we call \(a\) an ineffective attack item (e.g., $a_1$ and $a_2$ in Figure~\ref{fig:data}). The set of ineffective attack items can be defined as
\begin{equation}
A_{\text{ineff}} = A - A_{\text{eff}} = \{a_1, a_2, \dots, a_{j-1}\}.
\end{equation}
\begin{definition}[\textnormal{Attack Cost for Each Item}]
For each effective attack item \(a_i \in A_{\text{eff}}\), its attack cost is calculated as
\begin{equation}
\delta_i = \left(p-q\right)\cdot\left(n_{t_{k_i}} - n_{a_i}\right),
\end{equation}
where $p$ and $q$ are probability parameters in LDP protocols, the target item \(t_{k_i}\) is the item with the smallest frequency in the subset \(T_{a_i}\). And for each ineffective attack items, the cost is set to \(+\infty\).
\end{definition}

In particular, the attack cost for each item is obtained as follows.
For an effective attack item $a_i$, making its attacked frequency exceed $t_{k_i}$'s attacked frequency 
is the easiest, which means the fewest fake users need to be allocated.
In kRR, OUE, and OLH protocols, for each item $v$, its expected perturbed frequency $E[n_v']$ is linearly related to its frequency $n_v$, i.e.,
\begin{equation}
E[n_v'] =n \cdot q + n_v \cdot \left(p-q\right).
\end{equation}
Besides, OIA is a form of output poisoning attack, that is, the selected value is reported to aggregator without perturbation.
Thus, the fewest fake users need to be allocated for $a_i$ is
\begin{equation}
\delta_i= E[n_{t_{k_i}}'] - E[n_{a_i}'] = (p-q)(n_{t_{k_i}} - n_{a_i}).
\end{equation}

\begin{definition}[\textnormal{Optimal Attack Item}]
For all effective attack items, the cost array can be denoted as \(\Delta = [\delta_j, \delta_{j+1}, \dots, \delta_s]\), and the minimum value $\delta_* = \min\left\{\delta_j, \delta_{j+1}, \dots, \delta_s\right\}$ represents the lowest cost among them. Then, the corresponding effective attack item is defined as the optimal attack item, which is referred to as $a_{\text{opt}}$.
\end{definition}

Based on these concepts, tailored attack strategies are meticulously developed for the kRR, OUE, and OLH protocols.
The OIA attack primarily consists of two steps: \textbf{fake user allocation} and \textbf{fake data reporting}. Since fake data reporting is more straightforward, we will focus our discussion on fake user allocation.

\begin{figure}
\begin{center}
\includegraphics[width=\linewidth]{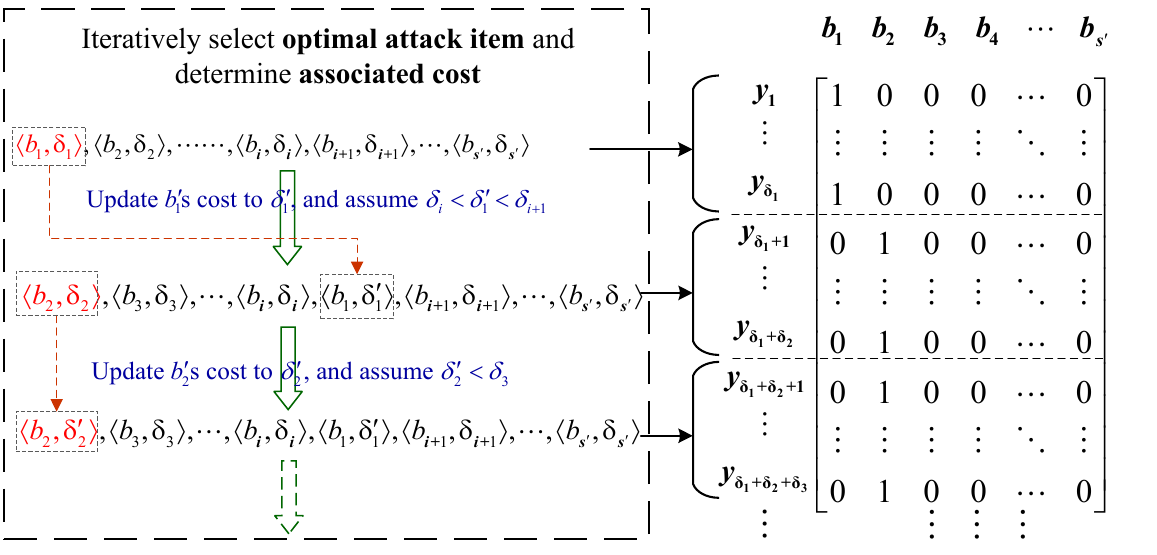}
\end{center}
\caption{\label{fig:Attacking kRR}An example of attacking kRR}
\end{figure}

\subsection{Attacking kRR} \label{section:krr}
In the kRR protocol, where $S\left(y\right) = \left\{y\right\}$,
each user's perturbed value {y} can only support at most one effective attack item, denoted as $\sum_{a_i \in A_{\text{eff}}} \mathbbm{1}_{S\left(y\right)}(a_i) \leq 1
$, and $\sum_{a_i \in A_{\text{eff}}} \mathbbm{1}_{S\left(y\right)}(a_i) = 1
$ only when $y$ is an effective attack item. Therefore, to attack kRR effectively, we iteratively choose optimal attack items and assign appropriate fake users based on their costs. Algorithm~\ref{alg:kRR} provides the details.

The algorithm takes as input a set of effective attack items \(A_{\text{eff}}\), the subset of target items \(T_{a_i}\) corresponding to each effective attack item \(a_i\), an initial cost array \(\Delta\) and the number of fake users \(m\). First, we initialize a zero vector \(M\) to record the allocation of fake users (Line~\ref{alg:kRR:1}). In each round, we select the optimal attack item \(a_{\text{opt}}\) with the minimum cost \(\delta_*\) and allocate the corresponding number of fake users to this item (Lines~\ref{alg:kRR:6}--\ref{alg:kRR:7}). Then, we update \( T_{a_{\text{opt}}} \) to \( \{t_1, t_2, \cdots, t_{k_{\text{opt}}-1}\} \). If \( T_{a_{\text{opt}}} \) is empty at this point, set the cost of \( a_{\text{opt}} \) to \( +\infty \); otherwise, update the cost corresponding to \( a_{\text{opt}} \) to \( \Delta[a_{\text{opt}}] = \lceil E[\tilde{n}_{t_{k_{\text{opt}}-1}}] - E[\tilde{n}_{a_{\text{opt}}}] \rceil \) (Line~\ref{alg:kRR:8}).
The algorithm continues until all the fake users are allocated.
Finally, we obtain a fake user allocation vector $M$. Figure~\ref{fig:Attacking kRR} illustrates the process of attacking kRR.

\begin{algorithm}[t]
  \caption{Attacking OUE}
  \label{alg:OUE}
  \begin{algorithmic}[1]
    \Require 
      $A_{\text{eff}}$: effective attack item set; $T_{a_i}$: target item subset; $\Delta$: cost array; $m$: number of fake users; $E_1$: number of 1s in the perturbed value of genuine users; $B$: effective attack item sequence
    \Ensure 
      $M$: fake user allocation result
    \State $M = 0_{m \times d}, m_{\text{start}} = 1$
    \label{alg:OUE:1}
    \While {$m > 0$}
      \If {$E_1 \leq s'$}
        \State $\delta_* = \min(\Delta)$
        \label{alg:OUE:4}
        \Comment{find the minimum attack cost}
        \State $B_{\text{opt}} = [b_1, b_2, \dots, b_{E_1}]$ 
        \label{alg:OUE:5}
        \State $a_{\text{opt}} = b_1$ 
        \label{alg:OUE:6}
        \If {$m > \delta_*$}
        \label{alg:OUE:7}
          \State $m_i = \delta_*$ 
          \label{alg:OUE:8}
          \Comment{determine the number of fake users}
          \State $M[k][b_j] = 1, k \in [m_{\text{start}}, m_{\text{start}} + m_i - 1], j \in [1, E_1]$ 
          \label{alg:OUE:9}
          \State $m_{\text{start}} = m_{\text{start}} + m_i$
          \State \text{Update} $T_{a_{\text{opt}}}$, $\Delta[a_{\text{opt}}]$, and $B_{\text{opt}}$
          \label{alg:OUE:11}
          \State $m = m - m_i$ \Comment{update the number of fake users}
        \Else
          \State $M[k][b_j] = 1, \forall i \in [m_{\text{start}}, \dots, m_{\text{start}} + m - 1], \forall j \in [1, 2, \dots, E_1]$
          \label{alg:OUE:14}
          \State $m = 0$
        \EndIf
      \Else
        \State $M[k][b_j] = 1, \forall i \in [m_{\text{start}}, \dots, m_{\text{start}} + m - 1], \forall j \in [1, 2, \dots, s']$ 
        \label{alg:OUE:18}
        \State all perturbed values also support random $E_1 - s'$ ineffective attack items;
        \label{alg:OUE:19}
        \State $m = 0$
      \EndIf
    \EndWhile
    \State \Return $M$
  \end{algorithmic}
\end{algorithm}

\subsection{Attacking OUE}
In the OUE protocol, each user's perturbed value \( y \) is represented as a binary vector. Since \( S(y) = \{v \mid v \in [d] \text{ and } y_v = 1\} \), each perturbed value \( y \) can support up to all effective attack items, whose number is $s'$, i.e., \( \sum_{a_i \in A_{\text{eff}}} \mathbbm{1}_{S(y)}(a_i) \leq s' \), and only when all positions corresponding to the effective attack items in the \( y \) vector are set to 1, do we have \( \sum_{a_i \in A_{\text{eff}}} \mathbbm{1}_{S(y)}(a_i) = s' \).
Thus,  to maximize the overall gain in attacking OUE, all fake users can simply choose perturbed values as follows: the \( s' \) bits corresponding to the effective attack items, denoted by \([j, j+1, \dots, s']\), are set to 1, and the bits for target items are set to 0. As ineffective attack items do not influence the attack, they can be set to either 0 or 1.

Although the above scheme yields the best attack effect, the perturbed values chosen by different fake users might be highly similar, and the number of 1s in these perturbed values may differ significantly from those in the genuine users' perturbed values, making them susceptible to detection by the server. To avoid detection, we can set the number of 1s in the perturbed values equal to the expected number of 1s in the genuine users' perturbed values, denoted as \( E_1 \), where \( E_1 = p + (d-1)q \), with \( p = \frac{1}{2} \) and \( q = \frac{1}{e^\epsilon + 1} \).
Usually, $ E_1 < s'$.
Therefore, to attack OUE effectively, we iteratively select optimal attack item sets. Additionally, we take into account the incremental changes in item frequencies across various selected sets. Details can be found in Algorithm~\ref{alg:OUE}.

%

\begin{figure}[t]
\begin{center}
\includegraphics[width=\linewidth]{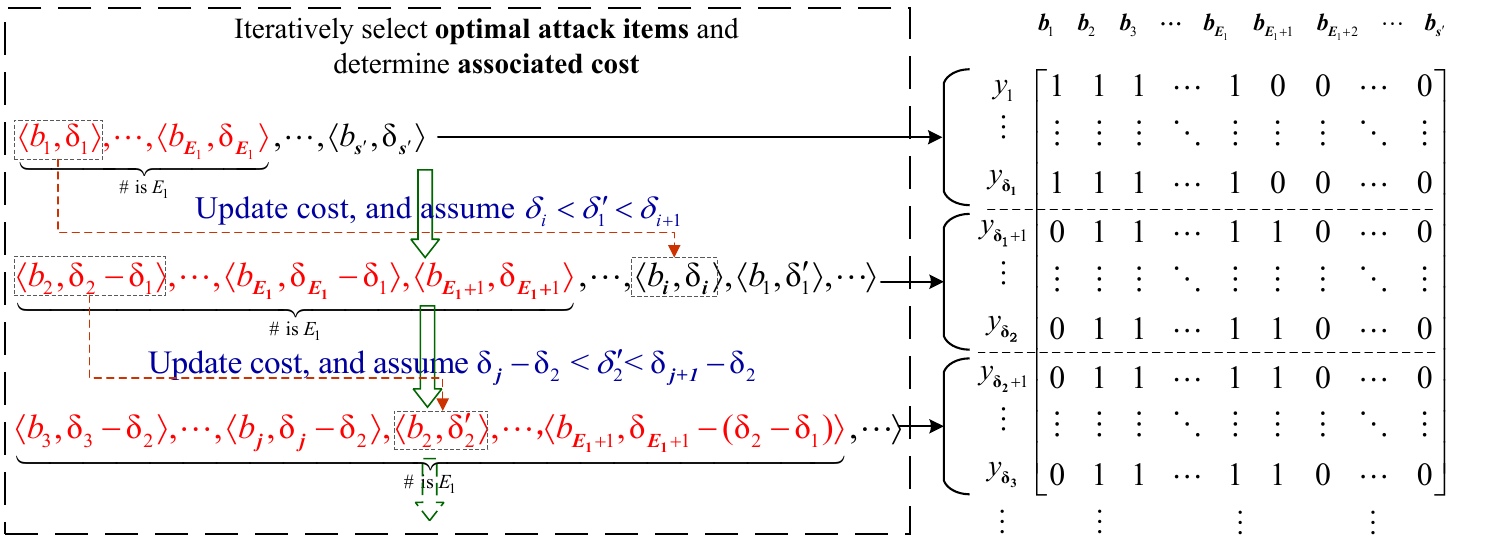}
\end{center}
\caption{\label{fig:Attacking OUE}An example of attacking OUE.}
\end{figure}

The algorithm for attacking OUE takes the same inputs as attacking kRR, plus \( E_1 \) (expected number of 1s in genuine users' perturbed values) and \( B \) (effective attack item sequence).
At the beginning, we initialize a zero matrix \( M \) to record the allocation of fake users (Line~\ref{alg:OUE:1}), where each row represents the perturbed value chosen by a fake user.
Furthermore, we obtain the effective attack item array \( \textbf{B} = [b_1, b_2, \dots, b_{s'}] \) in ascending order based on their costs.
In each iteration, we select the first \( E_1 \) items in $\textbf{B}$ to form the optimal attack item set \( B_{\text{opt}} \), and allocate \( \delta_* \) fake users to this set, where $\delta_*$ denotes the cost of $b_1$ (Lines~\ref{alg:OUE:5}--\ref{alg:OUE:9}).
That is, we let these $\delta_*$ fake users set the perturbed value as follows: the $E_1$ bits corresponding to the items in $B_{\text{opt}}$ be 1, and the other bits be 0.
Then, we update the set of $T_{b_1}$ that contains all $k$ target items whose frequencies are higher than $b_1$, and the frequency of all items in $B_{\text{opt}}$ (Line~\ref{alg:OUE:11}).
Specifically,
\begin{equation}
T_{b_1} = T_{b_1}-\left\{t_k\right\} = \left\{t_1,\ldots, t_{k-1}\right\},
\end{equation}
\begin{equation}
{\delta _{{b_i}}} = \begin{cases}
{n_{{t_{k-1}}}}-\left({n_{{b_i}}} + {\delta _ * } \right),&\text{ if }i = 1,\\
{\delta _{{b_i}}} - {\delta _ * },&\text{ if }2 \le i \le {E_1}.
\end{cases}
\end{equation}
Next, re-sort the elements in $\textbf{B}$ according to their updated costs.
The algorithm continues until all the fake users are allocated.
Finally, we obtain a fake user allocation matrix $M$. Figure~\ref{fig:Attacking OUE} illustrates the process of attacking OUE.
%




\subsection{Attacking OLH} \label{section:olh}
In the OLH protocol, the perturbed value for each user is $y = \left(H,h\right)$. Given that $S\left(y\right) = \left\{v\mid v \in \left[d\right] \text{ and } H(v) = h \right\}$, each perturbed value $y$ can support up to all $s'$ effective attack items, i.e., $\sum_{a_i \in A_{\text{eff}}} \mathbbm{1}_{S\left(y\right)}(a_i) \leq s'$, and only when all $a_i \in A_{\text{eff}}$ satisfy $H(a_i) = h$, the equality $\sum_{a_i \in A_{\text{eff}}} \mathbbm{1}_{S\left(y\right)}(a_i) = s'$ holds.
Thus, to maximize the attack effect in attacking OLH, an ideal solution would be to let all fake users choose a perturbed value \(y = (H, h)\) that satisfies:
\begin{equation}
    \forall a_i \in A_{\text{eff}},\ H(a_i) = h \text{ and } \forall t \in T,\ H(t) \neq h.
\end{equation}
However, due to the inherent randomness of the hash function, it is often challenging or even impossible to find such a perturbed value.
Therefore, our objective is to select an optimal perturbed value $y=<H,h>$ that satisfies the following three criteria: 1) no mapping to target items, 2) maximal coverage of effective attack items, and 3) minimal associated attack costs.
This requires a \textit{scoring function} to quantify perturbation effectiveness, where lower values indicate better attack performance.

Let $l$ denote the number of hash functions, and $d'$ denote the range (or domain) size of each hash function.
Assume the perturbed value \(y_{(i,j)} = (H_i, h_j)\) (\(i \in \{1, 2, \dots, l\}\), \(j \in \{1, 2, \dots, d'\}\)) supports an effective attack item set $A_{(i,j)} \subseteq A_{\text{eff}}$ and $A_{(i,j)} \ne \emptyset$.
It is found that the \textit{scoring function} can be defined as:
\begin{equation}
C_{\left(i,j\right)}=\frac{1}{\sum_{a_k\in A_{(i,j)}}\frac{1}{\delta_k}}.
\end{equation}
This \textit{harmonic scoring function} exhibits a key property:
Monotonic decrease with either more attack items or lower individual costs.
Thus, minimizing $C_{\left(i,j\right)}$ maximizes marginal attack contribution, perfectly aligning with submodular optimization principles.
For detailed discussions, please refer to the Appendix~\ref{app:discussions}.
Based on this \textit{scoring function}, we introduce an approach for attacking the OLH protocol,
in which we iteratively select perturbed values that have high quality and allocate suitable fake users.
The details are described in Algorithm~\ref{alg:OLH}.
\begin{algorithm}[t]
  \caption{Attacking OLH}
  \label{alg:OLH}
  \begin{algorithmic}[1]
    \Require 
      $A_{\text{eff}}$: effective attack item set; $T_{a_i}$: target item subset; $\Delta$: cost array; $m$: number of fake users
    \Ensure 
      $M$: fake user allocation result
    \State $M = 0, C = \{\}, m_{\text{start}} = 1$  
    \label{alg:OLH:1}
    \While {$m > 0$}
      \For {$i,j = 0; i < l, j < d'; i++, j++$}
        \State $C_{(i,j)} = \frac{1}{\sum_{a_k \in A_{(i,j)}} \frac{1}{\delta_k}}$  
        \label{alg:OLH:4}
      \EndFor
      \State $C_{(i_1, j_1)} = \min(C)$  \Comment{find the minimum cost}
      \label{alg:OLH:6}
      \State select ${(H_{{\text{i}}_1}, h_{{\text{j}}_1})}$ based on $C_{(i_1, j_1)}$
      \label{alg:OLH:7}
      \State ${\delta _{*}} = \mathop {\min }\limits_{v \in A} \delta (v), \text{ where } A = \{ v|{H_{{\text{i}}_1}(v)  =  {h_{{\text{j}}_1}}}\} $  
      \label{alg:OLH:8}
      \State select $a_{\text{opt}}$ based on $\delta _{*}$
      \label{alg:OLH:9}
      \If {$m > \delta_*$}
        \State $m_k = \delta_*$  \Comment{determine the number of fake users}
        \label{alg:OLH:11}
        \State put $\left\langle {(H_{{\text{i}}_1}, h_{{\text{j}}_1}), m_k}\right\rangle$ into $M$
        \label{alg:OLH:12}
        \State $m_{\text{start}} = m_{\text{start}} + m_i$  
        \State \text{Update} $T_{a_{\text{opt}}}$, $\Delta$, and $C$
        \State $m = m - m_i$  \Comment{update the number of fake users}
      \Else
        \State put $\left\langle {(H_{{\text{i}}_1}, h_{{\text{j}}_1}), m}\right\rangle$ into $M$
        \State $m = 0$
      \EndIf
    \EndWhile
    \State \Return $M$
  \end{algorithmic}
\end{algorithm}

\begin{figure}[t]
\begin{center}
\includegraphics[width=\linewidth]{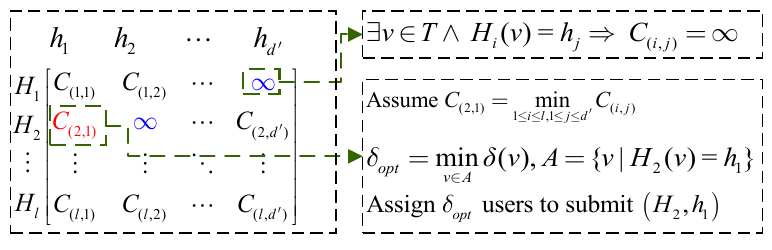}
\end{center}
\caption{\label{fig:Attacking OLH}An example of attacking OLH}
\end{figure}

The algorithm for attacking OLH takes the same inputs as attacking kRR.
At the beginning, we initialize a vector $M$ to record the allocation of fake users (Line~\ref{alg:OLH:1}). Each element in $M$ can be treated as $\left\langle {(H_i, h_j),m_k } \right\rangle$, where $(H_i, h_j)$ denotes the selected perturbed value in each iteration, and $m_k$ denotes the number of allocated fake users to this perturbed value.
In the first iteration, we calculate the cost of each perturbed value $(H_i, h_j)$ (Line~\ref{alg:OLH:4}).
In particular, if there exists $t\in T$ satisfying $H_i\left(t\right) = h_j$, then the cost of ${(H_i, h_j),m_k }$ is set to $\infty$.
Then, we select the perturbed value that has the smallest cost.
Assume
\begin{equation}
{C_{(i_1,j_1)}} = \mathop {\min }\limits_{1 \le i \le l,1 \le j \le d'} {C_{(i,j)}},
\end{equation}
and let
\begin{equation}
{\delta _{*}} = \mathop {\min }\limits_{v \in A} \delta (v), \text{ where } A = \{ v|{H_{i_1}}(v)  =  {h_{j_1}}\}.
\end{equation}
We will assign ${\delta _{*}}$ fake users to report the perturbed value $\left({H_{i_1}}, h_{j_1}\right)$.
Thus, we put element $\left\langle (H_{i_1}, h_{j_1}),\delta _{*} \right\rangle$ into $M$ (Lines~\ref{alg:OLH:6}--\ref{alg:OLH:12}).
Next, we update the cost of each effective attack item $a\in\left\{x|{H_{{i_1}}}\left( x \right) = {h_{{j_1}}}\right\}$, and the cost of each perturbed value $(H_{i_2}, h_{j_2})$ if it satisfies
\begin{equation}
\begin{split}
&\forall\ t \in T,{H_{{i_2}}}\left( t \right) \ne {h_{{j_2}}},\text{ and,}\\
&\exists\ a \in {A_{\text{eff}}},{H_{{i_1}}}\left( a \right) = {h_{{j_1}}} \wedge {H_{{i_2}}}\left( a \right) = {h_{{j_2}}}.
\end{split}
\end{equation}
The algorithm continues until all the fake users are allocated.
Finally, we obtain a fake user allocation vector $M$. Figure~\ref{fig:Attacking OLH} illustrates the process of attacking 
OLH.

\subsection{\raggedright More Precise Optimal Item Attack $\textbf{(MPOIA)}$}
\label{sec:MPOIA}
To mitigate the effect of random errors introduced by perturbation, we propose a method that utilizes confidence intervals to enhance the success probability of each round of attack, termed \textit{More precise optimal item attack (MPOIA)}.

As described in Section~\ref{sec:OIA}, we use the expected value $E[n'_{v}]$ of the perturbed frequency to estimate the random variable $n'_{v}$, and determine the number of fake users that allocated to the selected effective items.
In fact, the perturbed frequencies of each item can be viewed as a random variable.
To enhance accuracy, we compute confidence intervals for the random variable $n'_v$ at different confidence levels to develop a more precise optimal attack strategy.

Given the large number of random variables involved, the central limit theorem allows us to approximate the distributions of these perturbed frequencies by normal distributions. Let $n'_a$ and $n'_t$ denote the frequencies of the effective attack item $a$ and the target item $t$ after perturbation. In each iteration of the attacks, our goal is to determine the minimum number of fake users that need to be allocated to ensure the frequency of $a$ exceed that of $t$. Assuming $m_i$ fake users choose $a$ as their perturbed value, since $n'_a$ and $n'_t$ approximately follow normal distributions, we have:
\begin{equation}
    n_a' \sim N\left( E[n_a'] + m_i, \text{Var}(n_a') \right)
\end{equation}
and
\begin{equation}
    n_t' \sim N\left( E[n_t'], \text{Var}(n_t') \right).
\end{equation}
where $Var(n'_{v})$ denotes the variance of  $n'_{v}$, which is given by:
\begin{equation} \label{eq:variance}
\text{Var}(n_{v}') = n_v \cdot p \cdot (1-p) + (N - n_v) \cdot q \cdot (1-q).
\end{equation}
For different protocols, the variance of the perturbed frequency counts depends only on the values of $p$ and $q$.
According to the properties of normal distribution, the difference random variable $D = n'_a+m_i-n'_t$ also follows a normal distribution, with the expected value given by:
\begin{equation}
    E[D] = E[n_a']+m-E[n_t'],
\end{equation}
and the variance is:
\begin{equation}
    Var(D)=Var(n_t')+Var(n_a'). 
\end{equation}

We derive the one-sided confidence interval for the mean of $D$:$(E[D] - z_{\alpha} \times SE, +\infty)$. To ensure \( P(D > 0) \geq 1 - \alpha \), the required number of fake users is:
\begin{equation}
    m_i = \left \lceil E[n'_t] - E[n'_a] + z_\alpha \cdot SE \right \rceil,
\end{equation}
where \( SE = \sqrt{Var(D)} \).

It is worth noting that, for the attacks introduced in Sections~\ref{section:krr}-\ref{section:olh}, we can consider the confidence level to be 0.5. 
To demonstrate MPOIA's core principles and implementation, we provide complete pseudocode in Algorithm~\ref{alg:MPOIA wiht kRR} (Appendix~\ref{app:MPOIA with kRR}), using the kRR attack as a representative case study.

\subsection{Attack's Effectiveness Analysis}
\label{sec:analysis}
In the previous discussion, we used the expected values of frequency counts (Sections~\ref{section:krr}-\ref{section:olh}) or confidence intervals (Section~\ref{sec:MPOIA}) to determine the number of fake users needed to be allocated for each selected effective items,  with the objective of achieving a maximal gain per iteration (or round). The overall gain $G$ is influenced by the number of attack rounds $x$ and the probability $P$ of success in each round of attack. The larger the number of attack rounds and the higher the success probability per round, the greater the overall gain $G$. Assuming a success probability $P$ for each attack round, we have:
\begin{equation}
    G = x \cdot P.
\end{equation}
Once the dataset is fixed, $x$ is ultimately determined by the total number of fake users $m$, privacy budget $\epsilon$, and significance level $\alpha$, while $P$ is solely determined by the significance level $\alpha$.

\noindent\textbf{Analysis of Attack Rounds.}
Assuming that $m_i$ fake users are injected in the $i$-th round of allocation, the following relationship is satisfied:
\begin{equation}
    m = m_1 + m_2 + \dots +m_x,
\end{equation}
this shows that the larger $m$ or the smaller each $m_i$, the greater the number of $x$. Considering a special case, when the fake users injected in each round of attack are the same and all of them are $m_i$, there is $m = x \cdot m_i$, that is $x = \frac{m}{m_i}$.
In OIA attacks, the number of fake users that need to be injected in the $i$-th round is (excluding rounding):
\begin{equation}
m_i = E[n_t'] - E[n_a'] = (n_t - n_a) \cdot (p - q),
\end{equation}
where $p = \frac{e^\epsilon}{e^\epsilon+d-1}$, $q = \frac{1}{e^\epsilon+d-1}$. Then, we can learn that:
\begin{equation}
\begin{split}
    m_i &= \frac{(n_t - n_a) \times (e^\epsilon - 1)}{e^\epsilon + d - 1} \\
&= (n_t - n_a) \left( 1 - \frac{d}{e^\epsilon + d - 1} \right)
\end{split}.
\end{equation}
When the dataset is fixed, with $d$, $n_t$, and $n_a$ being constants, $m_i$ becomes solely dependent on $\epsilon$. By defining the constants as $A = n_t - n_a$ and $B = d$, we can express $m_i$ as:
\begin{equation}
m_i = A \left( 1 - \frac{B}{e^\epsilon + B - 1} \right).
\end{equation}
It is clear that $m_i$ is positively correlated with $\epsilon$, that is, as $\epsilon$ increases, the number of fake users $m_i$ required for injection per round of attack also increases. This is because a smaller $\epsilon$ leads to perturbed data being more uniformly distributed than the original distribution, resulting in smaller discrepancies in frequency counts among items. Consequently, injecting fewer fake users achieves the desired gain. Conversely, since $x$ is inversely related to $m_i$, $x$ exhibits a negative correlation with $\epsilon$.


\noindent\textbf{Success Probability Analysis}.
In MPOIA, the number of fake users to inject in the $i$-th round of attacks is (excluding rounding):
\begin{equation}
    m_i = E[n'_t] - E[n'_a] + z_{\alpha} \times SE,
\end{equation}
and the probability $P$ of successful attack in each round is determined by the confidence level $1 - \alpha$, satisfying:
\begin{equation}
    P = 1- \alpha,
\end{equation}
as $\alpha$ increases, $P$ decreases. The confidence level should satisfy $1 - \alpha \leq 0.5$. When $1 - \alpha = 0.5$, $z_{\alpha} = 0$, which is equivalent to using OIA.
To analyze the effect of changing $\alpha$, we fix $\epsilon$, where $E[n'_t] - E[n'_a]$ and $SE$ are both constant values. We set the constant $C = E[n'_t] - E[n'_a]$, $D = SE$, and obtain:
\begin{equation}
    m_i = C + D \cdot SE,
\end{equation}
as $\alpha$ increases, $z_{\alpha}$ decreases, leading to a smaller $m_i$ and thus a greater number of attack rounds $x$. For simplicity, we assume the same number of fake users is injected in each round, resulting in:
\begin{equation}
    x = \frac{m}{C + D \cdot z_\alpha},
\end{equation}
\begin{equation}
    G = x \cdot P = \frac{m (1 - \alpha)}{C + D \cdot z_\alpha}.
\end{equation}

As seen, as $\alpha$ increases, both the numerator and denominator in $G$ decrease simultaneously, and the specific trend of $G$ change depends on the coefficient. This is because increasing the success probability in each attack round, the confidence interval will be larger, and more fake users need to be injected to ensure a higher success rate, resulting in a reduction in the number of attack rounds.

In conclusion, the overall gain $G$ is determined by $m$, $\epsilon$, and $\alpha$. $G$ is positively correlated with $m$ and negatively correlated with $\epsilon$.

\section{Attacking Heavy Hitter Identification} \label{heavy}

The most advanced protocols for heavy hitter identification currently iteratively apply frequency estimation protocols. Thus, the three attack methods previously discussed for ranking estimation can also be applied to heavy hitter identification.
The detailed experimental results are included in Appendix~\ref{app:MPOIA}.

\section{Evaluation} \label{experiments}
\subsection{Experimental Setup}

\subsubsection{Datasets}\

We evaluate our attacks on three datasets, including a synthetic dataset and two real-world datasets, i.e., Adult and Fire. 

\begin{itemize}[left=0pt, labelsep=4pt, itemsep=0pt, parsep=0pt, topsep=0pt]
    \item \textbf{Synthetic}: Following the settings in~\cite{DBLP:conf/uss/CaoJG21, DBLP:conf/uss/WangBLJ17}, we created a synthetic dataset to evaluate our attack. The data in this dataset follows a Zipf distribution with 100 items and 100,000 users.
    \item \textbf{Adult}: The Adult dataset contains census data collected over several years, including 14 attributes. It was extracted by Barry Becker from the census database. We selected the age attribute as the item owned by each user. This dataset contains 74 items and 45,222 users.
    \item \textbf{Fire}: The Fire dataset was collected by the San Francisco Fire Department. We used the latest data from 2024, filtering for records where the type is ``Alarms'', and selecting the Unit ID attribute as the item owned by each user. This dataset contains 311 items and 740,728 users.
\end{itemize}

\begin{figure}[t]
\begin{center}
\includegraphics[width=\linewidth]{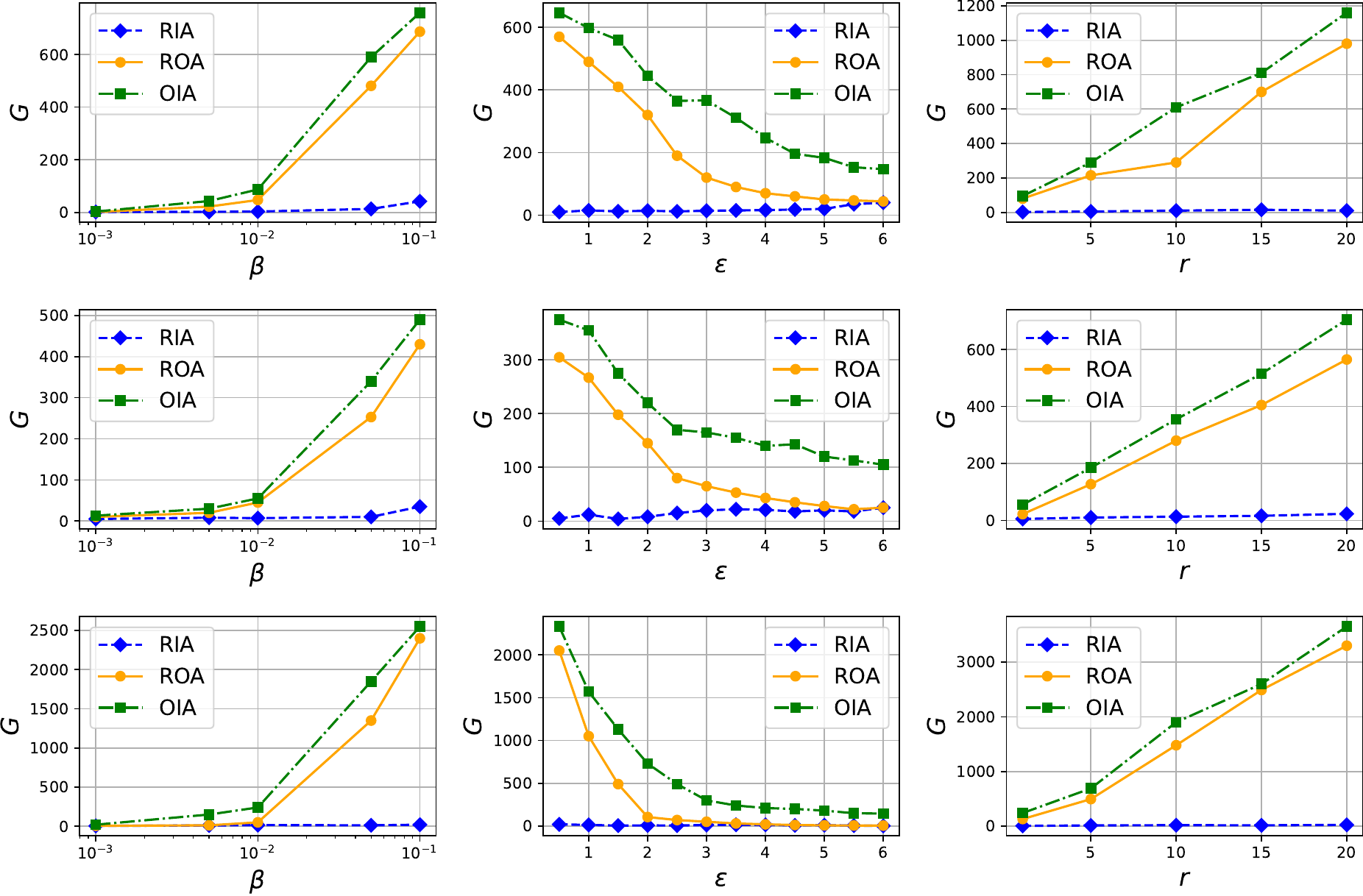}
\end{center}
\caption{\label{fig:kRR result} Impact of OIA with different parameters ($\beta$, $\epsilon$, $r$) on the overall gain for kRR under rank lowering scenario. The three rows are for Synthetic, Adult and Fire datasets, respectively.}
\end{figure}

\subsubsection{Parameter Settings}\

For ranking estimation, the overall gain obtained by the attack is primarily influenced by three parameters: $\epsilon$ (privacy budget), $\beta$ (proportion of fake users), and $r$ (number of target items). Unless otherwise specified, we set the default values of these parameters as follows: $\epsilon = 1$, $\beta = 0.05$, and $r = 10$. It is important to note that for attacks on the OUE protocol under rank lowering scenario, when $\epsilon$ is small, both OIA and ROA achieve optimal attack effects, making it difficult to distinguish their effectiveness as $\beta$ and $r$ vary. Therefore, we set $\epsilon = 3$ as the default value when attacking OUE under rank lowering scenario. Similarly, for attacks on the kRR protocol under rank elevation scenario, we observe comparable behavior. Therefore, we set $\epsilon = 4.5$ as the default value. In the experiments, we will examine the impact of each parameter while keeping the remaining parameters fixed at their default values.

\subsubsection{Experimental Foundation}\

The foundational implementations of the kRR, OUE, and OLH protocols used in this study are derived from the Multi-Freq-LDPy package~\cite{DBLP:conf/esorics/ArcoleziCGPZ22}. This package provides efficient and well-validated implementations of LDP protocols, ensuring the accuracy and reproducibility of our experiments and enabling us to focus on the innovative aspects of our research.

\begin{figure}[t]
\begin{center}
\includegraphics[width=\linewidth]{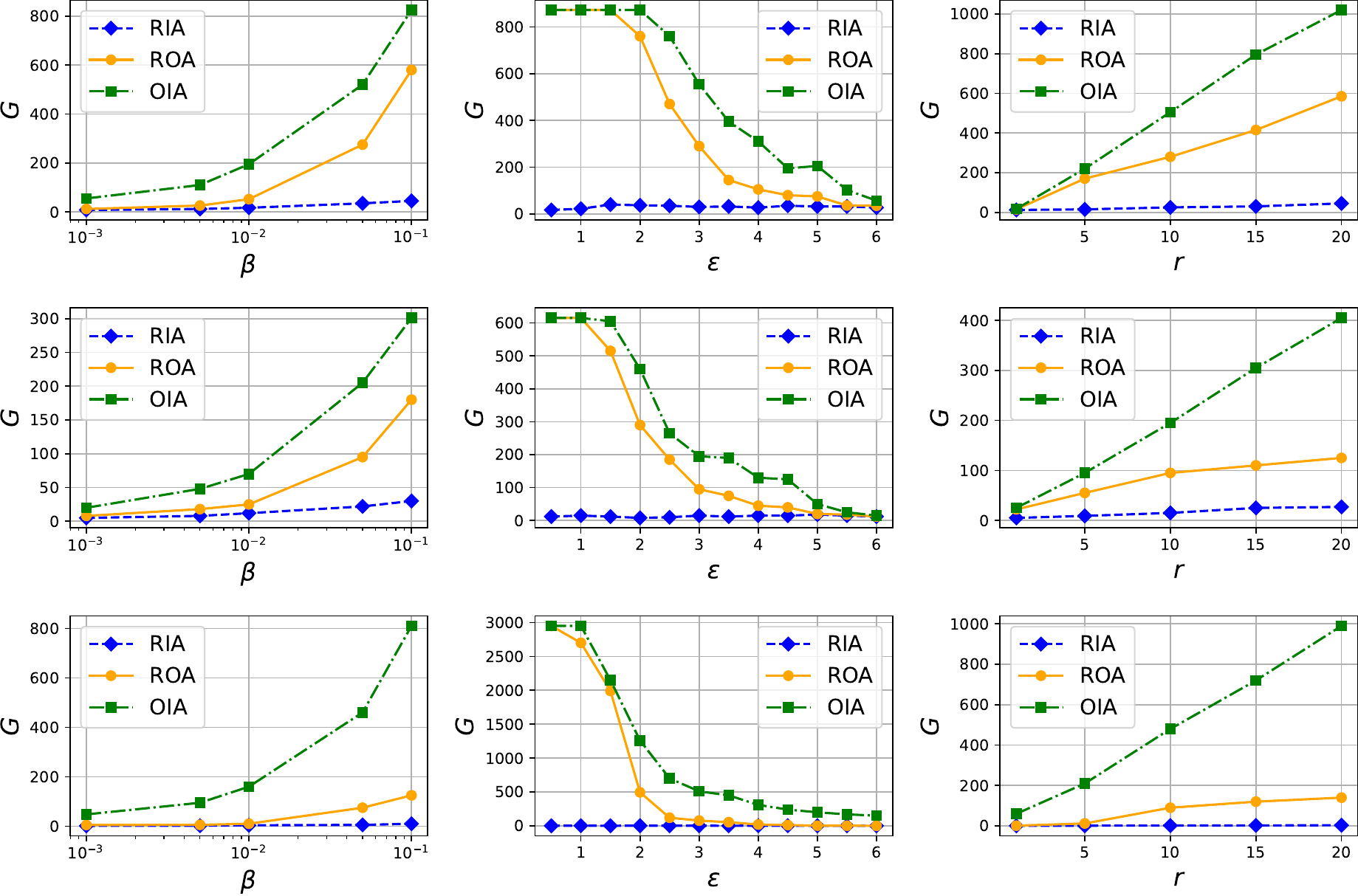}
\end{center}
\caption{\label{fig:OUE result} Impact of OIA with different parameters ($\beta$, $\epsilon$, $r$) on the overall gain for OUE under rank lowering scenario. The three rows are for Synthetic, Adult and Fire datasets, respectively. }
\end{figure}

\subsection{Results for Ranking Estimation}
\subsubsection{Rank Lowering Scenario}\ 

\begin{figure}[t]
\begin{center}
\includegraphics[width=\linewidth]{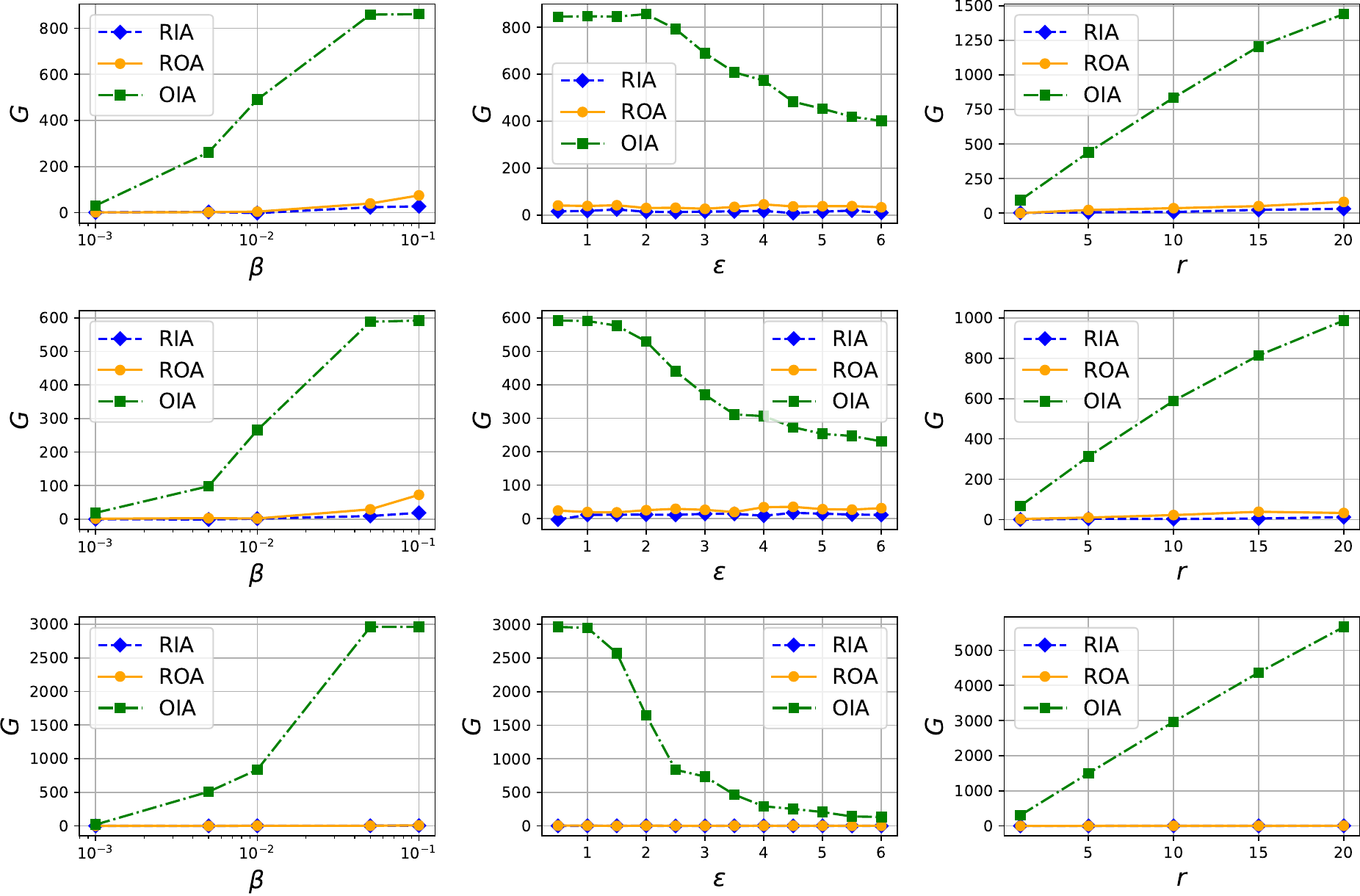}
\end{center}
\caption{\label{fig:OLH result} Impact of OIA with different parameters ($\beta$, $\epsilon$, $r$) on the overall gain for OLH under rank lowering scenario. The three rows are for Synthetic, Adult and Fire datasets, respectively. }
\end{figure}

Figures~\ref{fig:kRR result}--\ref{fig:OLH result} respectively show the impact of different parameters ($\beta$, $\epsilon$, $r$) on the overall gain $G$ under rank lowering scenario across the three datasets. We observe that OIA consistently outperforms ROA because OIA always selects the optimal attack item when crafting the perturbed values, whereas ROA randomly selects a non-target item as the perturbed value. Additionally, ROA generally outperforms RIA because RIA randomly selects a non-target item as the input value, and after perturbation, some of these values may be perturbed into target items, thereby increasing the frequency of the target items and reducing the attack effectiveness.

When attacking kRR and OUE, OIA and ROA achieve higher overall gain, while RIA generally results in a lower overall gain. In contrast, when attacking OLH, only OIA achieves a significant overall gain. This is because the perturbed values selected by ROA, although supporting some non-target items, may also inadvertently support certain target items, resulting in less effective attacks.

\noindent\textbf{Impact of $\beta$:} We increase the proportion of fake users $\beta$ from $10^{-3}$ to $10^{-1}$ to evaluate the impact of this parameter. We observe that for all three protocols, the overall gain of all three attacks increases to varying degrees as $\beta$ increases.

As discussed earlier, when attacking kRR and OUE, the overall gain of OIA and ROA shows a significant increase. However, when attacking OLH, only OIA maintains a substantial increase in gain.

\noindent\textbf{Impact of $\epsilon$:} We increase the privacy budget $\epsilon$ from 0.5 to 6, in increments of 0.5, to comprehensively evaluate the impact of this parameter. We observe that as $\epsilon$ increases, the overall gain of the three attacks gradually decreases (with the overall gain of RIA remaining consistently low). This occurs because the dataset's distribution is generally non-uniform; the smaller the $\epsilon$, the greater the degree of perturbation, and the dataset distribution after perturbation tends to approach a uniform distribution. In this scenario, the frequency difference between the target and non-target items is reduced, allowing fewer fake users to achieve a significant gain.

Specifically, when attacking the OUE protocol, if $\epsilon$ is small, it may satisfy $E_1 > s'$, making it easier to achieve maximum gain, which reduces the difference in effectiveness between ROA and OIA attacks.

\noindent\textbf{Impact of $r$:} We increase the number of target items $r$ from 1 to 20 to comprehensively evaluate the impact of this parameter. We observe that as $r$ increases, the overall gain of the three attacks gradually rises (with the increase in RIA consistently being small or nearly unchanged, and ROA showing similarly small or nearly unchanged growth when attacking OLH). This is because the more target items there are, the greater the number of potential effective attack items with smaller attack cost, thereby increasing the likelihood of achieving gains by injecting fewer fake users.

Consider an extreme case: suppose there is only one target item $t$. Among the effective attack items with a frequency lower than $t$, $a_i$ has the highest frequency, and its attack cost is $\delta_i = E[n_t'] - E[n_{a_i}']$. For the next effective attack item $a_{i+1}$, we have $\delta_{i+1} = E[n_t'] - E[n_{a_{i+1}}']$. It is clear that $\delta_i < \delta_{i+1} < \dots < \delta_s$, meaning that each subsequent round of attack requires injecting more fake users than the previous one. When the number of fake users is limited, the number of attack rounds decreases, resulting in a smaller overall gain.

\noindent\textbf{Security-privacy trade-off:} In this paper, we demonstrate that the OIA attack adheres to the security-privacy trade-off across all three protocols. This means that the smaller the privacy budget $\epsilon$ (i.e., stronger privacy), the larger the overall gain $G$ achieved by OIA (i.e., weaker security).

\subsubsection{Rank Elevation Scenario}\

\begin{figure}[t]
\begin{center}
\includegraphics[width=\linewidth]{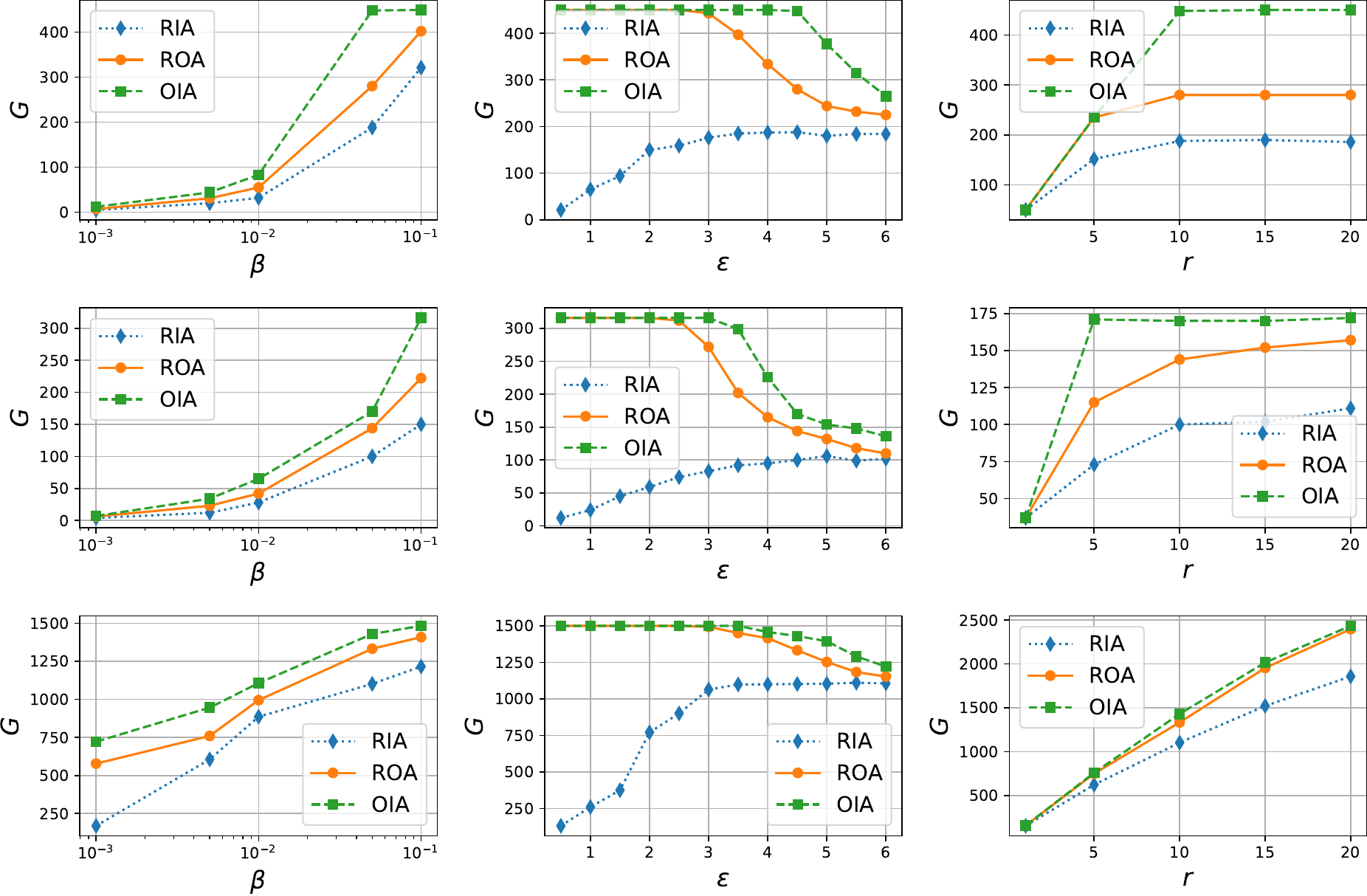}
\end{center}
\caption{\label{fig:rank_elevation} Impact of OIA with different parameters ($\beta$, $\epsilon$, $r$) on the overall gain for kRR under rank elevation scenario. The three rows are for Synthetic, Adult and Fire datasets, respectively. }
\end{figure}

In the rank elevation scenario, the attacker’s goal is to elevate the ranking of the target items. Since the number of target items is relatively small, this scenario is easier to attack compared to the rank lowering scenario. To further investigate whether OIA remains more effective than the two baseline attack methods, we take the attack on kRR as an example and test the impact of different parameters (\( \beta \), \( \epsilon \), \( r \)) on the overall gain \( G \) across
the three datasets, as shown in Figure~\ref{fig:rank_elevation}.  

\begin{figure}[t]
\begin{center}
\includegraphics[width=\linewidth]{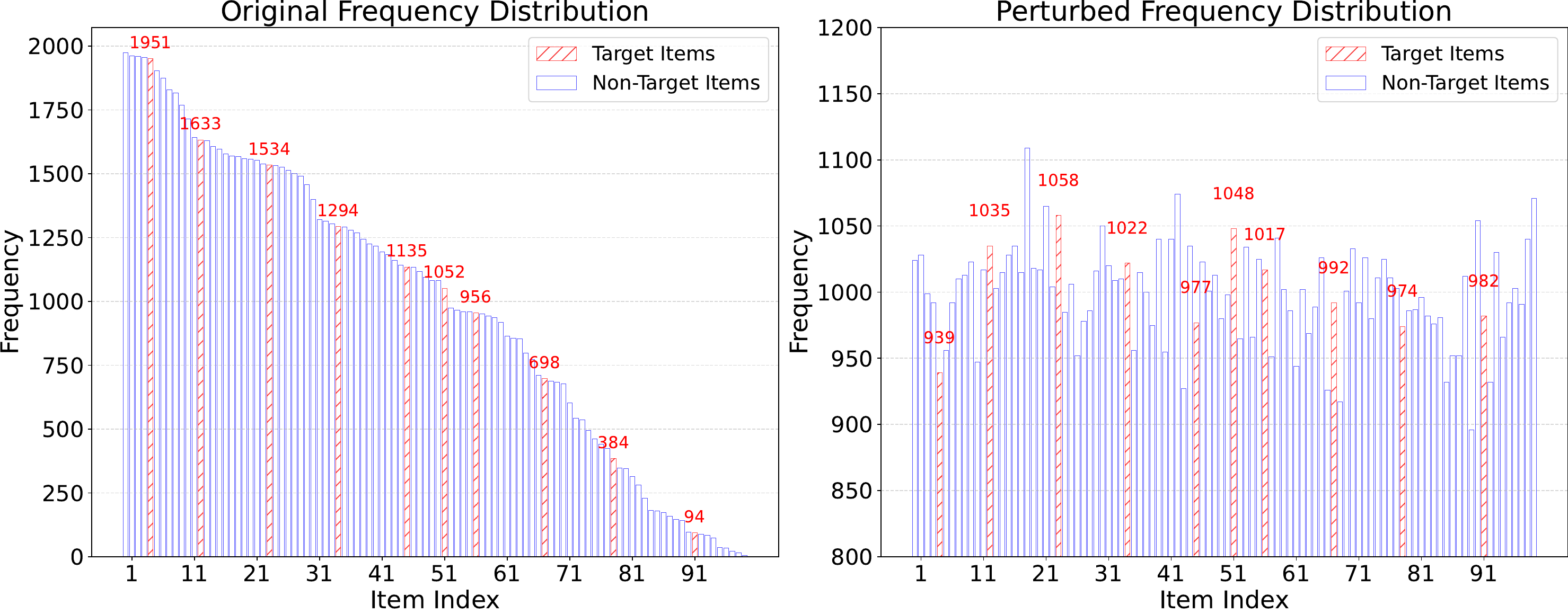}
\end{center}
\caption{\label{fig:fre_distribution} Frequency distribution before and after perturbation (synthetic dataset, $\epsilon=1$). }
\end{figure}
Overall, the trend of the overall gain \( G \) concerning parameter changes in the rank elevation scenario is similar to that in the rank lowering scenario, i.e., the attack effectiveness of OIA is better than that of ROA and RIA.
In addition, there exist two unique aspects of the rank elevation scenario:
\begin{enumerate}[leftmargin = *]
    \item \textbf{When $\epsilon$ takes small values (e.g., $\epsilon \leq 3$), the attack effectiveness of OIA and ROA becomes comparable:} This is because, as the perturbed data distribution approaches uniformity, the frequency differences between items become negligible. Additionally, the limited number of target items ($r=10$) allows all targets to achieve peak rankings regardless of the attack strategy. For example, as described in Figure~\ref{fig:fre_distribution}, when $\epsilon = 1$, the frequency distribution of the Synthetic dataset shifts from the original Zipf distribution to a more uniform distribution, and the target items show a mean frequency of approximately 1000, with a maximum frequency difference of $1109 - 1000 = 109$. With the default fake user ratio $\beta=5\%$ (yielding 5000 fake users), even uniform allocation among 10 targets provides about 500 increments per item, ensuring both OIA and ROA can elevate all target items to the highest rankings.
    \item \textbf{The overall gain of RIA increases with the increase of $\beta$, $\epsilon$, and $r$, but still always remains lower than that of ROA or OIA:} This is because when $\epsilon$ is small, the target items used as input values have a high probability of being perturbed into non-target items, resulting in poorer attack performance. As $\epsilon$ increases, the noise introduced by perturbation decreases, and the gain gradually increases and stabilizes, but it always remains lower than the gain brought by ROA or OIA. In addition, since $\epsilon$ is fixed at 4.5 when testing the variations of $\beta$ and $r$, the noise at this point is relatively small. Therefore, as $\beta$ or $r$ increases, the overall gain of RIA also increases. However, in the rank lowering scenario, due to the larger number of non-target items, the gain obtained by randomly assigning fake users is inherently low. As a result, even if the three parameters increase, the improvement in RIA's attack effectiveness is not significant.
\end{enumerate}

\subsection{Results for MPOIA}

The difference between MPOIA and OIA lies in the choice of the confidence level $1-\alpha$. Since OIA is known to outperform RIA and ROA in terms of attack effectiveness, we focus solely on comparing the performance of OIA and MPOIA in our experiments. Specifically, to explore the impact of varying confidence levels and privacy budgets on attack effectiveness, we evaluate the overall gain across three datasets as the confidence level changes from 0.5 to 0.99.

The experimental results demonstrate that the attack effectiveness of MPOIA and OIA remains consistently similar across different confidence levels. To better compare the differences in their attack effectiveness, we take OIA as the baseline and analyze the overall gain achieved by MPOIA relative to OIA. Even for the same dataset under the same protocol, the trend of MPOIA's overall gain varies with confidence levels under different privacy budgets. This observation aligns with our theoretical analysis, indicating that the overall gain does not exhibit a simple linear relationship with the confidence level. Therefore, using OIA (with a confidence level of 0.5) as our attack method is reasonable. Furthermore, this finding provides an important insight: the confidence level should generally not exceed 0.9. Due to space constraints, detailed experimental results are included in the full version.

\subsection{Attacking under Limited Knowledge}

\subsubsection{Attack Strategies with Limited Knowledge}\ 

\label{Limited Knowledge}
In the previous discussion, we assumed that the attacker had knowledge of the frequencies and proposed corresponding attack strategies. In this chapter, we will discuss how to adjust the attack strategies when the attacker's background knowledge is limited. Experimental validation shows that our attack strategies remain effective even when the attacker's background knowledge is restricted.

\noindent\textbf{Attack Based on Noisy Frequencies.}
In this scenario, the attacker no longer has direct access to the items' frequencies. Instead, the attacker learns the noisy frequencies that perhaps have been published by the aggregator previous statistics. When the privacy budget is low, the noisy frequencies deviate significantly from the true frequencies, making it challenging for the attacker to precisely estimate the frequency differences between items. Despite this, the fundamental attack strategy remains unchanged. The attacker can still apply the same methodology as in the full knowledge scenario, with the added noise only moderately diminishing the attack’s effectiveness.

Furthermore, we investigate how the accuracy of noisy frequency estimates affects attack effectiveness by systematically varying either: (1) the number of genuine users, or (2) the privacy budget during frequency estimation.
We conducted more in-depth tests from three perspectives:
1) Using only a portion of genuine user data for frequency estimation;  
2) Using different levels of noise for frequency estimation, where the noise level is determined by the privacy budget \(\epsilon'\) during estimation;  
3) Testing the impact of the privacy budget \(\epsilon\) during the attack on overall gain when \(\rho\) and \(\epsilon'\) are fixed.

\noindent\textbf{Attack Based on Ranking Knowledge.}
To further constrain the attacker's prior knowledge, we examine a restricted scenario where the attacker only possesses item rankings and cannot access either exact or noisy frequencies of individual items, relying exclusively on publicly available historical statistics to estimate the global frequency range~\cite{DBLP:journals/csur/RigakiG24} (i.e., the minimum and maximum frequency values).
Under this constraint, the attacker can only assume a uniform frequency distribution across all items, where the frequencies are equally spaced between the estimated bounds, and derive approximate values accordingly. This assumption leads to a simplified OIA attack strategy: in each attack round, the attacker randomly selects an item ranked immediately below the target as a candidate. However, this approach suffers from two inherent limitations. First, without precise frequency knowledge, the selected candidate may deviate from the truly optimal choice. Second, the uniform distribution assumption introduces systematic inaccuracies, as real-world frequency distributions rarely exhibit such perfect regularity. As a result, the attack effectiveness is inevitably degraded compared to cases where exact frequency information is available.

For estimating frequency bounds from historical rankings, attackers leverage public historical rankings to estimate extremal frequencies. Let upper bound of top-ranked items' frequencies be max frequency $f_{max}$, and let lower bound of bottom-ranked items be min frequency $f_{min}$.
For frequency construction and attack execution, with no further knowledge, attackers conservatively assume frequencies decrease linearly with rank:
\begin{equation}
f_i = f_{\max} - \frac{(i-1)(f_{\max} - f_{\min})}{d-1} \quad \text{for rank } i \in \{1,\dots,d\}.
\end{equation}
The candidate optimal attack items are always the immediate next-lower-ranked items, fake users randomly select from these candidates.

\begin{figure}[t]
\begin{center}
\includegraphics[width=\linewidth]{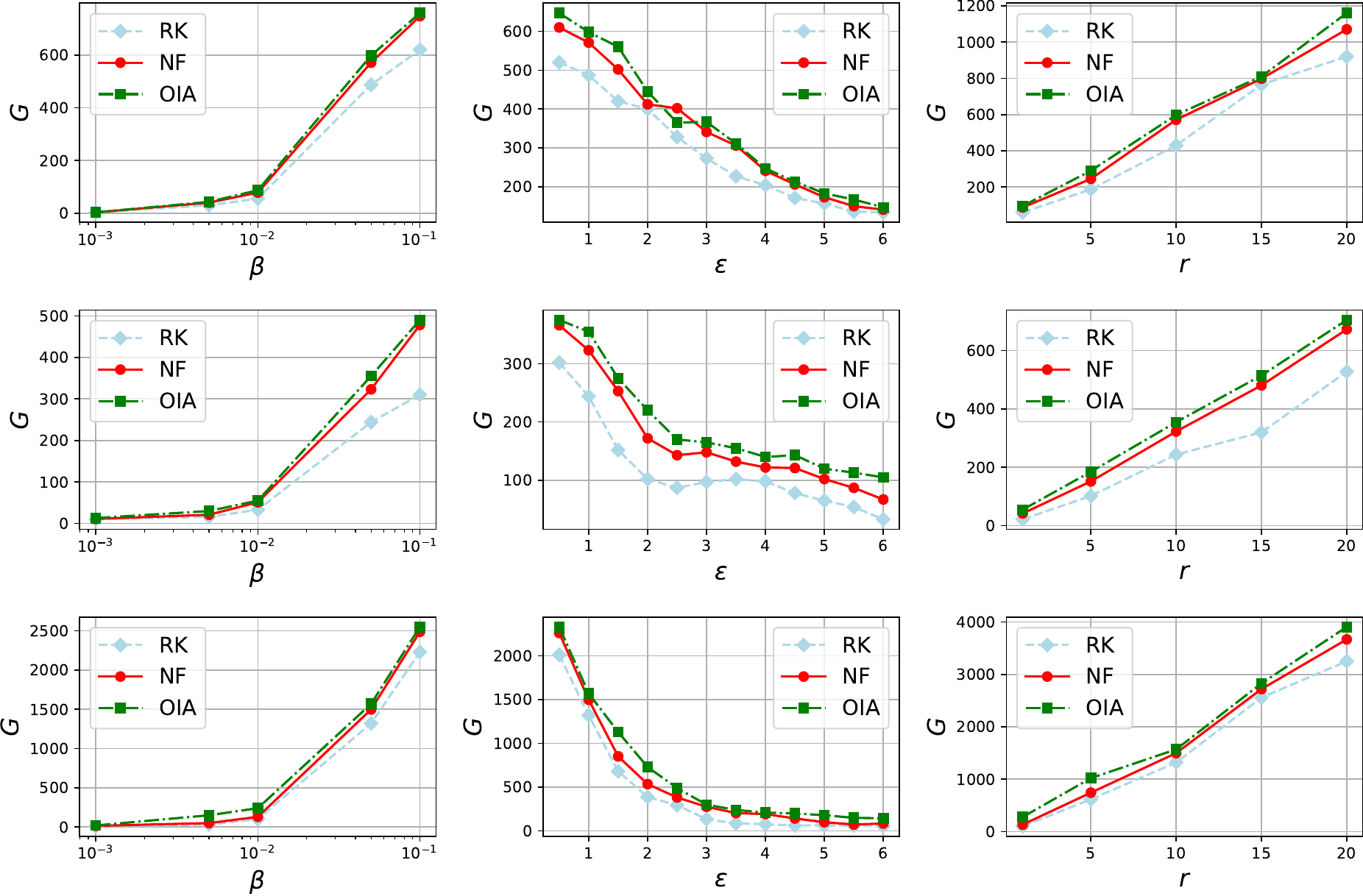}
\end{center}
\caption{\label{fig:limited_kRR} Attacking kRR under Limited Knowledge for RK, NF and OIA. The three rows are for Synthetic,
Adult and Fire datasets, respectively.}
\end{figure}

\subsubsection{Experimental Results}\

Figure~~\ref{fig:limited_kRR} illustrates the overall gain of attacking the kRR protocol under varying levels of background knowledge available to the attacker. Based on the amount of background knowledge, we compare three attack strategies: \textit{ranking knowledge-based attack (RK)}, \textit{noisy frequency-based attack (NF)}, and \textit{optimal item attack (OIA)}.
The results indicate that attack performance improves as the level of background knowledge increases; however, the differences in performance among the three attack strategies are relatively minor, with NF and OIA achieving nearly identical results. Specifically, regardless of the protocol being attacked or parameter variations, RK and NF achieve at least 80\% and 90\% of the performance of OIA, respectively. This demonstrates that even when the attacker has limited background knowledge, the proposed attack methods can still achieve strong performance across various scenarios. Additional experimental results on attacks against the OUE and OLH protocols are presented in Appendix~\ref{app:limit}.

Figure~\ref{fig:diff_noise} demonstrates the variation in attack effectiveness across different levels of noisy frequency estimation accuracy controlled by genuine user sampling ratio or the privacy budget.

\noindent\textbf{Impact of $\rho$:} We systematically vary the sampling ratio $\rho$ across the range $[0.01, 1]$ in 0.05 increments, while maintaining fixed privacy budgets ($\epsilon'=1$ for frequency estimation and $\epsilon=1$ for attack execution). For $\rho$, the larger the sampling ratio, the better the estimation accuracy, and the greater the overall gain. Experiments show that in multiple datasets, it is generally sufficient to sample only a small portion of genuine users to achieve good attack effectiveness (e.g., $\rho > 0.2$). 

\noindent\textbf{Impact of $\epsilon'$:} We vary the estimation privacy budget $\epsilon'$ over the range $[0.5, 6]$ in increments of 0.5, while maintaining a fixed sampling ratio ($\rho=1$) and attack privacy budget ($\epsilon=1$). For $\epsilon'$, the larger the $\epsilon'$, the smaller the amount of noise added during estimation, and the better the attack effectiveness of NF. Generally, when $\epsilon' \geq 2$, the attack effectiveness gradually approaches that of OIA and stabilizes. Even when the noise level is at its maximum ($\epsilon' = 0.5$), the attack effectiveness of NF is significantly better than that of random input/output attack algorithms across all datasets. 

\noindent\textbf{Impact of $\epsilon$:} We vary the attack privacy budget $\epsilon$ over the range $[0.5, 6]$ in increments of 0.5, while maintaining a fixed sampling ratio ($\rho=1$) and estimation privacy budget ($\epsilon'=1$). For $\epsilon$, as $\epsilon$ increases, the attack effectiveness of both OIA and NF decreases, with OIA consistently outperforming NF. This is consistent with our previous findings. However, in this case, we do not set $\epsilon' = \epsilon$, meaning that the privacy budgets during estimation and during the attack are different. Instead, we fix $\epsilon'$ at its default value of 1, maintaining a high noise level. Therefore, the attack effectiveness of NF is slightly reduced compared to the case where $\epsilon' = \epsilon$.

\begin{figure}[t]
\begin{center}
\includegraphics[width=\linewidth]{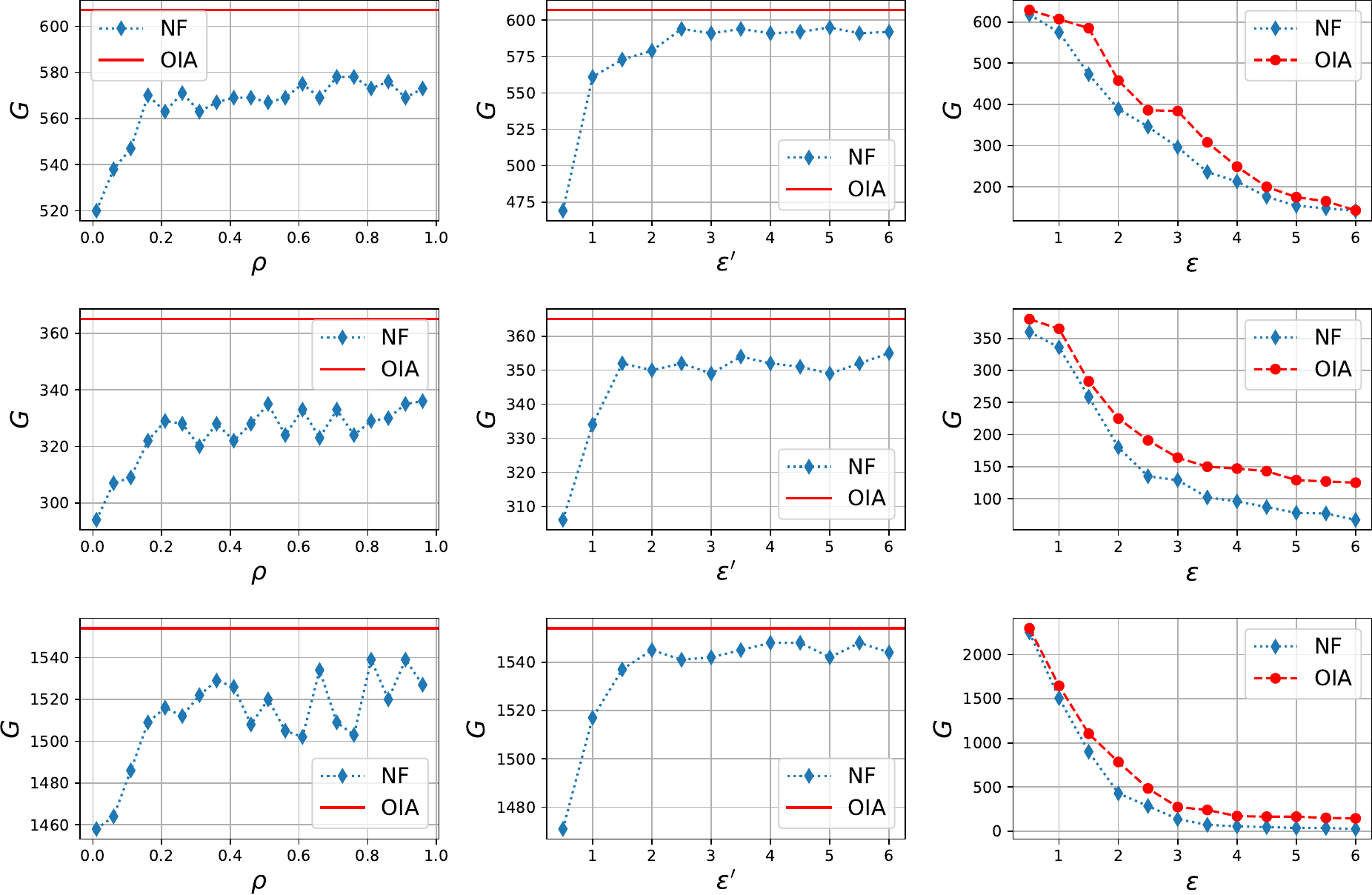}
\end{center}
\caption{\label{fig:diff_noise} Impact of NF with different parameters ($\rho$, $\epsilon'$, $\epsilon$) on the overall gain for kRR. The three rows are for Synthetic, Adult and Fire datasets, respectively.} 
\end{figure}

\section{Defenses} \label{sec:defense}
To explore possible defenses against our proposed attacks, we begin by evaluating two previously suggested strategies: normalization~\cite{DBLP:conf/uss/CaoJG21} and fake user detection \cite{DBLP:conf/uss/CaoJG21, DBLP:conf/uss/WuCJG22}. 
However, these traditional defenses either fail to counter our proposed ranking-based attacks (normalization) or perform suboptimally (fake user detection). Therefore, we adopt the state-of-the-art defense method against poisoning attacks on LDP frequency estimation protocols, \textbf{LDPRecover}~\cite{DBLP:conf/icde/Sun0HDW0Y24}, which can recover accurate aggregated frequencies despite poisoning attacks.
We also employ an enhanced version, \textbf{LDPRecover*}, to assess its effectiveness in mitigating our proposed attacks.

\subsection{Normalization and Fake User Detection}
In the normalization method, the aggregator identifies the minimal estimated item frequency $\tilde{f}_{\text{min}}$, and calibrates the estimated frequency for each item $v$ as $\bar{f}_v = \frac{\tilde{f}_v - \tilde{f}_{\text{min}}}{\sum_v (\tilde{f}_v - \tilde{f}_{\text{min}})}$, where $\bar{f}_v$ is the calibrated frequency. Normalization can mitigate attacks that increase target item frequencies but is ineffective against ranking attacks and heavy hitter identification.

Fake user detection identifies statistical anomalies in perturbed values, often using \textit{frequent itemset mining}. Taking OUE as an example, the server first identifies the itemsets where all bits are set to 1 in the perturbed binary vectors of a large number of users. If the number of users with these itemsets exceeds a predefined threshold, those users are considered fake users. This method can filter out some fake users to a certain extent, but in our scheme, the perturbed values selected in each iteration differ significantly, and the large number of iterations further reduces the method's effectiveness.

\subsection{LDPRecover and LDPRecover*}
\label{subsec:LDPRecover}

LDPRecover is a method designed to recover accurate aggregated frequencies after poisoning attacks, without requiring the server to know the specific details of the attacks. Experimental results~\cite{DBLP:conf/icde/Sun0HDW0Y24} demonstrate its significant effectiveness against poisoning attacks on frequency estimation protocols, particularly for MGA, across three popular LDP protocols: kRR, OUE, and OLH. The method first establishes a genuine frequency estimator to guide the server in recovering the true frequencies from the poisoned ones. It then introduces an adaptive attack, which leverages the statistical properties of the frequency estimation protocols to estimate the statistics of fake frequencies. Based on the genuine frequency estimator and the fake frequency statistics, the problem is formulated as a \textit{constraint inference (CI) problem} to recover the true frequencies.

Specifically, in our problem scenario, since the fake users only select non-target items and the server is unaware of any attack details, it can only treat the items with frequencies less than or equal to 0 in the poisoned frequencies as target items. We denote this set as $D_0$, and the remaining items form the set $D_1$. Let the observation value of item $v$ be $c_v$. For the kRR protocol, the server can only assume that $m$ fake users are evenly distributed across $|D_1|$ non-target items. Thus, the true frequency of target items can be estimated as: 
\begin{equation}
\tilde{n}_v^{0} = \frac{c_v - n q}{p - q},
\end{equation}
while the true frequency of the non-target items can be estimated as:
\begin{equation}
\tilde{n}_v^{1} = \frac{c_v - \frac{m}{|D_1|} - n q}{p - q}
\label{eq:defense_nv}
\end{equation}
For the OUE protocol, in our attack strategy, each fake user can support $E_1$ items. As a result, the original $m$ in Eq.\eqref{eq:defense_nv} becomes $E_1 \times m$. In the OLH protocol, due to the varying number of items that each pair of hash functions and hash values can support, the upper bound for the number of items that each fake user can support is $E_1$. We can estimate it based on the hash function properties. Finally, we use optimal solution techniques to solve the CI problem, achieving high-precision frequency recovery.

In addition, by detecting statistical anomalies in the historical frequency data of each item, the target items can be identified. In this case, LDPRecover* can achieve more accurate aggregated frequency recovery. Here, the server can explicitly distinguish between target and non-target items, and the size of $|D_1|$ becomes an exact value. Using Eq.~\eqref{eq:defense_nv}, the server can recover more precise true frequencies.

\subsection{Experimental Results}

We conducted experiments to evaluate the effectiveness of these two defense strategies and compared the results with the overall gain achieved by OIA. All parameters were set to their default values as described earlier, and the total number of items that fake users could support was adjusted according to the design in Section~\ref{subsec:LDPRecover} for different protocols. 

\begin{figure}[t]
\begin{center}
\includegraphics[width=\linewidth]{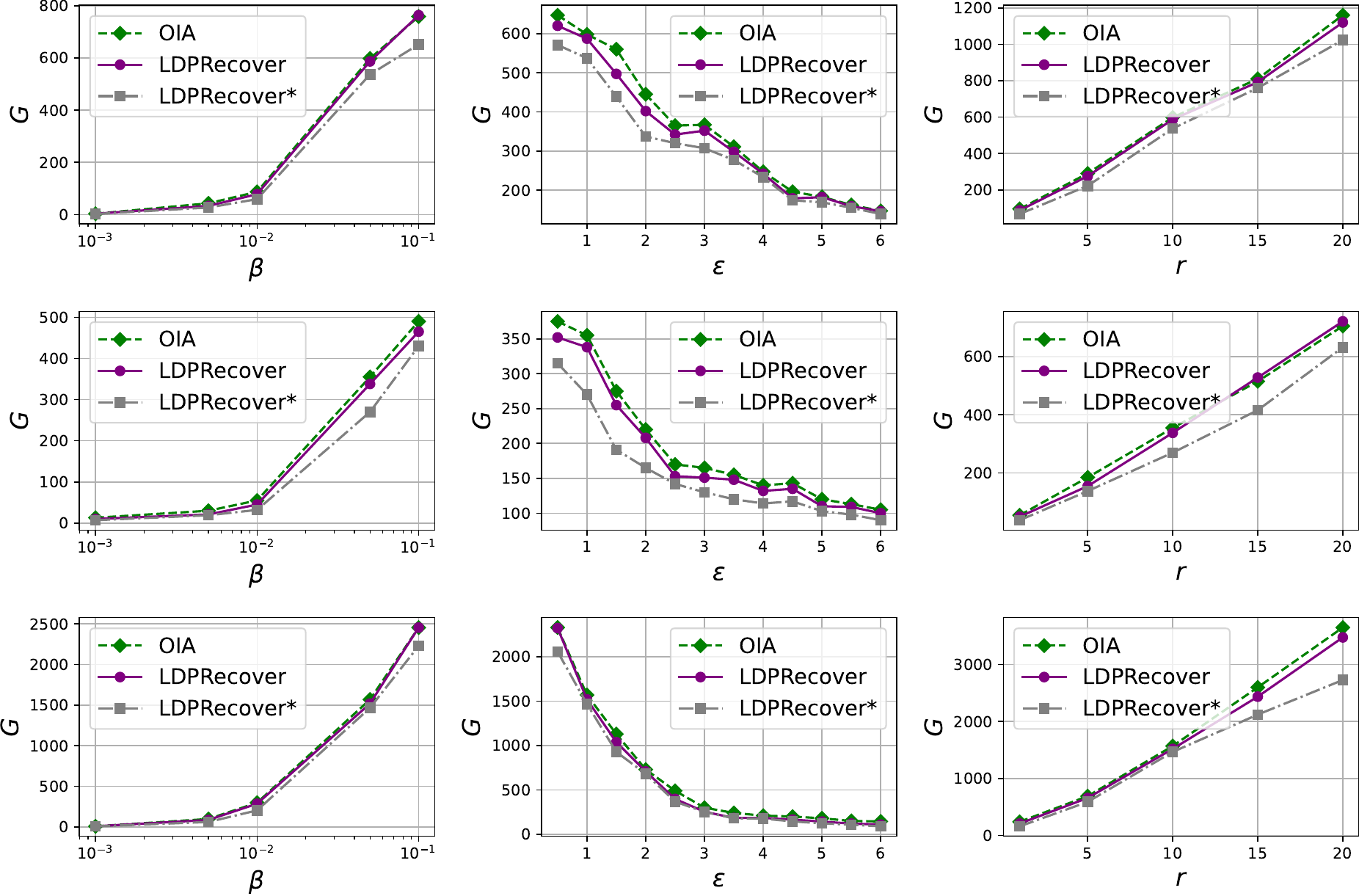}
\end{center}
\caption{\label{fig:defense_kRR} Impact of defense methods with different parameters ($\beta$, $\epsilon$, $r$) on the overall gain for kRR. The three rows are for Synthetic, Adult and Fire datasets, respectively.}
\end{figure}

Taking the kRR protocol as an example, Figure~\ref{fig:defense_kRR} shows that although these two defense strategies are effective against frequency attacks, their defense performance against OIA in ranking attack scenarios is insufficient. Specifically, LDPRecover can only slightly reduce the overall gain, mainly due to the following reasons:
\begin{itemize}[left=0pt, labelsep=4pt, itemsep=0pt, parsep=0pt, topsep=0pt]
    \item Our attack strategy iteratively selects the optimal attack items, which often have high support. Meanwhile, some other non-target items may not even be selected by fake users, creating a significant discrepancy with the assumptions of the defense and leading to inaccurate recovery.
    \item Frequency recovery may only narrow the frequency gap between target and non-target items, which is insufficient to restore the correct ranking of the items.
    \item Without any knowledge of the attack details, the target and non-target item sets identified by the server, as well as their sizes, are inaccurate.
\end{itemize}

Furthermore, even if the server has precise knowledge of the target item set, LDPRecover* can only partially reduce the overall gain. Our attack still causes significant damage to all protocols. This highlights the need for further research into defense mechanisms specifically designed to counter ranking attacks. Experimental results for the OUE and OLH protocols are provided in Appendix~\ref{app:defenses}.

\section{Conclusion} \label{conclusions}
In this paper, we conduct a comprehensive analysis of poisoning attacks on LDP protocols for ranking estimation, introducing the concepts of attack cost and optimal attack item (set) within the LDP framework. We propose tailored strategies for the kRR, OUE, and OLH protocols. For kRR, we meticulously select optimal attack items and allocate suitable fake users. In the case of OUE, we methodically determine optimal attack item sets, accounting for incremental changes in item frequencies across different sets. Concerning OLH, we devise scoring functions based on the pre-image of a hash to select those supporting a larger number of effective attack items. Additionally, we present an attack strategy based on confidence levels to precisely quantify the likelihood of a successful attack and the necessary number of attack iterations. Through theoretical and empirical evidence, we demonstrate the efficacy of our approaches, underscoring the critical need for robust defenses against such attacks.

\begin{acks}
We thank the reviewers for their valuable comments which significantly improved this paper. The work was supported by National Natural Science Foundation of China under Grant No. 62002203, No. 62072136, No. 62372268, Major Scientific and Technological Innovation Projects of Shandong Province, China No. 2024CXGC010114, Shandong Provincial Natural Science Foundation, China No. ZR2020QF045, No. ZR2022LZH013, No. ZR2021LZH007, Fundamental Research Funds for the Central Universities No. 3072020CFT2402, Young Scholars Program of Shandong University, Department of Science \&Technology of Shandong Province grant No. SYS202201, and Quan Cheng Laboratory grant No. QCLZD202302.
\end{acks}

\bibliographystyle{ACM-Reference-Format}
\bibliography{main}


\begin{thebibliography}{53}


\ifx \showCODEN    \undefined \def \showCODEN     #1{\unskip}     \fi
\ifx \showISBNx    \undefined \def \showISBNx     #1{\unskip}     \fi
\ifx \showISBNxiii \undefined \def \showISBNxiii  #1{\unskip}     \fi
\ifx \showISSN     \undefined \def \showISSN      #1{\unskip}     \fi
\ifx \showLCCN     \undefined \def \showLCCN      #1{\unskip}     \fi
\ifx \shownote     \undefined \def \shownote      #1{#1}          \fi
\ifx \showarticletitle \undefined \def \showarticletitle #1{#1}   \fi
\ifx \showURL      \undefined \def \showURL       {\relax}        \fi
\providecommand\bibfield[2]{#2}
\providecommand\bibinfo[2]{#2}
\providecommand\natexlab[1]{#1}
\providecommand\showeprint[2][]{arXiv:#2}

\bibitem[Acharya et~al\mbox{.}(2020)]%
        {DBLP:conf/icml/AcharyaBKRS20}
\bibfield{author}{\bibinfo{person}{Jayadev Acharya}, \bibinfo{person}{Kallista~A. Bonawitz}, \bibinfo{person}{Peter Kairouz}, \bibinfo{person}{Daniel Ramage}, {and} \bibinfo{person}{Ziteng Sun}.} \bibinfo{year}{2020}\natexlab{}.
\newblock \showarticletitle{Context Aware Local Differential Privacy}. In \bibinfo{booktitle}{\emph{{ICML}}} \emph{(\bibinfo{series}{Proceedings of Machine Learning Research}, Vol.~\bibinfo{volume}{119})}. \bibinfo{publisher}{{PMLR}}, \bibinfo{pages}{52--62}.
\newblock


\bibitem[Alfeld et~al\mbox{.}(2016)]%
        {DBLP:conf/aaai/AlfeldZB16}
\bibfield{author}{\bibinfo{person}{Scott Alfeld}, \bibinfo{person}{Xiaojin Zhu}, {and} \bibinfo{person}{Paul Barford}.} \bibinfo{year}{2016}\natexlab{}.
\newblock \showarticletitle{Data Poisoning Attacks against Autoregressive Models}. In \bibinfo{booktitle}{\emph{{AAAI}}}. \bibinfo{publisher}{{AAAI} Press}, \bibinfo{pages}{1452--1458}.
\newblock


\bibitem[Arcolezi et~al\mbox{.}(2022)]%
        {DBLP:conf/esorics/ArcoleziCGPZ22}
\bibfield{author}{\bibinfo{person}{H{\'{e}}ber~Hwang Arcolezi}, \bibinfo{person}{Jean{-}Fran{\c{c}}ois Couchot}, \bibinfo{person}{S{\'{e}}bastien Gambs}, \bibinfo{person}{Catuscia Palamidessi}, {and} \bibinfo{person}{Majid Zolfaghari}.} \bibinfo{year}{2022}\natexlab{}.
\newblock \showarticletitle{Multi-Freq-LDPy: Multiple Frequency Estimation Under Local Differential Privacy in Python}. In \bibinfo{booktitle}{\emph{{ESORICS} {(3)}}} \emph{(\bibinfo{series}{Lecture Notes in Computer Science}, Vol.~\bibinfo{volume}{13556})}. \bibinfo{publisher}{Springer}, \bibinfo{pages}{770--775}.
\newblock


\bibitem[Bagdasaryan et~al\mbox{.}(2020)]%
        {DBLP:conf/aistats/BagdasaryanVHES20}
\bibfield{author}{\bibinfo{person}{Eugene Bagdasaryan}, \bibinfo{person}{Andreas Veit}, \bibinfo{person}{Yiqing Hua}, \bibinfo{person}{Deborah Estrin}, {and} \bibinfo{person}{Vitaly Shmatikov}.} \bibinfo{year}{2020}\natexlab{}.
\newblock \showarticletitle{How To Backdoor Federated Learning}. In \bibinfo{booktitle}{\emph{{AISTATS}}} \emph{(\bibinfo{series}{Proceedings of Machine Learning Research}, Vol.~\bibinfo{volume}{108})}. \bibinfo{publisher}{{PMLR}}, \bibinfo{pages}{2938--2948}.
\newblock


\bibitem[Barroso et~al\mbox{.}(2020)]%
        {DBLP:journals/inffus/BarrosoSJRMGLVH20}
\bibfield{author}{\bibinfo{person}{Nuria~Rodr{\'{\i}}guez Barroso}, \bibinfo{person}{Goran Stipcich}, \bibinfo{person}{Daniel Jim{\'{e}}nez{-}L{\'{o}}pez}, \bibinfo{person}{Jos{\'{e}}~Antonio Ruiz{-}Mill{\'{a}}n}, \bibinfo{person}{Eugenio Mart{\'{\i}}nez{-}C{\'{a}}mara}, \bibinfo{person}{Gerardo Gonz{\'{a}}lez{-}Seco}, \bibinfo{person}{Mar{\'{\i}}a~Victoria Luz{\'{o}}n}, \bibinfo{person}{Miguel~Angel Veganzones}, {and} \bibinfo{person}{Francisco Herrera}.} \bibinfo{year}{2020}\natexlab{}.
\newblock \showarticletitle{Federated Learning and Differential Privacy: Software tools analysis, the Sherpa.ai {FL} framework and methodological guidelines for preserving data privacy}.
\newblock \bibinfo{journal}{\emph{Inf. Fusion}}  \bibinfo{volume}{64} (\bibinfo{year}{2020}), \bibinfo{pages}{270--292}.
\newblock


\bibitem[Bassily et~al\mbox{.}(2017)]%
        {DBLP:conf/nips/BassilyNST17}
\bibfield{author}{\bibinfo{person}{Raef Bassily}, \bibinfo{person}{Kobbi Nissim}, \bibinfo{person}{Uri Stemmer}, {and} \bibinfo{person}{Abhradeep~Guha Thakurta}.} \bibinfo{year}{2017}\natexlab{}.
\newblock \showarticletitle{Practical Locally Private Heavy Hitters}. In \bibinfo{booktitle}{\emph{{NIPS}}}. \bibinfo{pages}{2288--2296}.
\newblock


\bibitem[Bassily and Smith(2015)]%
        {DBLP:conf/stoc/BassilyS15}
\bibfield{author}{\bibinfo{person}{Raef Bassily} {and} \bibinfo{person}{Adam~D. Smith}.} \bibinfo{year}{2015}\natexlab{}.
\newblock \showarticletitle{Local, Private, Efficient Protocols for Succinct Histograms}. In \bibinfo{booktitle}{\emph{{STOC}}}. \bibinfo{publisher}{{ACM}}, \bibinfo{pages}{127--135}.
\newblock


\bibitem[Bhagoji et~al\mbox{.}(2019)]%
        {DBLP:conf/icml/BhagojiCMC19}
\bibfield{author}{\bibinfo{person}{Arjun~Nitin Bhagoji}, \bibinfo{person}{Supriyo Chakraborty}, \bibinfo{person}{Prateek Mittal}, {and} \bibinfo{person}{Seraphin~B. Calo}.} \bibinfo{year}{2019}\natexlab{}.
\newblock \showarticletitle{Analyzing Federated Learning through an Adversarial Lens}. In \bibinfo{booktitle}{\emph{{ICML}}} \emph{(\bibinfo{series}{Proceedings of Machine Learning Research}, Vol.~\bibinfo{volume}{97})}. \bibinfo{publisher}{{PMLR}}, \bibinfo{pages}{634--643}.
\newblock


\bibitem[Biggio et~al\mbox{.}(2012)]%
        {DBLP:conf/icml/BiggioNL12}
\bibfield{author}{\bibinfo{person}{Battista Biggio}, \bibinfo{person}{Blaine Nelson}, {and} \bibinfo{person}{Pavel Laskov}.} \bibinfo{year}{2012}\natexlab{}.
\newblock \showarticletitle{Poisoning Attacks against Support Vector Machines}. In \bibinfo{booktitle}{\emph{{ICML}}}. \bibinfo{publisher}{icml.cc / Omnipress}.
\newblock


\bibitem[Cao et~al\mbox{.}(2021)]%
        {DBLP:conf/uss/CaoJG21}
\bibfield{author}{\bibinfo{person}{Xiaoyu Cao}, \bibinfo{person}{Jinyuan Jia}, {and} \bibinfo{person}{Neil~Zhenqiang Gong}.} \bibinfo{year}{2021}\natexlab{}.
\newblock \showarticletitle{Data Poisoning Attacks to Local Differential Privacy Protocols}. In \bibinfo{booktitle}{\emph{{USENIX} Security Symposium}}. \bibinfo{publisher}{{USENIX} Association}, \bibinfo{pages}{947--964}.
\newblock


\bibitem[Carlini(2021)]%
        {DBLP:conf/uss/Carlini21}
\bibfield{author}{\bibinfo{person}{Nicholas Carlini}.} \bibinfo{year}{2021}\natexlab{}.
\newblock \showarticletitle{Poisoning the Unlabeled Dataset of Semi-Supervised Learning}. In \bibinfo{booktitle}{\emph{{USENIX} Security Symposium}}. \bibinfo{publisher}{{USENIX} Association}, \bibinfo{pages}{1577--1592}.
\newblock


\bibitem[Chen et~al\mbox{.}(2017)]%
        {DBLP:journals/corr/abs-1712-05526}
\bibfield{author}{\bibinfo{person}{Xinyun Chen}, \bibinfo{person}{Chang Liu}, \bibinfo{person}{Bo Li}, \bibinfo{person}{Kimberly Lu}, {and} \bibinfo{person}{Dawn Song}.} \bibinfo{year}{2017}\natexlab{}.
\newblock \showarticletitle{Targeted Backdoor Attacks on Deep Learning Systems Using Data Poisoning}.
\newblock \bibinfo{journal}{\emph{CoRR}}  \bibinfo{volume}{abs/1712.05526} (\bibinfo{year}{2017}).
\newblock


\bibitem[Cheu et~al\mbox{.}(2019)]%
        {DBLP:journals/corr/abs-1909-09630}
\bibfield{author}{\bibinfo{person}{Albert Cheu}, \bibinfo{person}{Adam~D. Smith}, {and} \bibinfo{person}{Jonathan~R. Ullman}.} \bibinfo{year}{2019}\natexlab{}.
\newblock \showarticletitle{Manipulation Attacks in Local Differential Privacy}.
\newblock \bibinfo{journal}{\emph{CoRR}}  \bibinfo{volume}{abs/1909.09630} (\bibinfo{year}{2019}).
\newblock


\bibitem[Differential Privacy~Team(2017)]%
        {abadi2017learning}
\bibfield{author}{\bibinfo{person}{Apple Differential Privacy~Team}.} \bibinfo{year}{2017}\natexlab{}.
\newblock \bibinfo{title}{Learning with privacy at scale}.
\newblock


\bibitem[Ding et~al\mbox{.}(2017)]%
        {ding2017collecting}
\bibfield{author}{\bibinfo{person}{Bolin Ding}, \bibinfo{person}{Janardhan Kulkarni}, {and} \bibinfo{person}{Sergey Yekhanin}.} \bibinfo{year}{2017}\natexlab{}.
\newblock \showarticletitle{Collecting Telemetry Data Privately}. In \bibinfo{booktitle}{\emph{{NIPS}}}. \bibinfo{pages}{3571--3580}.
\newblock


\bibitem[Duchi et~al\mbox{.}(2013)]%
        {DBLP:conf/allerton/DuchiJW13}
\bibfield{author}{\bibinfo{person}{John~C. Duchi}, \bibinfo{person}{Michael~I. Jordan}, {and} \bibinfo{person}{Martin~J. Wainwright}.} \bibinfo{year}{2013}\natexlab{}.
\newblock \showarticletitle{Local privacy and statistical minimax rates}. In \bibinfo{booktitle}{\emph{Allerton}}. \bibinfo{publisher}{{IEEE}}, \bibinfo{pages}{1592}.
\newblock


\bibitem[Dwork(2006)]%
        {dwork2006differential}
\bibfield{author}{\bibinfo{person}{Cynthia Dwork}.} \bibinfo{year}{2006}\natexlab{}.
\newblock \showarticletitle{Differential Privacy}. In \bibinfo{booktitle}{\emph{{ICALP} {(2)}}} \emph{(\bibinfo{series}{Lecture Notes in Computer Science}, Vol.~\bibinfo{volume}{4052})}. \bibinfo{publisher}{Springer}, \bibinfo{pages}{1--12}.
\newblock


\bibitem[Dwork et~al\mbox{.}(2006)]%
        {dwork2006calibrating}
\bibfield{author}{\bibinfo{person}{Cynthia Dwork}, \bibinfo{person}{Frank McSherry}, \bibinfo{person}{Kobbi Nissim}, {and} \bibinfo{person}{Adam~D. Smith}.} \bibinfo{year}{2006}\natexlab{}.
\newblock \showarticletitle{Calibrating Noise to Sensitivity in Private Data Analysis}. In \bibinfo{booktitle}{\emph{{TCC}}} \emph{(\bibinfo{series}{Lecture Notes in Computer Science}, Vol.~\bibinfo{volume}{3876})}. \bibinfo{publisher}{Springer}, \bibinfo{pages}{265--284}.
\newblock


\bibitem[Erlingsson et~al\mbox{.}(2014)]%
        {DBLP:conf/ccs/ErlingssonPK14}
\bibfield{author}{\bibinfo{person}{{\'{U}}lfar Erlingsson}, \bibinfo{person}{Vasyl Pihur}, {and} \bibinfo{person}{Aleksandra Korolova}.} \bibinfo{year}{2014}\natexlab{}.
\newblock \showarticletitle{{RAPPOR:} Randomized Aggregatable Privacy-Preserving Ordinal Response}. In \bibinfo{booktitle}{\emph{{CCS}}}. \bibinfo{publisher}{{ACM}}, \bibinfo{pages}{1054--1067}.
\newblock


\bibitem[Fang et~al\mbox{.}(2020a)]%
        {DBLP:conf/uss/FangCJG20}
\bibfield{author}{\bibinfo{person}{Minghong Fang}, \bibinfo{person}{Xiaoyu Cao}, \bibinfo{person}{Jinyuan Jia}, {and} \bibinfo{person}{Neil~Zhenqiang Gong}.} \bibinfo{year}{2020}\natexlab{a}.
\newblock \showarticletitle{Local Model Poisoning Attacks to Byzantine-Robust Federated Learning}. In \bibinfo{booktitle}{\emph{{USENIX} Security Symposium}}. \bibinfo{publisher}{{USENIX} Association}, \bibinfo{pages}{1605--1622}.
\newblock


\bibitem[Fang et~al\mbox{.}(2020b)]%
        {DBLP:conf/www/FangG020}
\bibfield{author}{\bibinfo{person}{Minghong Fang}, \bibinfo{person}{Neil~Zhenqiang Gong}, {and} \bibinfo{person}{Jia Liu}.} \bibinfo{year}{2020}\natexlab{b}.
\newblock \showarticletitle{Influence Function based Data Poisoning Attacks to Top-N Recommender Systems}. In \bibinfo{booktitle}{\emph{{WWW}}}. \bibinfo{publisher}{{ACM} / {IW3C2}}, \bibinfo{pages}{3019--3025}.
\newblock


\bibitem[Fang et~al\mbox{.}(2018)]%
        {DBLP:conf/acsac/FangYGL18}
\bibfield{author}{\bibinfo{person}{Minghong Fang}, \bibinfo{person}{Guolei Yang}, \bibinfo{person}{Neil~Zhenqiang Gong}, {and} \bibinfo{person}{Jia Liu}.} \bibinfo{year}{2018}\natexlab{}.
\newblock \showarticletitle{Poisoning Attacks to Graph-Based Recommender Systems}. In \bibinfo{booktitle}{\emph{{ACSAC}}}. \bibinfo{publisher}{{ACM}}, \bibinfo{pages}{381--392}.
\newblock


\bibitem[Gu et~al\mbox{.}(2017)]%
        {DBLP:journals/corr/abs-1708-06733}
\bibfield{author}{\bibinfo{person}{Tianyu Gu}, \bibinfo{person}{Brendan Dolan{-}Gavitt}, {and} \bibinfo{person}{Siddharth Garg}.} \bibinfo{year}{2017}\natexlab{}.
\newblock \showarticletitle{BadNets: Identifying Vulnerabilities in the Machine Learning Model Supply Chain}.
\newblock \bibinfo{journal}{\emph{CoRR}}  \bibinfo{volume}{abs/1708.06733} (\bibinfo{year}{2017}).
\newblock


\bibitem[Hassan et~al\mbox{.}(2020)]%
        {DBLP:journals/jpdc/HassanRC20}
\bibfield{author}{\bibinfo{person}{Muneeb~Ul Hassan}, \bibinfo{person}{Mubashir~Husain Rehmani}, {and} \bibinfo{person}{Jinjun Chen}.} \bibinfo{year}{2020}\natexlab{}.
\newblock \showarticletitle{Differential privacy in blockchain technology: {A} futuristic approach}.
\newblock \bibinfo{journal}{\emph{J. Parallel Distributed Comput.}}  \bibinfo{volume}{145} (\bibinfo{year}{2020}), \bibinfo{pages}{50--74}.
\newblock


\bibitem[Hidano et~al\mbox{.}(2020)]%
        {DBLP:journals/popets/HidanoMKKH20}
\bibfield{author}{\bibinfo{person}{Seira Hidano}, \bibinfo{person}{Takao Murakami}, \bibinfo{person}{Shuichi Katsumata}, \bibinfo{person}{Shinsaku Kiyomoto}, {and} \bibinfo{person}{Goichiro Hanaoka}.} \bibinfo{year}{2020}\natexlab{}.
\newblock \showarticletitle{Exposing Private User Behaviors of Collaborative Filtering via Model Inversion Techniques}.
\newblock \bibinfo{journal}{\emph{Proc. Priv. Enhancing Technol.}} \bibinfo{volume}{2020}, \bibinfo{number}{3} (\bibinfo{year}{2020}), \bibinfo{pages}{264--283}.
\newblock


\bibitem[Huang et~al\mbox{.}(2021)]%
        {DBLP:conf/ndss/HuangMGL0X21}
\bibfield{author}{\bibinfo{person}{Hai Huang}, \bibinfo{person}{Jiaming Mu}, \bibinfo{person}{Neil~Zhenqiang Gong}, \bibinfo{person}{Qi Li}, \bibinfo{person}{Bin Liu}, {and} \bibinfo{person}{Mingwei Xu}.} \bibinfo{year}{2021}\natexlab{}.
\newblock \showarticletitle{Data Poisoning Attacks to Deep Learning Based Recommender Systems}. In \bibinfo{booktitle}{\emph{{NDSS}}}. \bibinfo{publisher}{The Internet Society}.
\newblock


\bibitem[Jagielski et~al\mbox{.}(2018)]%
        {DBLP:conf/sp/JagielskiOBLNL18}
\bibfield{author}{\bibinfo{person}{Matthew Jagielski}, \bibinfo{person}{Alina Oprea}, \bibinfo{person}{Battista Biggio}, \bibinfo{person}{Chang Liu}, \bibinfo{person}{Cristina Nita{-}Rotaru}, {and} \bibinfo{person}{Bo Li}.} \bibinfo{year}{2018}\natexlab{}.
\newblock \showarticletitle{Manipulating Machine Learning: Poisoning Attacks and Countermeasures for Regression Learning}. In \bibinfo{booktitle}{\emph{{IEEE} Symposium on Security and Privacy}}. \bibinfo{publisher}{{IEEE} Computer Society}, \bibinfo{pages}{19--35}.
\newblock


\bibitem[Ji et~al\mbox{.}(2018)]%
        {DBLP:conf/ccs/JiZJLW18}
\bibfield{author}{\bibinfo{person}{Yujie Ji}, \bibinfo{person}{Xinyang Zhang}, \bibinfo{person}{Shouling Ji}, \bibinfo{person}{Xiapu Luo}, {and} \bibinfo{person}{Ting Wang}.} \bibinfo{year}{2018}\natexlab{}.
\newblock \showarticletitle{Model-Reuse Attacks on Deep Learning Systems}. In \bibinfo{booktitle}{\emph{{CCS}}}. \bibinfo{publisher}{{ACM}}, \bibinfo{pages}{349--363}.
\newblock


\bibitem[Ji et~al\mbox{.}(2017)]%
        {DBLP:conf/cns/JiZW17}
\bibfield{author}{\bibinfo{person}{Yujie Ji}, \bibinfo{person}{Xinyang Zhang}, {and} \bibinfo{person}{Ting Wang}.} \bibinfo{year}{2017}\natexlab{}.
\newblock \showarticletitle{Backdoor attacks against learning systems}. In \bibinfo{booktitle}{\emph{{CNS}}}. \bibinfo{publisher}{{IEEE}}, \bibinfo{pages}{1--9}.
\newblock


\bibitem[Jia et~al\mbox{.}(2021)]%
        {DBLP:conf/aaai/JiaCG21}
\bibfield{author}{\bibinfo{person}{Jinyuan Jia}, \bibinfo{person}{Xiaoyu Cao}, {and} \bibinfo{person}{Neil~Zhenqiang Gong}.} \bibinfo{year}{2021}\natexlab{}.
\newblock \showarticletitle{Intrinsic Certified Robustness of Bagging against Data Poisoning Attacks}. In \bibinfo{booktitle}{\emph{{AAAI}}}. \bibinfo{publisher}{{AAAI} Press}, \bibinfo{pages}{7961--7969}.
\newblock


\bibitem[Joseph et~al\mbox{.}(2020)]%
        {DBLP:journals/jpc/JosephRUW20}
\bibfield{author}{\bibinfo{person}{Matthew Joseph}, \bibinfo{person}{Aaron Roth}, \bibinfo{person}{Jonathan~R. Ullman}, {and} \bibinfo{person}{Bo Waggoner}.} \bibinfo{year}{2020}\natexlab{}.
\newblock \showarticletitle{Local Differential Privacy for Evolving Data}.
\newblock \bibinfo{journal}{\emph{J. Priv. Confidentiality}} \bibinfo{volume}{10}, \bibinfo{number}{1} (\bibinfo{year}{2020}).
\newblock


\bibitem[Kermani et~al\mbox{.}(2015)]%
        {DBLP:journals/titb/KermaniSRJ15}
\bibfield{author}{\bibinfo{person}{Mehran~Mozaffari Kermani}, \bibinfo{person}{Susmita Sur{-}Kolay}, \bibinfo{person}{Anand Raghunathan}, {and} \bibinfo{person}{Niraj~K. Jha}.} \bibinfo{year}{2015}\natexlab{}.
\newblock \showarticletitle{Systematic Poisoning Attacks on and Defenses for Machine Learning in Healthcare}.
\newblock \bibinfo{journal}{\emph{{IEEE} J. Biomed. Health Informatics}} \bibinfo{volume}{19}, \bibinfo{number}{6} (\bibinfo{year}{2015}), \bibinfo{pages}{1893--1905}.
\newblock


\bibitem[Li et~al\mbox{.}(2016)]%
        {DBLP:conf/nips/LiWSV16}
\bibfield{author}{\bibinfo{person}{Bo Li}, \bibinfo{person}{Yining Wang}, \bibinfo{person}{Aarti Singh}, {and} \bibinfo{person}{Yevgeniy Vorobeychik}.} \bibinfo{year}{2016}\natexlab{}.
\newblock \showarticletitle{Data Poisoning Attacks on Factorization-Based Collaborative Filtering}. In \bibinfo{booktitle}{\emph{{NIPS}}}. \bibinfo{pages}{1885--1893}.
\newblock


\bibitem[Li et~al\mbox{.}(2023)]%
        {DBLP:conf/uss/LiLSG023}
\bibfield{author}{\bibinfo{person}{Xiaoguang Li}, \bibinfo{person}{Ninghui Li}, \bibinfo{person}{Wenhai Sun}, \bibinfo{person}{Neil~Zhenqiang Gong}, {and} \bibinfo{person}{Hui Li}.} \bibinfo{year}{2023}\natexlab{}.
\newblock \showarticletitle{Fine-grained Poisoning Attack to Local Differential Privacy Protocols for Mean and Variance Estimation}. In \bibinfo{booktitle}{\emph{{USENIX} Security Symposium}}. \bibinfo{publisher}{{USENIX} Association}, \bibinfo{pages}{1739--1756}.
\newblock


\bibitem[Li et~al\mbox{.}(2024)]%
        {DBLP:journals/corr/abs-2403-19510}
\bibfield{author}{\bibinfo{person}{Xiaoguang Li}, \bibinfo{person}{Zitao Li}, \bibinfo{person}{Ninghui Li}, {and} \bibinfo{person}{Wenhai Sun}.} \bibinfo{year}{2024}\natexlab{}.
\newblock \showarticletitle{On the Robustness of {LDP} Protocols for Numerical Attributes under Data Poisoning Attacks}.
\newblock \bibinfo{journal}{\emph{CoRR}}  \bibinfo{volume}{abs/2403.19510} (\bibinfo{year}{2024}).
\newblock


\bibitem[Liu et~al\mbox{.}(2018)]%
        {DBLP:conf/ndss/LiuMALZW018}
\bibfield{author}{\bibinfo{person}{Yingqi Liu}, \bibinfo{person}{Shiqing Ma}, \bibinfo{person}{Yousra Aafer}, \bibinfo{person}{Wen{-}Chuan Lee}, \bibinfo{person}{Juan Zhai}, \bibinfo{person}{Weihang Wang}, {and} \bibinfo{person}{Xiangyu Zhang}.} \bibinfo{year}{2018}\natexlab{}.
\newblock \showarticletitle{Trojaning Attack on Neural Networks}. In \bibinfo{booktitle}{\emph{{NDSS}}}. \bibinfo{publisher}{The Internet Society}.
\newblock


\bibitem[Mei and Zhu(2015)]%
        {DBLP:conf/aaai/MeiZ15}
\bibfield{author}{\bibinfo{person}{Shike Mei} {and} \bibinfo{person}{Xiaojin Zhu}.} \bibinfo{year}{2015}\natexlab{}.
\newblock \showarticletitle{Using Machine Teaching to Identify Optimal Training-Set Attacks on Machine Learners}. In \bibinfo{booktitle}{\emph{{AAAI}}}. \bibinfo{publisher}{{AAAI} Press}, \bibinfo{pages}{2871--2877}.
\newblock


\bibitem[Mu{\~{n}}oz{-}Gonz{\'{a}}lez et~al\mbox{.}(2017)]%
        {DBLP:conf/ccs/Munoz-GonzalezB17}
\bibfield{author}{\bibinfo{person}{Luis Mu{\~{n}}oz{-}Gonz{\'{a}}lez}, \bibinfo{person}{Battista Biggio}, \bibinfo{person}{Ambra Demontis}, \bibinfo{person}{Andrea Paudice}, \bibinfo{person}{Vasin Wongrassamee}, \bibinfo{person}{Emil~C. Lupu}, {and} \bibinfo{person}{Fabio Roli}.} \bibinfo{year}{2017}\natexlab{}.
\newblock \showarticletitle{Towards Poisoning of Deep Learning Algorithms with Back-gradient Optimization}. In \bibinfo{booktitle}{\emph{AISec@CCS}}. \bibinfo{publisher}{{ACM}}, \bibinfo{pages}{27--38}.
\newblock


\bibitem[Nelson et~al\mbox{.}(2008)]%
        {DBLP:conf/nsdi/NelsonBCJRSSTX08}
\bibfield{author}{\bibinfo{person}{Blaine Nelson}, \bibinfo{person}{Marco Barreno}, \bibinfo{person}{Fuching~Jack Chi}, \bibinfo{person}{Anthony~D. Joseph}, \bibinfo{person}{Benjamin I.~P. Rubinstein}, \bibinfo{person}{Udam Saini}, \bibinfo{person}{Charles Sutton}, \bibinfo{person}{J.~Doug Tygar}, {and} \bibinfo{person}{Kai Xia}.} \bibinfo{year}{2008}\natexlab{}.
\newblock \showarticletitle{Exploiting Machine Learning to Subvert Your Spam Filter}. In \bibinfo{booktitle}{\emph{{LEET}}}. \bibinfo{publisher}{{USENIX} Association}.
\newblock


\bibitem[Newell et~al\mbox{.}(2014)]%
        {DBLP:conf/ccs/NewellPXN14}
\bibfield{author}{\bibinfo{person}{Andrew Newell}, \bibinfo{person}{Rahul Potharaju}, \bibinfo{person}{Luojie Xiang}, {and} \bibinfo{person}{Cristina Nita{-}Rotaru}.} \bibinfo{year}{2014}\natexlab{}.
\newblock \showarticletitle{On the Practicality of Integrity Attacks on Document-Level Sentiment Analysis}. In \bibinfo{booktitle}{\emph{AISec@CCS}}. \bibinfo{publisher}{{ACM}}, \bibinfo{pages}{83--93}.
\newblock


\bibitem[Ren et~al\mbox{.}(2016)]%
        {DBLP:journals/sensors/RenLLHDZ16}
\bibfield{author}{\bibinfo{person}{Hao Ren}, \bibinfo{person}{Hongwei Li}, \bibinfo{person}{Xiaohui Liang}, \bibinfo{person}{Shibo He}, \bibinfo{person}{Yuanshun Dai}, {and} \bibinfo{person}{Lian Zhao}.} \bibinfo{year}{2016}\natexlab{}.
\newblock \showarticletitle{Privacy-Enhanced and Multifunctional Health Data Aggregation under Differential Privacy Guarantees}.
\newblock \bibinfo{journal}{\emph{Sensors}} \bibinfo{volume}{16}, \bibinfo{number}{9} (\bibinfo{year}{2016}), \bibinfo{pages}{1463}.
\newblock


\bibitem[Rigaki and Garc{\'{\i}}a(2024)]%
        {DBLP:journals/csur/RigakiG24}
\bibfield{author}{\bibinfo{person}{Maria Rigaki} {and} \bibinfo{person}{Sebasti{\'{a}}n Garc{\'{\i}}a}.} \bibinfo{year}{2024}\natexlab{}.
\newblock \showarticletitle{A Survey of Privacy Attacks in Machine Learning}.
\newblock \bibinfo{journal}{\emph{{ACM} Comput. Surv.}} \bibinfo{volume}{56}, \bibinfo{number}{4} (\bibinfo{year}{2024}), \bibinfo{pages}{101:1--101:34}.
\newblock


\bibitem[Rubinstein et~al\mbox{.}(2009)]%
        {DBLP:conf/imc/RubinsteinNHJLRTT09}
\bibfield{author}{\bibinfo{person}{Benjamin I.~P. Rubinstein}, \bibinfo{person}{Blaine Nelson}, \bibinfo{person}{Ling Huang}, \bibinfo{person}{Anthony~D. Joseph}, \bibinfo{person}{Shing{-}hon Lau}, \bibinfo{person}{Satish Rao}, \bibinfo{person}{Nina Taft}, {and} \bibinfo{person}{J.~D. Tygar}.} \bibinfo{year}{2009}\natexlab{}.
\newblock \showarticletitle{{ANTIDOTE:} understanding and defending against poisoning of anomaly detectors}. In \bibinfo{booktitle}{\emph{Internet Measurement Conference}}. \bibinfo{publisher}{{ACM}}, \bibinfo{pages}{1--14}.
\newblock


\bibitem[Shafahi et~al\mbox{.}(2018)]%
        {DBLP:conf/nips/ShafahiHNSSDG18}
\bibfield{author}{\bibinfo{person}{Ali Shafahi}, \bibinfo{person}{W.~Ronny Huang}, \bibinfo{person}{Mahyar Najibi}, \bibinfo{person}{Octavian Suciu}, \bibinfo{person}{Christoph Studer}, \bibinfo{person}{Tudor Dumitras}, {and} \bibinfo{person}{Tom Goldstein}.} \bibinfo{year}{2018}\natexlab{}.
\newblock \showarticletitle{Poison Frogs! Targeted Clean-Label Poisoning Attacks on Neural Networks}. In \bibinfo{booktitle}{\emph{NeurIPS}}. \bibinfo{pages}{6106--6116}.
\newblock


\bibitem[Song et~al\mbox{.}(2024)]%
        {DBLP:journals/jksucis/SongSZHXWZ24}
\bibfield{author}{\bibinfo{person}{Haina Song}, \bibinfo{person}{Hua Shen}, \bibinfo{person}{Nan Zhao}, \bibinfo{person}{Zhangqing He}, \bibinfo{person}{Wei Xiong}, \bibinfo{person}{Minghu Wu}, {and} \bibinfo{person}{Mingwu Zhang}.} \bibinfo{year}{2024}\natexlab{}.
\newblock \showarticletitle{Adaptive personalized privacy-preserving data collection scheme with local differential privacy}.
\newblock \bibinfo{journal}{\emph{J. King Saud Univ. Comput. Inf. Sci.}} \bibinfo{volume}{36}, \bibinfo{number}{5} (\bibinfo{year}{2024}), \bibinfo{pages}{102042}.
\newblock


\bibitem[Song et~al\mbox{.}(2023)]%
        {DBLP:journals/tifs/SongXZ23}
\bibfield{author}{\bibinfo{person}{Shaorui Song}, \bibinfo{person}{Lei Xu}, {and} \bibinfo{person}{Liehuang Zhu}.} \bibinfo{year}{2023}\natexlab{}.
\newblock \showarticletitle{Efficient Defenses Against Output Poisoning Attacks on Local Differential Privacy}.
\newblock \bibinfo{journal}{\emph{{IEEE} Trans. Inf. Forensics Secur.}}  \bibinfo{volume}{18} (\bibinfo{year}{2023}), \bibinfo{pages}{5506--5521}.
\newblock


\bibitem[Sun et~al\mbox{.}(2024)]%
        {DBLP:conf/icde/Sun0HDW0Y24}
\bibfield{author}{\bibinfo{person}{Xinyue Sun}, \bibinfo{person}{Qingqing Ye}, \bibinfo{person}{Haibo Hu}, \bibinfo{person}{Jiawei Duan}, \bibinfo{person}{Tianyu Wo}, \bibinfo{person}{Jie Xu}, {and} \bibinfo{person}{Renyu Yang}.} \bibinfo{year}{2024}\natexlab{}.
\newblock \showarticletitle{LDPRecover: Recovering Frequencies from Poisoning Attacks Against Local Differential Privacy}. In \bibinfo{booktitle}{\emph{{ICDE}}}. \bibinfo{publisher}{{IEEE}}, \bibinfo{pages}{1619--1631}.
\newblock


\bibitem[Thomas et~al\mbox{.}(2013)]%
        {DBLP:conf/uss/ThomasMGKP13}
\bibfield{author}{\bibinfo{person}{Kurt Thomas}, \bibinfo{person}{Damon McCoy}, \bibinfo{person}{Chris Grier}, \bibinfo{person}{Alek Kolcz}, {and} \bibinfo{person}{Vern Paxson}.} \bibinfo{year}{2013}\natexlab{}.
\newblock \showarticletitle{Trafficking Fraudulent Accounts: The Role of the Underground Market in Twitter Spam and Abuse}. In \bibinfo{booktitle}{\emph{{USENIX} Security Symposium}}. \bibinfo{publisher}{{USENIX} Association}, \bibinfo{pages}{195--210}.
\newblock


\bibitem[Wang et~al\mbox{.}(2014)]%
        {DBLP:conf/uss/WangWZZ14}
\bibfield{author}{\bibinfo{person}{Gang Wang}, \bibinfo{person}{Tianyi Wang}, \bibinfo{person}{Haitao Zheng}, {and} \bibinfo{person}{Ben~Y. Zhao}.} \bibinfo{year}{2014}\natexlab{}.
\newblock \showarticletitle{Man vs. Machine: Practical Adversarial Detection of Malicious Crowdsourcing Workers}. In \bibinfo{booktitle}{\emph{{USENIX} Security Symposium}}. \bibinfo{publisher}{{USENIX} Association}, \bibinfo{pages}{239--254}.
\newblock


\bibitem[Wang et~al\mbox{.}(2017)]%
        {DBLP:conf/uss/WangBLJ17}
\bibfield{author}{\bibinfo{person}{Tianhao Wang}, \bibinfo{person}{Jeremiah Blocki}, \bibinfo{person}{Ninghui Li}, {and} \bibinfo{person}{Somesh Jha}.} \bibinfo{year}{2017}\natexlab{}.
\newblock \showarticletitle{Locally Differentially Private Protocols for Frequency Estimation}. In \bibinfo{booktitle}{\emph{{USENIX} Security Symposium}}. \bibinfo{publisher}{{USENIX} Association}, \bibinfo{pages}{729--745}.
\newblock


\bibitem[Wang et~al\mbox{.}(2021)]%
        {DBLP:journals/tdsc/0001LJ21}
\bibfield{author}{\bibinfo{person}{Tianhao Wang}, \bibinfo{person}{Ninghui Li}, {and} \bibinfo{person}{Somesh Jha}.} \bibinfo{year}{2021}\natexlab{}.
\newblock \showarticletitle{Locally Differentially Private Heavy Hitter Identification}.
\newblock \bibinfo{journal}{\emph{{IEEE} Trans. Dependable Secur. Comput.}} \bibinfo{volume}{18}, \bibinfo{number}{2} (\bibinfo{year}{2021}), \bibinfo{pages}{982--993}.
\newblock


\bibitem[Wu et~al\mbox{.}(2022)]%
        {DBLP:conf/uss/WuCJG22}
\bibfield{author}{\bibinfo{person}{Yongji Wu}, \bibinfo{person}{Xiaoyu Cao}, \bibinfo{person}{Jinyuan Jia}, {and} \bibinfo{person}{Neil~Zhenqiang Gong}.} \bibinfo{year}{2022}\natexlab{}.
\newblock \showarticletitle{Poisoning Attacks to Local Differential Privacy Protocols for Key-Value Data}. In \bibinfo{booktitle}{\emph{{USENIX} Security Symposium}}. \bibinfo{publisher}{{USENIX} Association}, \bibinfo{pages}{519--536}.
\newblock


\bibitem[Yang et~al\mbox{.}(2017)]%
        {DBLP:conf/ndss/YangGC17}
\bibfield{author}{\bibinfo{person}{Guolei Yang}, \bibinfo{person}{Neil~Zhenqiang Gong}, {and} \bibinfo{person}{Ying Cai}.} \bibinfo{year}{2017}\natexlab{}.
\newblock \showarticletitle{Fake Co-visitation Injection Attacks to Recommender Systems}. In \bibinfo{booktitle}{\emph{{NDSS}}}. \bibinfo{publisher}{The Internet Society}.
\newblock


\end{thebibliography}


\appendix

\section{Detailed discussion of OIA is Near-optimal}
To prove OIA is near-optimal, we adopt a four-phase methodological approach: (1) presenting key observations, (2) proposing a solution based on these observations, (3) theoretically validating the solution's correctness, which guarantees obtaining the optimal solution, and (4) introducing computational approximations to reduce algorithmic complexity and get our OIA approach.

\textbf{(1) Presenting key observations.}
For ranking attacks (especially ranking lowering), we establish the following critical observations:
\begin{itemize}
\item
For a target item $t_j$ with multiple consecutively ranked effective attack items ${a_{i},..., a_{i+k}}$, increasing the frequency of the immediately adjacent attack item $a_{i}$ proves more effective for demoting ${t_j}$'s ranking than modifying other attack items. We therefore designate $a_{i}$ as the candidate optimal attack item. This principle generalizes to all $r$ target items, where each $t_j$ has a corresponding candidate optimal attack item. This can reduce the selection space for optimal attack items from all effective attacks to these r candidates during each iteration. For example, as shown in Figure~\ref{fig:comment2-1}, for the target item $t_1$, there are consecutively ranked effective attack items $a_1$ and $a_2$. Among them, the attack item $a_1$, which is immediately adjacent to the target item, is the candidate optimal attack item.
\begin{figure}[h]
\begin{center}
\includegraphics[width=0.65\linewidth]{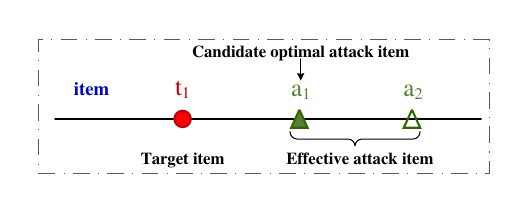}
\end{center}
\caption{\label{fig:comment2-1} Illustration of candidate optimal attack item.}
\end{figure}
\item
In each iteration, when selecting from the $r$ candidate optimal attack items, for each candidate $a_i$ with predecessor $a_{i-1}$ (the closest another candidate optimal attack item), we only need to increase the frequency of $a_i$' up to surpass the items between $a_i$ and $a_{i-1}$. Based on Observation 1, surpassing earlier targets can be achieved in subsequent iterations by increasing $a_{i-1}$. For example, as shown in Figure~\ref{fig:comment2-2}, for the candidate optimal attack item $a_2$ with predecessor $a_1$, we only need to consider the target items $t_3, t_4, t_5$, without considering $t_1$ and $t_2$ in this iteration.
\begin{figure}[h]
\begin{center}
\includegraphics[width=0.9\linewidth]{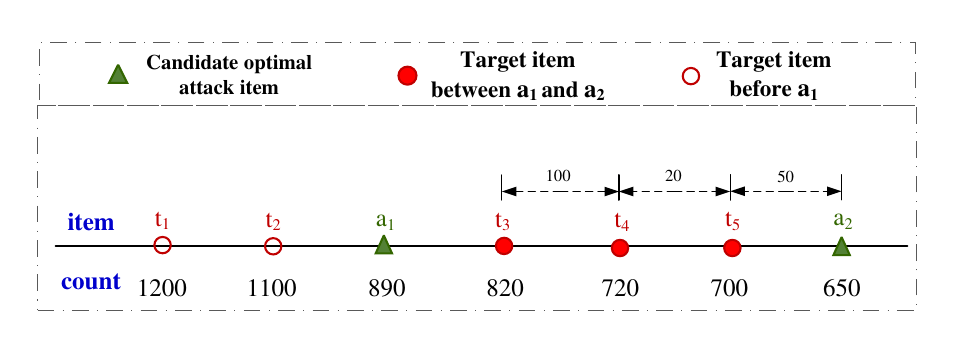}
\end{center}
\caption{\label{fig:comment2-2} Illustration of target items between consecutive candidate optimal attack items.}
\end{figure}
\end{itemize}

\textbf{(2) Proposing a solution.}
Based on the above observations, we take the kRR attack as an example and propose a solution for the defined optimization problem.

In this solution, we iteratively select the optimal attack item and allocate a corresponding number of fake users. The procedure is implemented as follows:
\begin{itemize}
\item Candidate Attack Item Selection:
Identify $r$ candidate optimal attack items denoted as $\{a_1, ..., a_r\}$.
\item Fake User Value Calculation:
For each candidate attack item $a_i$, locate its nearest preceding valid candidate attack item $a_{i-1}$.
	Let $\{t_j, ..., t_k\}$ denote the target items between $a_{i-1}$ and $a_{i}$.
	Define the distance (frequency difference) between adjacent items in $\{t_j, ..., t_k, a_{i}\}$ as $d_{i,j-k+1}, ..., d_{i,2},d_{i,1} $.
\item Compute the value of fake users for $a_{i}$ as:
\begin{equation}
\mathcal{V}_i=\mathop {\max }\limits_{1 \le l \le j - k + 1} \left\{ {\frac{l}{{\sum\nolimits_{h = 1}^l {d_{i,l} } }}} \right\}.
\end{equation}
\begin{figure}[h]
\begin{center}
\includegraphics[width=0.9\linewidth]{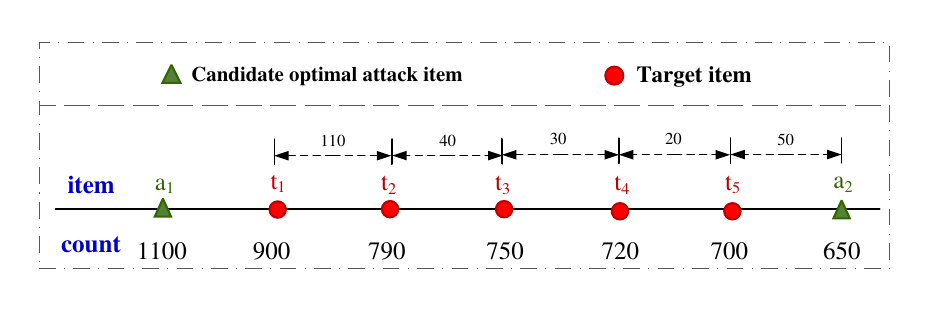}
\end{center}
\caption{\label{fig:comment2-3} Illustration of fake user value calculation.}
\end{figure}
For example, as shown in Figure~\ref{fig:comment2-3}, the value of fake users for $a_2$ can be calculated as:
\begin{displaymath}
\begin{split}
 \mathcal{V}_2 &= \left\{ {\frac{1}{{50}},\frac{2}{{70}},\frac{3}{{100}},\frac{4}{{140}},\frac{5}{{250}}} \right\} \\ 
  &= \frac{3}{{100}}. \\ 
\end{split}
\end{displaymath}
\item Optimal Attack Item Selection: 
	Compare the values $V_i$ across all candidates and select the item with maximum value:
	\begin{equation}
    i_o=argmax{\left\{\mathcal{V}_i\right\}}.
    \end{equation}
	Determine $l_o$ as the $l$ that maximizes $\mathcal{V}_{i_o}$, then allocate $\sum\limits_{h=1}^{lo}d_{io,h}$ fake users to $a_{i_o}$.
\end{itemize}

\textbf{(3) Theoretically validating the solution's correctness.}
We now formally demonstrate the correctness and optimality of the proposed scheme, which guarantees obtaining the optimal solution. The proof consists of two key aspects:
\begin{itemize}
\item Optimal Attack Item Selection:
The correctness of selecting the optimal attack item from candidate items during iteration is self-evident. To decrease the ranking of a target item, choosing the first effective attack item ranked below the target item inherently requires the minimum number of fake users. This selection criterion ensures efficiency in fake user allocation.

\item Monotonicity of Fake User Value:
For any given iteration, the computed value of fake users will always exceed that calculated in subsequent iterations. This property holds because: for each candidate optimal attack item $a_i$ with current iteration value $\mathcal{V}_{i}$ satisfies
\begin{displaymath}
\mathcal{V}_{i}=\frac{{l_o }}{{\sum\nolimits_{h = 1}^{l_o } {d_{i,h} } }} = \mathop {\max }\limits_{1 \le l \le j - k + 1} \left\{ {\frac{l}{{\sum\nolimits_{h = 1}^l {d_{i,h} } }}} \right\}.
\end{displaymath}
Then, we can prove that,
\begin{displaymath}
\mathcal{V}_i=\frac{{l_o }}{{\sum\nolimits_{h = 1}^{l_o } {d_{i,h} } }} \ge \mathop {\max }\limits_{l_o  + 1 \le l \le j - k + 1} \left\{ {\frac{{l - l_o }}{{\sum\nolimits_{h = l_o  + 1}^l {d_{i,h} } }}} \right\}.
\end{displaymath}
For example, returning to Figure~\ref{fig:comment2-3}, the current value of fake users is $\frac{3}{{100}}$, and
\begin{displaymath}
\frac{3}{{100}} > \max\left\{ {\frac{{4 - 3}}{{140 - 100}},\frac{{5 - 3}}{{250 - 100}}} \right\} = \frac{1}{{40}}.
\end{displaymath}
In addition, we have $V_{i_o }  \ge V_{i \ne i_o }$ in the current iteration.
Thus, the computed value of fake users $V_{i_o }$ in the current iteration will always exceed that calculated in subsequent iterations.
\end{itemize}

At each iteration, our allocation strategy ensures that:
(1) The assigned fake data yields the maximal value, and
(2) The value dominates all subsequent allocations.
Consequently, the solution guarantees global maximization of the fake data value.
Hence, the proof is completed.

\textbf{(4) Introducing computational approximations.}
To reduce algorithmic complexity while maintaining near-optimal, we compute the value of fake users for $a_{i}$ as:
\begin{equation}
\mathcal{V}_i={\frac{1}{d_{i,1} } }.
\end{equation}
The justification for this primarily includes the following reasons:
\begin{itemize}
\item 
As demonstrated in Figure~\ref{fig:fre_distribution}, the perturbed data distribution in practical ranking attacks shows significantly greater uniformity compared to the pre-perturbation state, which makes
\begin{displaymath}
{\frac{1}{d_{i,1} } } \approx \mathop {\max }\limits_{1 \le l \le j - k + 1} \left\{ {\frac{l}{{\sum\nolimits_{h = 1}^l {d_{i,l} } }}} \right\}.
\end{displaymath}
For clarity, we refer to $d_{i,1}$ as the attack cost of the effective attack $a_i$ in the OIA scheme.
\item
The number of target items is significantly smaller than that of effective attack items.
Consequently, there generally exists only one target item between any two candidate optimal attack items.
\end{itemize}

Based on the above analysis, we can demonstrate that OIA is near-optimal.

\section{Numerical Example for Eq.~\eqref{Eq.optimal}} \label{app:example}
In the OUE protocol, suppose there exist $3$ fake users, the input domain size $d=6$, the set of target items $T = \{t_1, t_2\} = \{1, 2\}$, and the set of non-target items are $A = \{a_1, a_2, a_3, a_4\} = \{3, 4,5,6\}$.
Let the perturbed values of the $3$ fake users be
\begin{displaymath}
y_{fake_1 }=001111,\ y_{fake_2 }=001100,\ y_{fake_3 }=100110.
\end{displaymath}
Then, we learn that,
\begin{displaymath}
S\left( {y_{fake_1 } } \right)=\{3, 4, 5, 6\},\ S\left( {y_{fake_2 } } \right)=\{3, 4\},\ S\left( {y_{fake_3 } } \right)=\{1, 4, 5\},
\end{displaymath}
and
\begin{displaymath}
{\left| {S\left( {y_{fake_1 } } \right)} \right|} =4,\ {\left| {S\left( {y_{fake_2 } } \right)} \right|} =2,\ {\left| {S\left( {y_{fake_3 } } \right)} \right|} =3.
\end{displaymath}
Meanwhile, we obtain that,
\begin{displaymath}
\begin{split}
&Alloc_{3}= \{y_{fake_1}, y_{fake_2}\},\ Alloc_{4}= \{y_{fake_1}, y_{fake_2}, y_{fake_3}\},\\
&Alloc_{5}= \{y_{fake_1}, y_{fake_3}\},\ Alloc_{6}= \{y_{fake_1}\},
\end{split}
\end{displaymath}
and
\begin{displaymath}
{\left| {Alloc_{3} } \right|}=2,\ {\left| {Alloc_{4} } \right|}=3,\ {\left| {Alloc_{5} } \right|}=2,\ {\left| {Alloc_{6} } \right|}=1.
\end{displaymath}
Thus, we have
\begin{displaymath}
\sum\limits_{j = 1}^4 {\left| {Alloc_{a_j } } \right|} =8 \le 9 = \sum\limits_{i = 1}^3 {\left| {S\left( {y_{fake_i } } \right)} \right|}.
\end{displaymath}
We note that the equality $\sum\limits_{j = 1}^4 {\left| {Alloc_{a_j } } \right|} = \sum\limits_{i = 1}^3 {\left| {S\left( {y_{fake_i } } \right)} \right|}$ holds when any $S\left( {y_{fake_1 } } \right)$ contains only effective attack items.

\section{Detailed Discussions of Scoring Function in Attacking OLH} \label{app:discussions}
When attacking OLH, we aim to select an optimal perturbed value $y=<H,h>$ that must satisfy three criteria: 1) no mapping to target items, 2) maximal coverage of effective attack items, and 3) minimal associated attack costs.
This requires a \textit{scoring function} to quantify perturbation effectiveness, where lower values indicate better attack performance.
We initially consider the summation form
\begin{equation}
C_{\left(i,j\right)}=\sum_{a_k\in A_{(i,j)}}\delta_k,
\end{equation}
where $A_{(i,j)}$ denotes the set of effective attack items and $\delta_k$ denotes the attack cost.
While decreasing with individual cost $\delta_k$ reduction, this function increases with larger attack sets $A_{(i,j)}$, contradicting our objectives.
To address this, we examine the arithmetic mean
\begin{equation}
C_{\left(i,j\right)}=\frac{\sum_{a_k\in A_{(i,j)}}\delta_k}{\left|A_{(i,j)}\right|}.
\end{equation}
However, concurrent growth of numerator and denominator leads to ambiguous optimization directions.
The harmonic mean variant $C_{\left(i,j\right)}=\frac{\left|A_{(i,j)}\right|}{\sum_{a_k\in A_{(i,j)}}\frac{1}{\delta_k}}$ similarly fails to provide consistent trends with expanding attack sets.

Our analysis reveals the \textit{harmonic scoring function} $C_{\left(i,j\right)}=\frac{1}{\sum_{a_k\in A_{(i,j)}}\frac{1}{\delta_k}}$ exhibits a key property:
Monotonic decrease with either more attack items or lower individual costs.
Thus, minimizing $C_{\left(i,j\right)}$ maximizes marginal attack contribution, perfectly aligning with submodular optimization principles.

\section{Complexity Analysis of Attacking kRR, OUE, and OLH} \label{app:complexity}
kRR attack involves three steps: 1) sorting items by frequency and calculating attack costs at the beginning, 2) iteratively selecting optimal attack item $a_{opt}$ from the immediate effective attack items following each target item, and 3) updating $a_{opt}$'s attack cost.
Step 1 has a complexity of $O(d\log d)$, but sorting is performed only once.
In subsequent iterations, selecting the optimal attack item $a_{opt}$ has a time complexity of $O(|T|)$, where $|T|$ denotes the number of target items (typically much smaller than d). Updating the attack cost for this $a_{opt}$ takes $O(1)$ time. Given $\frac{m}{\delta}$ iterations, where $m$ denotes the number of fake users, and $\delta$ denotes the average attack cost, the overall iterative process achieves $\frac{|T|m}{\delta}$ complexity.
Thus, overall time complexity is $O(d\log d+\frac{|T|m}{\delta})$, and $\frac{|T|m}{\delta}$ is usually much smaller than $O(d\log d)$, yielding a manageable overall complexity.

In OUE attack, attackers iteratively selecting optimal attack item sets $B_{opt}$ from the immediate effective attack items following each target item.
The size of $B_{opt}$ is $E_1$, the expected number of 1s in the genuine users' perturbed values, where \( E_1 = p + (d-1)q \), with \( p = \frac{1}{2} \) and \( q = \frac{1}{e^\varepsilon + 1} \).
Notably, $E_1$ is smaller than $d$.
By leveraging a priority queue, selecting the optimal attack item set has a time complexity of $O(E_1 \log |T|)$. Updating the attack costs for this set takes $O(E_1)$ time.
Thus, the overall iterative process achieves $\frac{E_1 m \log |T|}{\delta}$ complexity, also yielding a manageable overall complexity.

\begin{figure}[ht]
\begin{center}
\includegraphics[width=\linewidth]{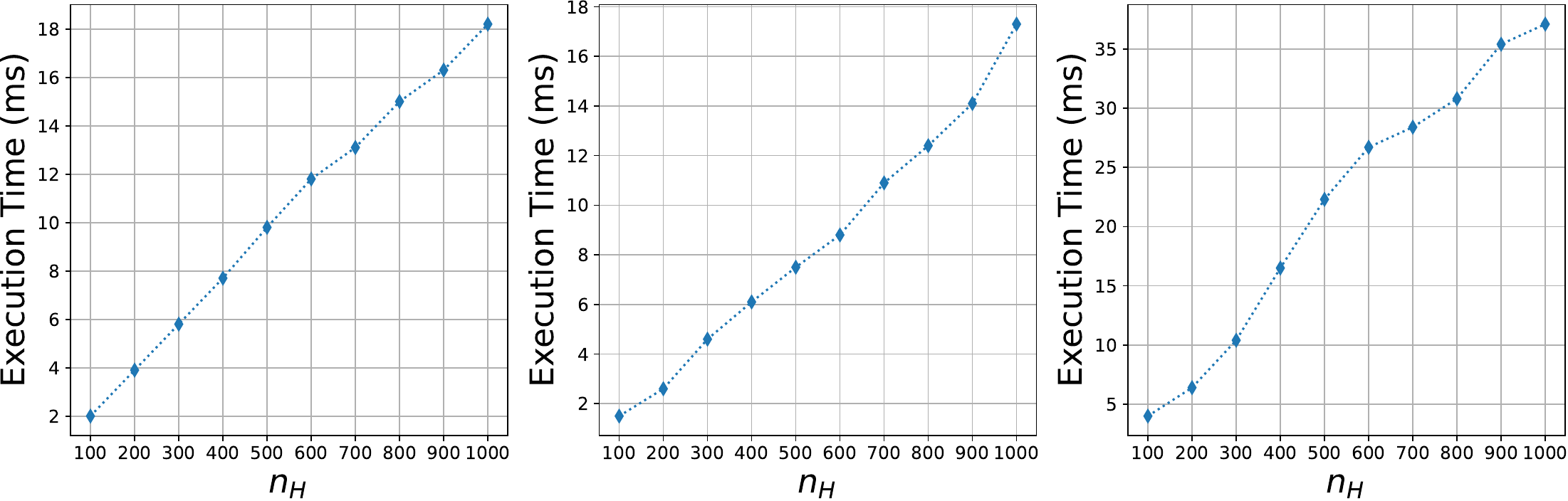}
\end{center}
\caption{\label{fig:hash_select} Optimal hash function selection time for $n_H \in [100, 1000]$ across Synthetic, Adult, and Fire Datasets.}
\end{figure}

OLH attack involves selecting hash function, i.e., selecting a high-quality hash function from a universal hash function family based on a scoring function.
However, direct selection from the universal hash function family proves computationally infeasible.
For LDP poisoning attacks, the standard methodology proceeds in two phases: (1) random sampling of $n_H$ candidate items, followed by (2) optimal selection from this subset. Empirical results confirm that $n_H=100$ achieves optimal attack efficacy, a parameter value we consequently adopt. Moreover, benchmark tests verify that the associated hash function selection completes within $40$ ms when $n_H \in [100,1000]$, as shown in Figure~\ref{fig:hash_select}.
We may also consider parallelization to reduce further the time required for hash function selection.

\section{Feasibility and Detection Resistance in Fake User Injection} \label{app:feasibility}
Injecting a controlled number of fake users is feasible, as demonstrated by existing work~\cite{DBLP:conf/uss/ThomasMGKP13}. Attackers can easily obtain fake accounts from popular web services like Twitter, Google, and Hotmail or purchase them at low prices from underground markets. For instance, a Hotmail account costs only \$0.004$\sim$\$0.03. Meanwhile, prior study~\cite{DBLP:conf/uss/ThomasMGKP13} have experimentally shown that attackers can reliably bypass some simple detection mechanisms. For example, against IP blacklisting, attackers can use dynamic IPs to circumvent restrictions, with many IPs originating from infected hosts worldwide, making them difficult to block via traditional blacklists. As for CAPTCHAs, while this mechanism effectively reduces registration success rates, it merely increases the cost of fake user registration—attackers can still persistently register accounts using automated tools (e.g., DeathByCAPTCHA) and low-rate requests (to avoid triggering alarms).

Admittedly, device fingerprinting, behavioral analysis, rate limiting, and multi-layered fraud prevention mechanisms can all impede attackers to varying degrees, forcing them to employ higher operational complexity and costs to inject fake users, such as generating diverse device fingerprints or combining human and automated efforts to evade detection. Furthermore, we might leverage algorithms or generative adversarial networks (GANs) to simulate genuine user behavior and employ more sophisticated device fingerprint spoofing techniques to mimic real device characteristics, making fake users even more stealthy.

\section{Attacking Adaptive LDP Protocols} \label{app:Adaptive}
The adaptive LDP mechanism incorporates two key perturbation functions: adaptive and dynamic. The adaptive perturbation function adjusts the privacy budget $\varepsilon$ according to data distribution or sensitivity, while the dynamic perturbation function modifies $\varepsilon$ over time or in response to external factors. Fundamentally, this shifts from the fixed-$\varepsilon$ approach of traditional LDP to a flexible, adjustable $\varepsilon$ framework. For instance, as demonstrated in~\cite{DBLP:journals/jksucis/SongSZHXWZ24}, users may tailor $\varepsilon$ to their individual privacy preferences.

If a ranking attack persists in using a fixed privacy budget (denoted as $\varepsilon$) against an adaptive LDP protocol with budget $\varepsilon_a$, two issues arise:
\begin{itemize}
  \item
  $\varepsilon<\varepsilon_a$. The attacker will overestimate inter-item frequency differences, disproportionately allocating fake users to attack targets and wasting resources.
  \item
  $\varepsilon>\varepsilon_a$. The attacker underestimates frequency differences, leading to inadequate fake user allocation and potential multi-attack failure.
\end{itemize}
Thus, to maintain attack efficacy against adaptive LDP, attackers should prioritize a conservative $\varepsilon$ (i.e., smaller than $\varepsilon_a$), even if mismatched, to mitigate estimation errors.

\section{MPOIA with kRR Algorithm} \label{app:MPOIA with kRR}
\begin{algorithm}[t]
  \caption{MPOIA with kRR}
  \label{alg:MPOIA wiht kRR}
  \begin{algorithmic}[1]
    \Require
      $A_{\text{eff}}$: effective attack item set; $T_{a_i}$: target item subset; $m$: number of fake users
    \Ensure
      $M$: fake user allocation vector
    \State $M = \left[0,0,\dots,0\right]$ \label{alg:MPOIA:initM}
    \State calculate $\mathrm{Var}(n_{t_{k_i}}')$ and $\mathrm{Var}(n_{a_i}')$ according to Eq.~\ref{eq:variance} \label{alg:MPOIA:var}
    \Comment{for each $a_i \in A_{\text{eff}}$, $t_{k_i} \in T_{a_i}$ is the closest target item to $a_i$}
    \State $SE = \sqrt{\mathrm{Var}(n_{t_{k_i}}')+\mathrm{Var}(n_{a_i}')}$ \label{alg:MPOIA:SE}
    \State $\delta_i = \left \lceil E[n_{t_{k_i}}'] - E[n_{a_i}'] + z_{\alpha} \times SE \right \rceil$ \label{alg:MPOIA:delta}
    \Comment{calculate cost based on confidence interval} 
    \State construct cost array $\Delta$ based on all $\delta_i$ values \label{alg:MPOIA:Delta}
    \While {$m > 0$} \label{alg:MPOIA:while}
      \State $\delta_* = min\left( \Delta\right)$
      \Comment{find the minimum attack cost}
      \State select $a_{opt}$ from $A_{\text{eff}}$ according to $\delta_*$;
      \If {$m>\delta_*$}
        \State $m_i = \delta_*$ 
        \Comment{determine the number of fake users}
        \State $M[a_{\text{opt}}] = M[a_{\text{opt}}] + m_i$ 
        \State \text{Update} $T_{a_{\text{opt}}}$, and $\Delta[a_{\text{opt}}]$
        \State $m = m - m_i$
        \Comment{update the number of fake users}
      \Else
        \State $M\left[a_{opt}\right] = M\left[a_{opt}\right] + m$
        \State $m = 0$
      \EndIf
    \EndWhile
    \State\Return $M$ \label{alg:MPOIA:return}
  \end{algorithmic}
\end{algorithm}

The algorithm takes as input a set of effective attack items \( A_{\text{eff}} \), the subset of target items \( T_{a_i} \) corresponding to each effective attack
item \( a_i \), and the number of fake users \( m \). First, the algorithm initializes the allocation vector \( M \) to all zeros (Line~\ref{alg:MPOIA:initM}). For each $a_i\in A_{\text{eff}}$ and $t_{k_i}\in T_{a_i}$, calculate the standard error $SE$ based on the variance (Lines~\ref{alg:MPOIA:var}--\ref{alg:MPOIA:SE}). Then, the attack cost for each effective attack item, \( \delta_i \), is computed using the expected frequencies difference between the estimated counts and adjusted by a confidence interval determined by \( z_{\alpha} \times SE \) (Line~\ref{alg:MPOIA:delta}). A cost array \( \Delta \) is constructed based on all \( \delta_i \) values (Lines~\ref{alg:MPOIA:Delta}). Compared with Attacking kRR (Algorithm~\ref{alg:kRR}), the difference of this algorithm lies in the calculation of $\Delta$, where the confidence interval adjusts the original difference in expected frequencies. Therefore, after calculating $\Delta$, the process of selecting the optimal attack item $a_{\text{opt}}$, updating parameters, allocating fake users, and other related steps (Lines~\ref{alg:MPOIA:while}--\ref{alg:MPOIA:return}) is consistent with Lines~\ref{alg:kRR:while}--\ref{alg:kRR:return} of Algorithm~\ref{alg:kRR}.

\section{Attacking Heavy Hitter Identification} \label{app:heavy}

\subsection{Heavy Hitter Identification}

The goal of Heavy Hitter Identification~\cite{DBLP:conf/nips/BassilyNST17, DBLP:conf/stoc/BassilyS15, DBLP:journals/tdsc/0001LJ21} is to identify the top-$k$ items with the highest frequency among the $n$ users. Researchers have developed a series of protocols that can identify frequent items without estimating the frequencies of all items~\cite{DBLP:conf/nips/BassilyNST17, DBLP:conf/stoc/BassilyS15, DBLP:journals/tdsc/0001LJ21}. The state-of-the-art protocol for frequent item identification is the \textit{Prefix Extending Method (PEM)}~\cite{DBLP:journals/tdsc/0001LJ21}. This protocol first groups users and then iteratively applies a frequency estimation protocol within each group to identify frequent prefixes.

In PEM, each user encodes its items as a $\gamma$-bit binary vector. Suppose users are evenly divided into $g$ groups. In the $j$-th iteration, users in the $j$-th group perturb the first $\lambda_j$ bits of their binary vectors using the OLH protocol, where 
\begin{equation}
\lambda_j = \lceil \log_2 k \rceil + \left\lceil \frac{j \cdot (\gamma - \lceil \log_2 k \rceil)}{g} \right\rceil,
\end{equation}
and send the perturbed bits to a aggregator. The aggregate step of the OLH protocol is used here to estimate the frequencies of the top-$k$ prefixes. For a vector of length $\lambda_j$, the total number of possible items is $2^{\lambda_j}$, and using OUE for frequency estimation would incur a high communication cost compared to OLH.
Specifically, the aggregator uses the aggregate step of OLH to estimate the frequencies of prefixes of length $\lambda_j$ in the set $R_{j-1} \times \{0,1\}^{\lambda_j - \lambda_{j-1}}$, where $R_{j-1}$ is the set of top-$k$ vectors of length $\lambda_{j-1}$ from the $(j-1)$-th iteration, and the $\times$ symbol denotes the Cartesian product. After estimating the frequencies of these $\lambda_j$-bit prefixes, the aggregator identifies the top-$k$ most frequent ones, denoted as the set $R_j$. This process is repeated for $g$ groups, and the top-$k$ set from the final iteration is identified as the top-$k$ frequent items.

\begin{figure}[h]
\begin{center}
\includegraphics[width=\linewidth]{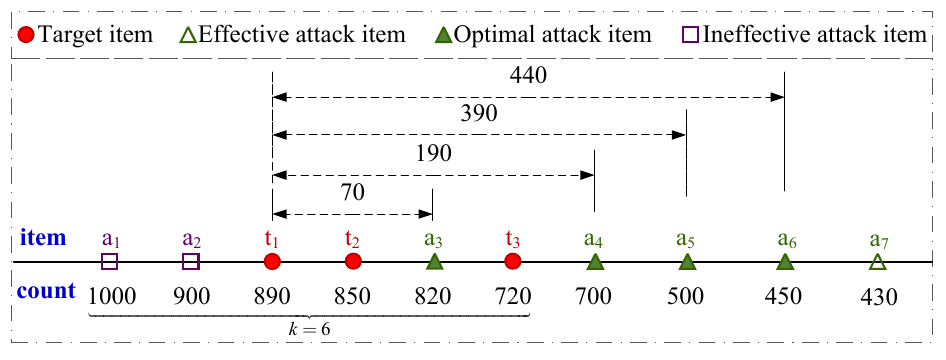}
\end{center}
\caption{\label{fig:PEM}Illustration of attacking heavy hitter identification}
\end{figure}

\subsection{Attacking Description}
In this scenario, we consider an attack that alters the ranking of a target item from being top-$k$ to not being top-$k$. We also use $T = \{t_1, t_2, \dots, t_r\}$ to represent the set of target items, and $A = \{a_1, a_2, \dots, a_s\}$ to represent the set of non-target items. We define the \textit{success rate} of the attack as the proportion of target items that become non-top-$k$ heavy hitters after the attack, expressed as:
\begin{equation}
\text{SR} = \frac{\sum_{t \in T} \mathbbm{1}(\tilde{r}_{t,2} > k)}{|T|},
\end{equation}
the attacker's goal is to maximize the success rate.

\subsection{Attack Method}
The most advanced protocols for heavy hitter identification currently iteratively apply frequency estimation protocols. Thus, the three attack methods previously discussed for ranking estimation can also be applied to heavy hitter identification.
Additionally, we demonstrate that when attacking heavy hitter identification protocols, each iteration of the attack is essentially a simplified version of the attack on ranking estimation. Specifically, fewer effective attack items need to be considered, and there is no need for iterative searches for the optimal attack item. We provide an example based on Figure~\ref{fig:PEM}.

Take $k=6$ as an example, with the target items $t_1$ to $t_3$ that need to be demoted. The highest success rate is achieved when all these target items are pushed out of the top-6 heavy hitters. First, we need to increase the frequency of the effective attack item within the top-6 ($a_3$) to surpass $t_1$ to $t_3$. Then, we must find three effective attack items outside the top-6 heavy hitters whose frequencies also exceed those of $t_1$ to $t_3$. It can be seen that selecting the three effective attack items $a_4$ to $a_6$ with the smallest frequency gap relative to $t_1$ to $t_3$ minimizes the number of fake users that need to be injected. At this point, it is unnecessary to consider effective attack items beyond $a_6$ to achieve the highest success rate.

In general, suppose the number of effective attack items within the top-$k$ is $x$, and the number of target items is $r$. The number of effective attack items outside the top-$k$ that need to be considered is also $r$. Therefore, the total number of effective attack items that need to be considered is $x + r$. Since the target items are all within the top-$k$, $r \leq k$, so $x + r \leq 2k$. Given that $k$ is usually small, the total number of effective attack items to consider is also small. Moreover, based on the analysis above, we only need to select the $x + r$ effective attack items with the highest frequencies.

Next, we illustrate an attack on a heavy hitter identification protocol using the PEM algorithm as an example. Due to communication cost concerns, the OLH protocol is generally preferred for frequency estimation in the PEM algorithm.

\noindent\textbf{Attacking PEM:} In the PEM protocol, each item is represented as a $\gamma$-bit binary vector, and users are randomly divided into $g$ groups. On average, each group contains $\frac{m}{n+m}$ fake users. In the $j$th iteration, PEM applies the OLH protocol to perturb the first $\lambda_j$ bits of the binary vector for the $j$th group of users and sends the result to the aggregator. The attacker treats the first $\lambda_j$ bits of the binary vectors corresponding to the target items as "target items" and then applies RIA, ROA, or OIA attacks against the OLH protocol to forge fake users' perturbed values. Given the small number of target and effective attack items, it is easier to find perturbed values with lower overall cost. By using this attack method, fewer of the target items' first $\lambda_j$ bits will be identified as top-$k$ prefixes in the $j$th iteration, leading to more target items being pushed out of the top-$k$.

\subsection{Experimental Results}

\begin{figure*}[h]
\begin{center}
\includegraphics[width=0.9\textwidth]{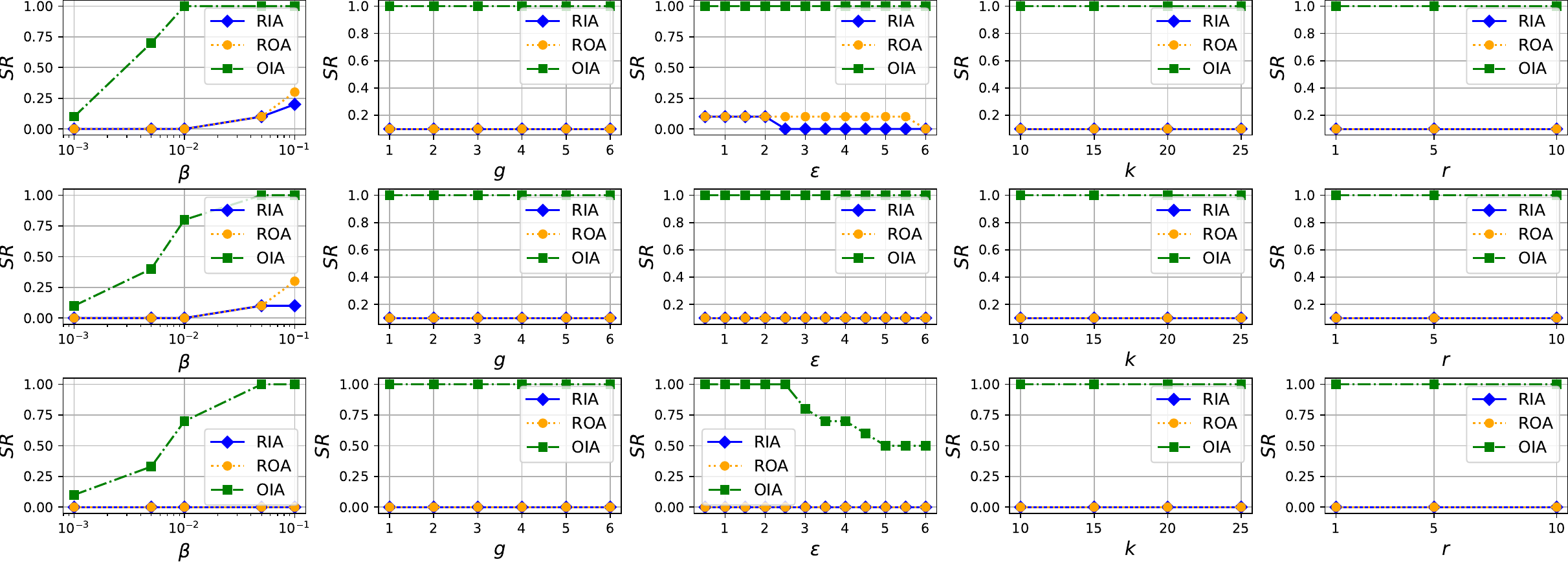}
\end{center}
\caption{\label{fig:PEM result} Impact of different parameters($\beta$, $g$, $\epsilon$, $k$, $r$) on the success rates of the three attacks for PEM. The three rows are for Synthetic, Adult and Fire datasets, respectively. }
\end{figure*}

Figure~\ref{fig:PEM result} shows the impact of different parameters($\beta$, $g$, $\epsilon$, $k$, $r$) on the success rates of the three attacks for PEM. By default, we follow prior work~\cite{DBLP:conf/uss/CaoJG21} to set the number of groups \(g = 10\) and the number of items identified as heavy hitters \(k = 20\), while the other parameter values are consistent with those used for attacking ranking estimation. We define the success rate of the attack as the proportion of target items that are successfully promoted to become top-\(k\) heavy hitters after the attack.

The experimental results demonstrate that OIA is highly effective in attacking the PEM protocol. For the synthetic dataset, only about 1\% of fake users are sufficient to achieve a 100\% success rate, while for the two real-world datasets, merely 5\% of fake users are required. This aligns with our previous analysis that each iteration of attacking PEM is essentially a simplified version of the attack on ranking estimation.

Considering the impact of different parameters, we observe that the success rate naturally increases as \(\beta\) grows. And the overall gain decreases as \(\epsilon\) increases, however, for both the synthetic dataset and the Adult dataset, a 100\% success rate is maintained even when \(\epsilon = 6\). This indicates that, under the default parameter settings, the rankings of all target items after the attack significantly exceed \(k\). Theoretically, as \(k\) increases, the number of effective attack items to consider also grows, which should lead to a lower success rate. Similarly, as \(g\) increases, the number of groups grows, reducing the number of fake users per group, making it harder to alter the rankings of target items and lowering the success rate. Nevertheless, experimental results show that regardless of how \(g\) or \(k\) varies, the success rate remains consistently at 100\%. This further demonstrates the strong disruptive power of our proposed attack against the PEM protocol.

\section{Additional Evaluation Results}

\subsection{Attacking under limited knowledge} \label{app:limit}

\begin{figure}[t]
\begin{center}
\includegraphics[width=\linewidth]{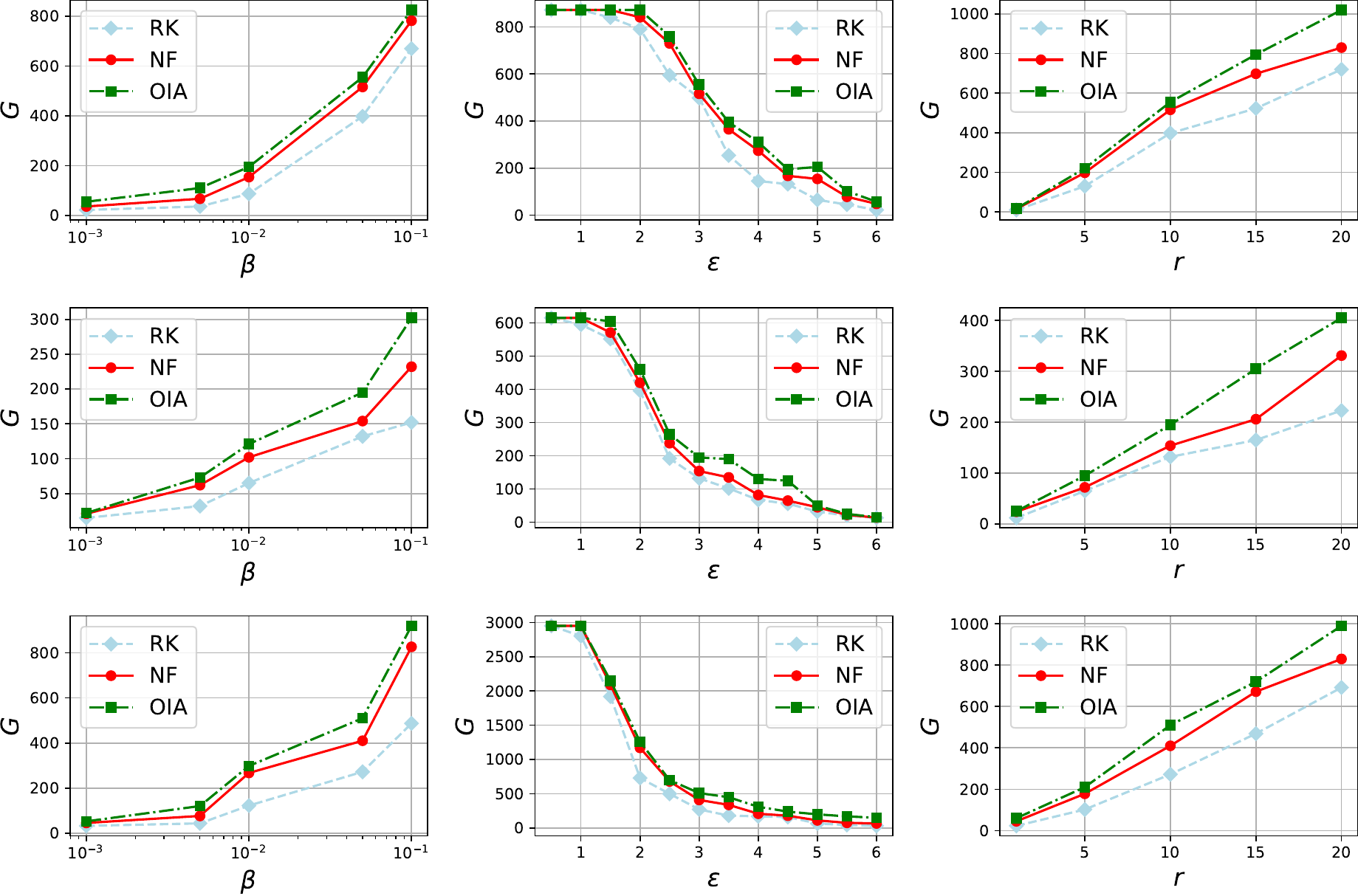}
\end{center}
\caption{\label{fig:limited_OUE} Attacking OUE under Limited Knowledge for RK, NF and OIA. The three rows are for Synthetic,
Adult and Fire datasets, respectively. }
\end{figure}

\begin{figure}[t]
\begin{center}
\includegraphics[width=\linewidth]{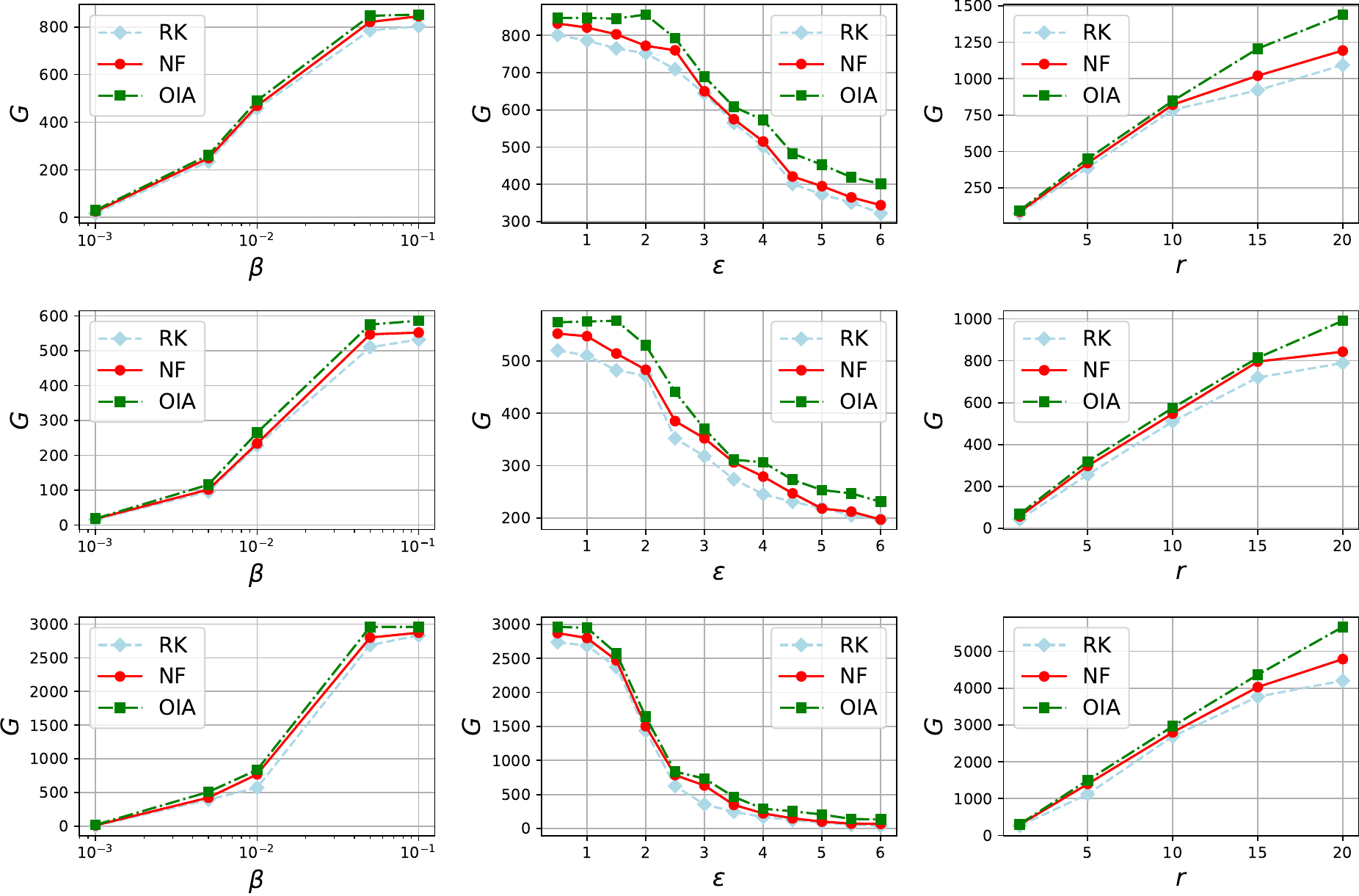}
\end{center}
\caption{\label{fig:limited_OLH} Attacking OLH under Limited Knowledge for RK, NF and OIA. The three rows are for Synthetic,
Adult and Fire datasets, respectively. }
\end{figure}

Figures~\ref{fig:limited_OUE}--\ref{fig:limited_OLH} present the supplementary results of attacking the OUE and OLH protocols under varying levels of background knowledge available to the attacker. Each figure contains three rows, corresponding to three different datasets. All parameters are set to their default values, and for attacking OUE, the default value of $\epsilon$ is set to 3 to better highlight the differences in effectiveness among the various attack strategies. Similar to the conclusions drawn from attacking kRR, the performance differences among the three attack strategies remain minimal when attacking OUE or OLH.

\subsection{Defenses} \label{app:defenses}

\begin{figure}[t]
\begin{center}
\includegraphics[width=\linewidth]{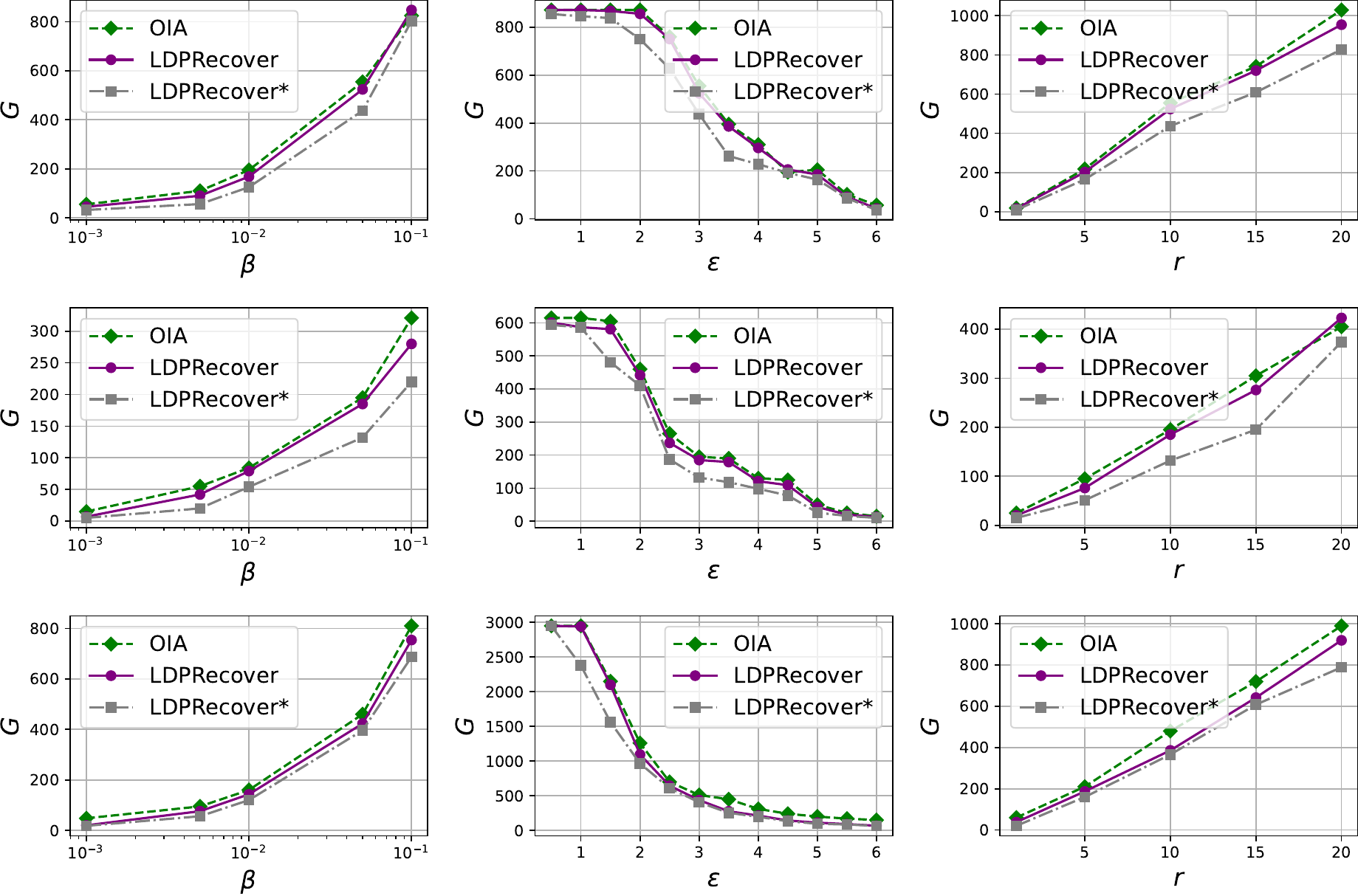}
\end{center}
\caption{\label{fig:defense:oue} Impact of defense methods with different parameters ($\beta$, $\epsilon$, $r$) on the overall gain for OUE. The three rows are for Synthetic, Adult and Fire datasets, respectively. }
\end{figure}

\begin{figure}[t]
\begin{center}
\includegraphics[width=\linewidth]{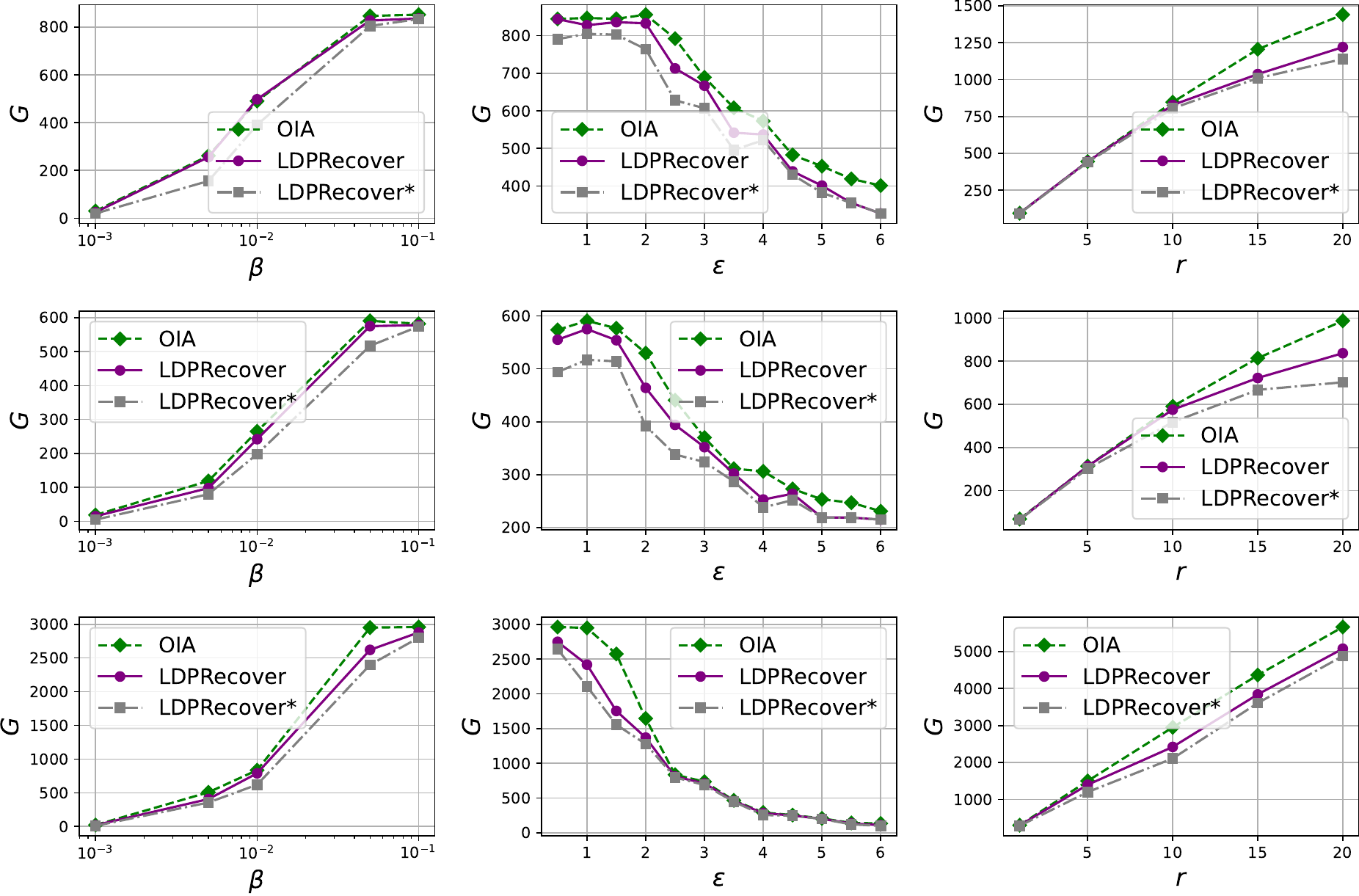}
\end{center}
\caption{\label{fig:defense:olh} Impact of defense methods with different parameters ($\beta$, $\epsilon$, $r$) on the overall gain for OLH. The three rows are for Synthetic, Adult and Fire datasets, respectively. }
\end{figure}

Figures~\ref{fig:defense:oue}--\ref{fig:defense:olh} provide supplementary results demonstrating the defense performance of two countermeasures against OUE and OLH protocols, compared with the attack effectiveness of OIA. Specifically, for the OLH protocol, the number of items that each fake user can support is set to $\frac{E_1}{2} \times m$. Experimental results indicate that the defense performance of these two methods is minimal across all protocols and datasets.

\subsection{MPOIA} \label{app:MPOIA}
\begin{figure}[t]
\begin{center}
\includegraphics[width=\linewidth]{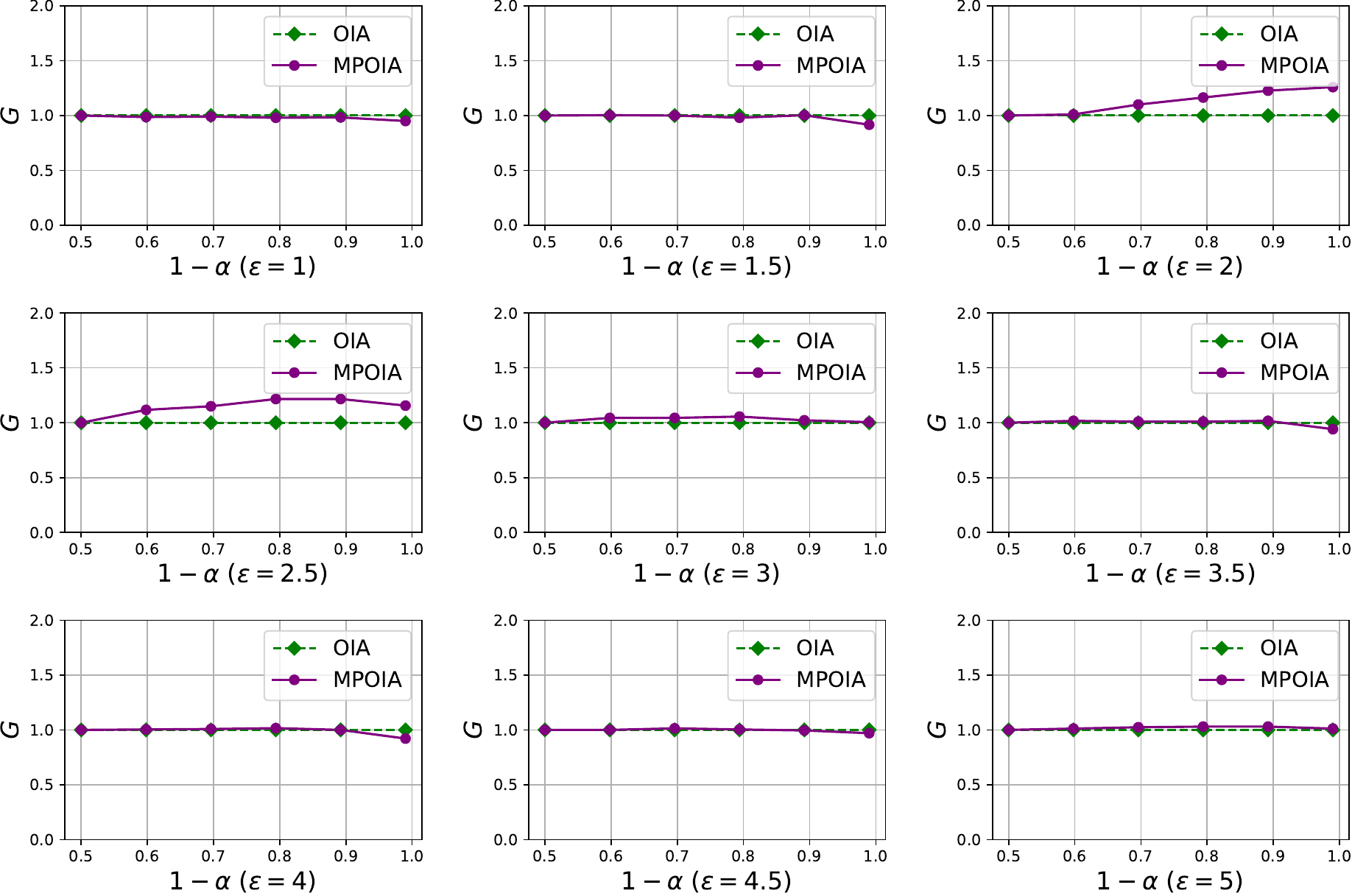}
\end{center}
\caption{\label{fig:MPOIA_kRR_Syn} Impact of MPOIA with different parameters ($\epsilon$, $1-\alpha$) on the overall gains for kRR on Synthetic dataset. }
\end{figure}

\begin{figure}[t]
\begin{center}
\includegraphics[width=\linewidth]{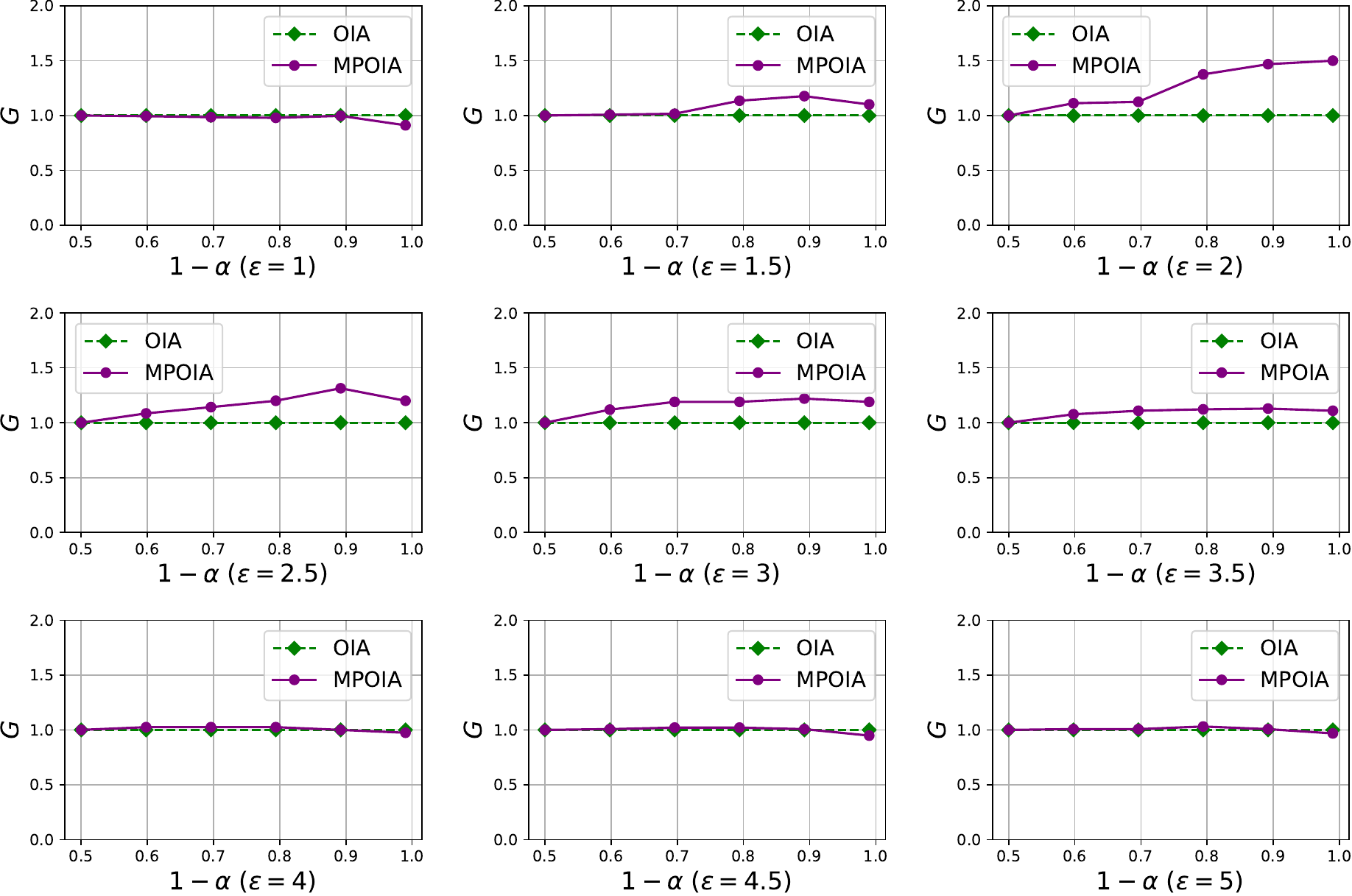}
\end{center}
\caption{\label{fig:MPOIA_kRR_Adult} Impact of MPOIA with different parameters ($\epsilon$, $1-\alpha$) on the overall gains for kRR on Adult dataset. }
\end{figure}

\begin{figure}[t]
\begin{center}
\includegraphics[width=\linewidth]{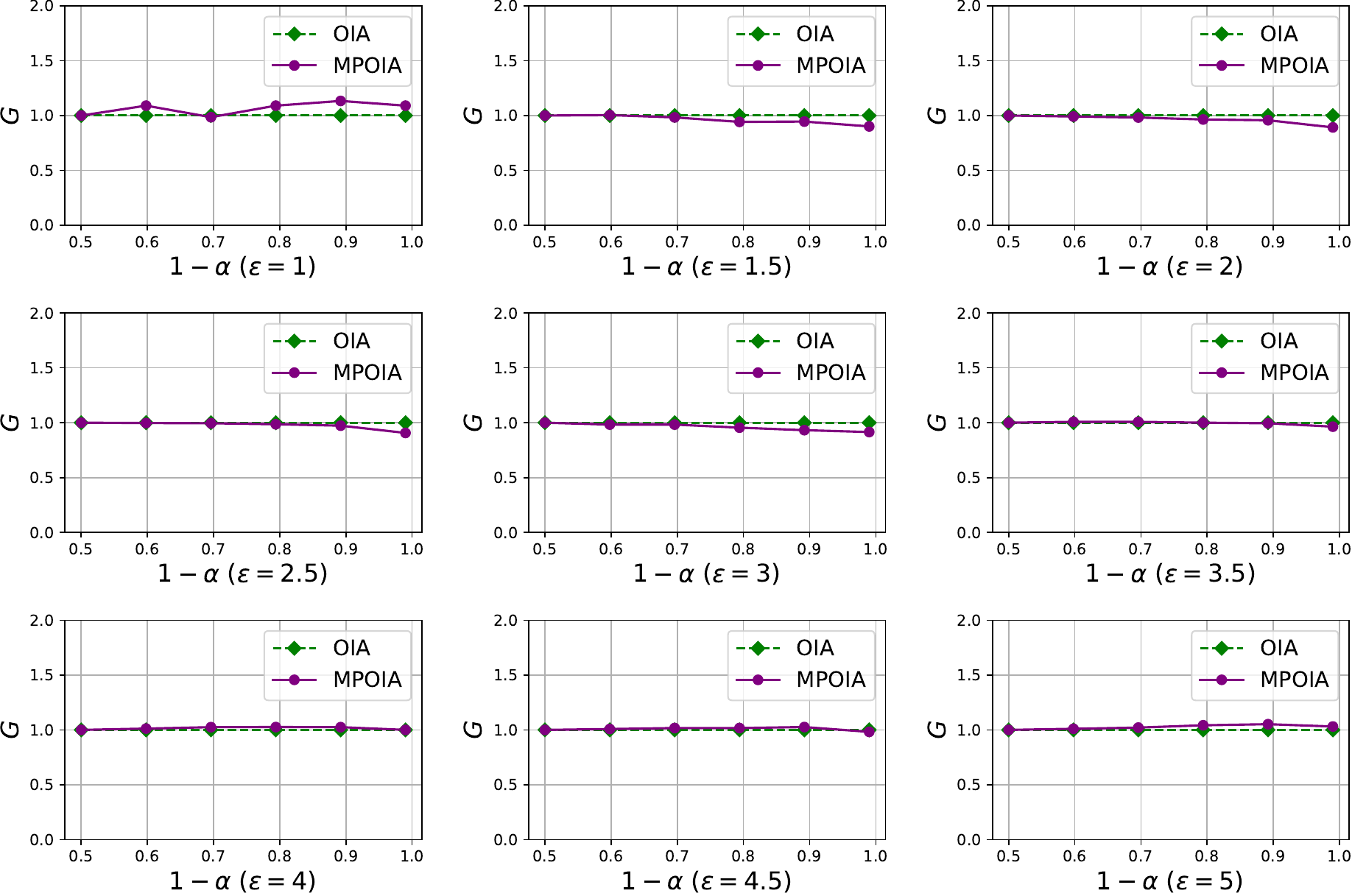}
\end{center}
\caption{\label{fig:MPOIA_kRR_Fire} Impact of MPOIA with different parameters ($\epsilon$, $1-\alpha$) on the overall gains for kRR on Fire dataset. }
\end{figure}

\begin{figure}[t]
\begin{center}
\includegraphics[width=\linewidth]{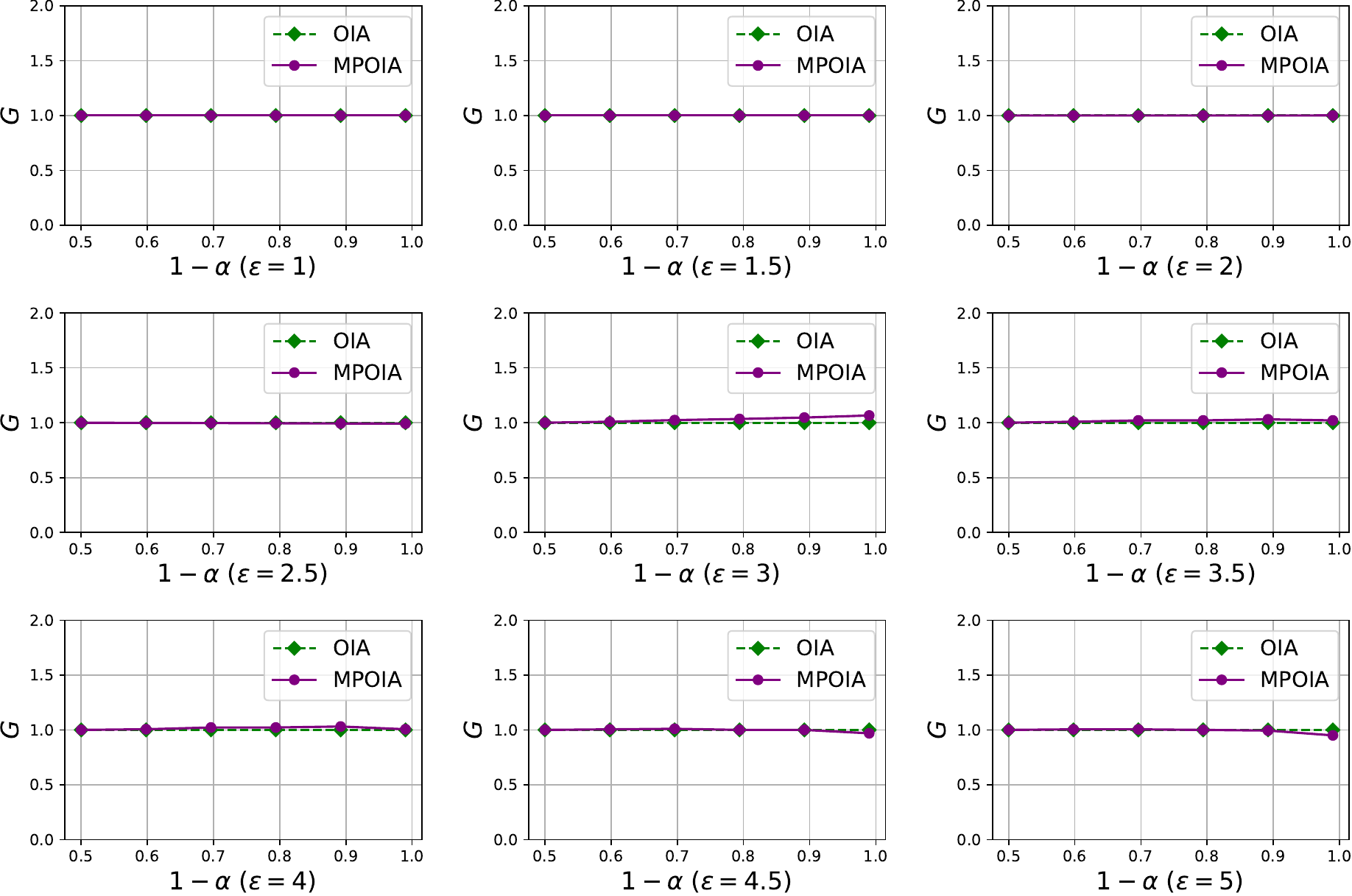}
\end{center}
\caption{\label{fig:MPOIA_OUE_Syn} Impact of MPOIA with different parameters ($\epsilon$, $1-\alpha$) on the overall gains for OUE on Synthetic dataset. }
\end{figure}

\begin{figure}[t]
\begin{center}
\includegraphics[width=\linewidth]{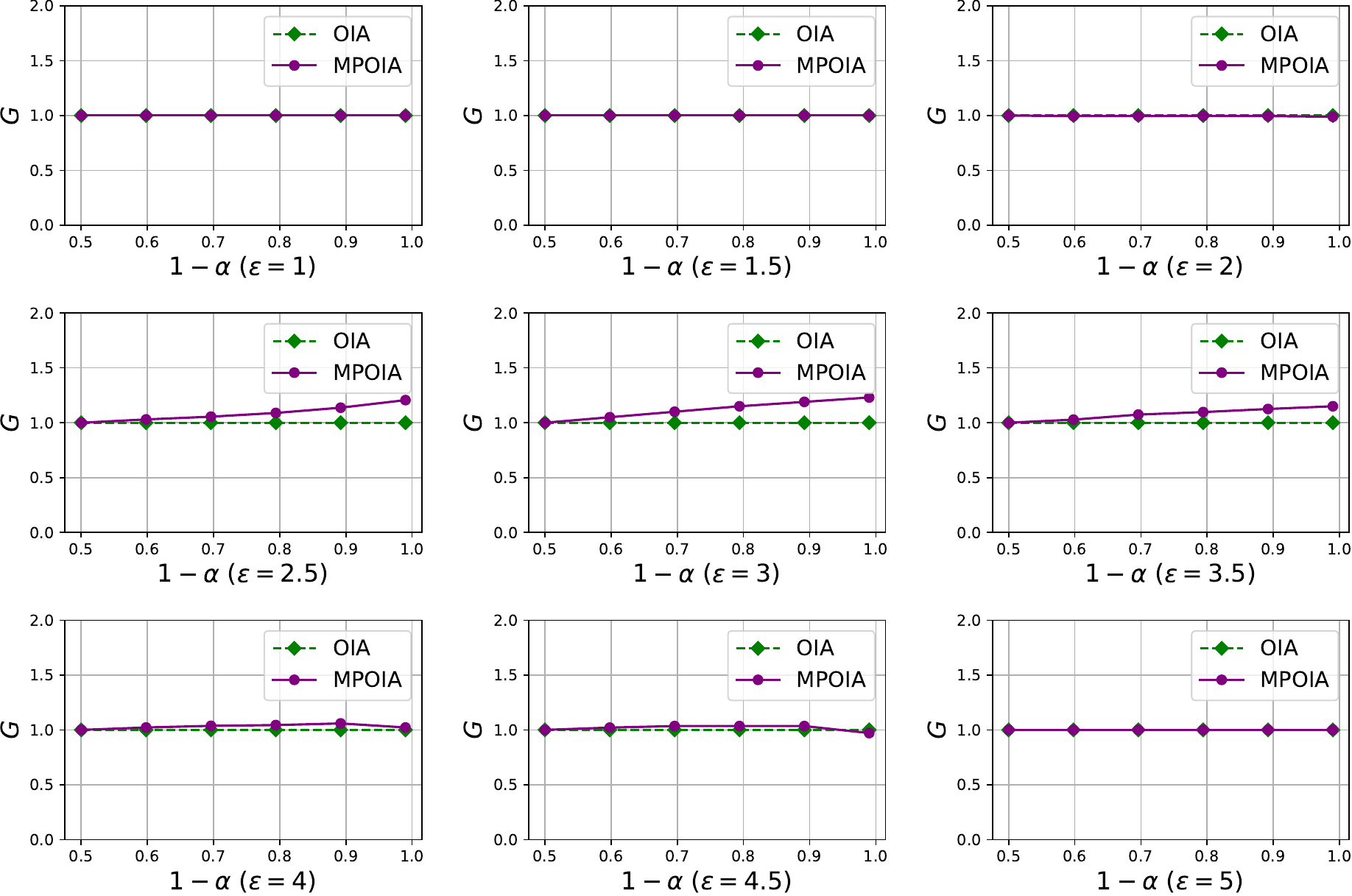}
\end{center}
\caption{\label{fig:MPOIA_OUE_Adult} Impact of MPOIA with different parameters ($\epsilon$, $1-\alpha$) on the overall gains for OUE on Adult dataset. }
\end{figure}

\begin{figure}[t]
\begin{center}
\includegraphics[width=\linewidth]{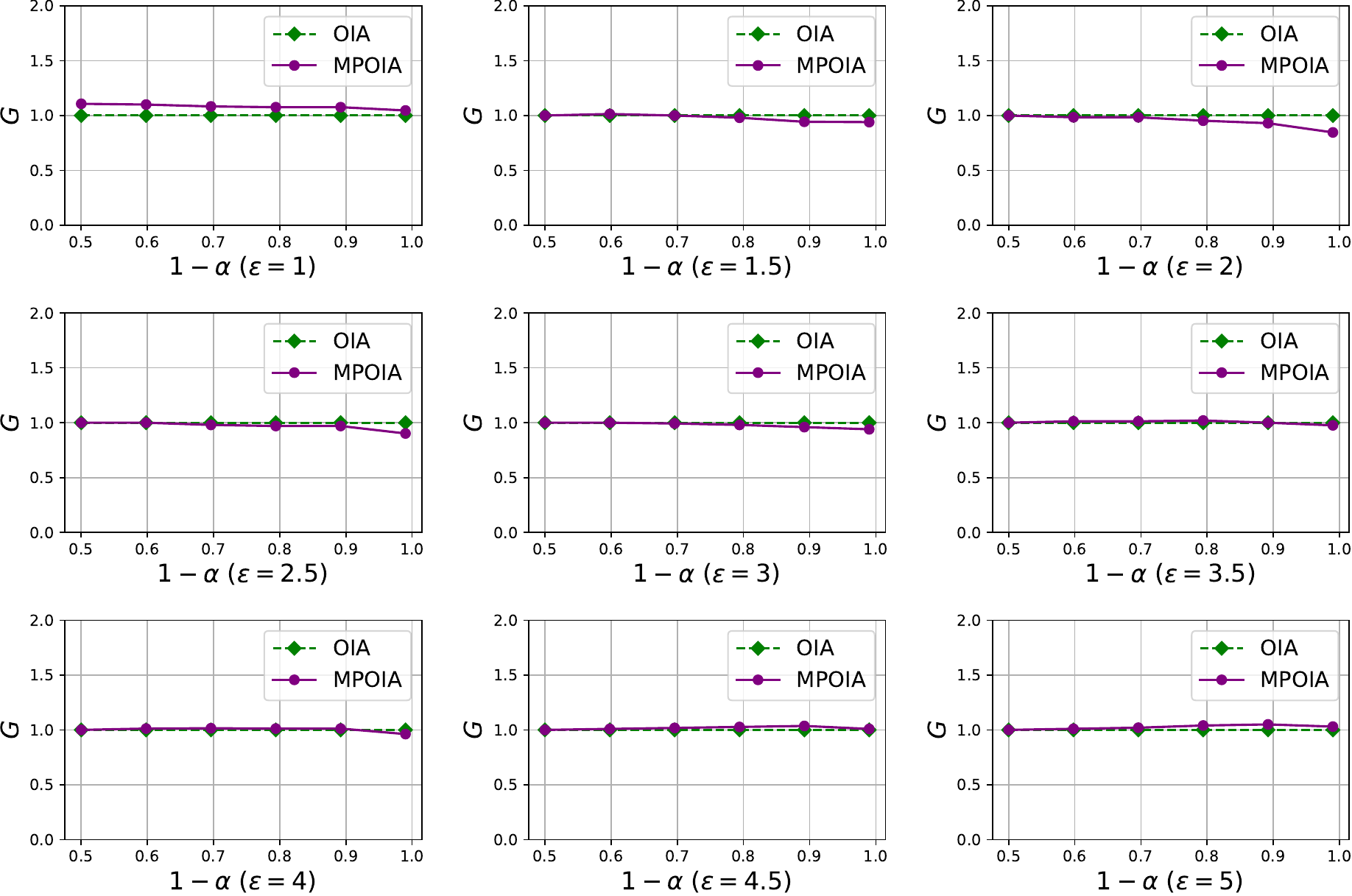}
\end{center}
\caption{\label{fig:MPOIA_OUE_Fire} Impact of MPOIA with different parameters ($\epsilon$, $1-\alpha$) on the overall gains for OUE on Fire dataset. }
\end{figure}

\begin{figure}[t]
\begin{center}
\includegraphics[width=\linewidth]{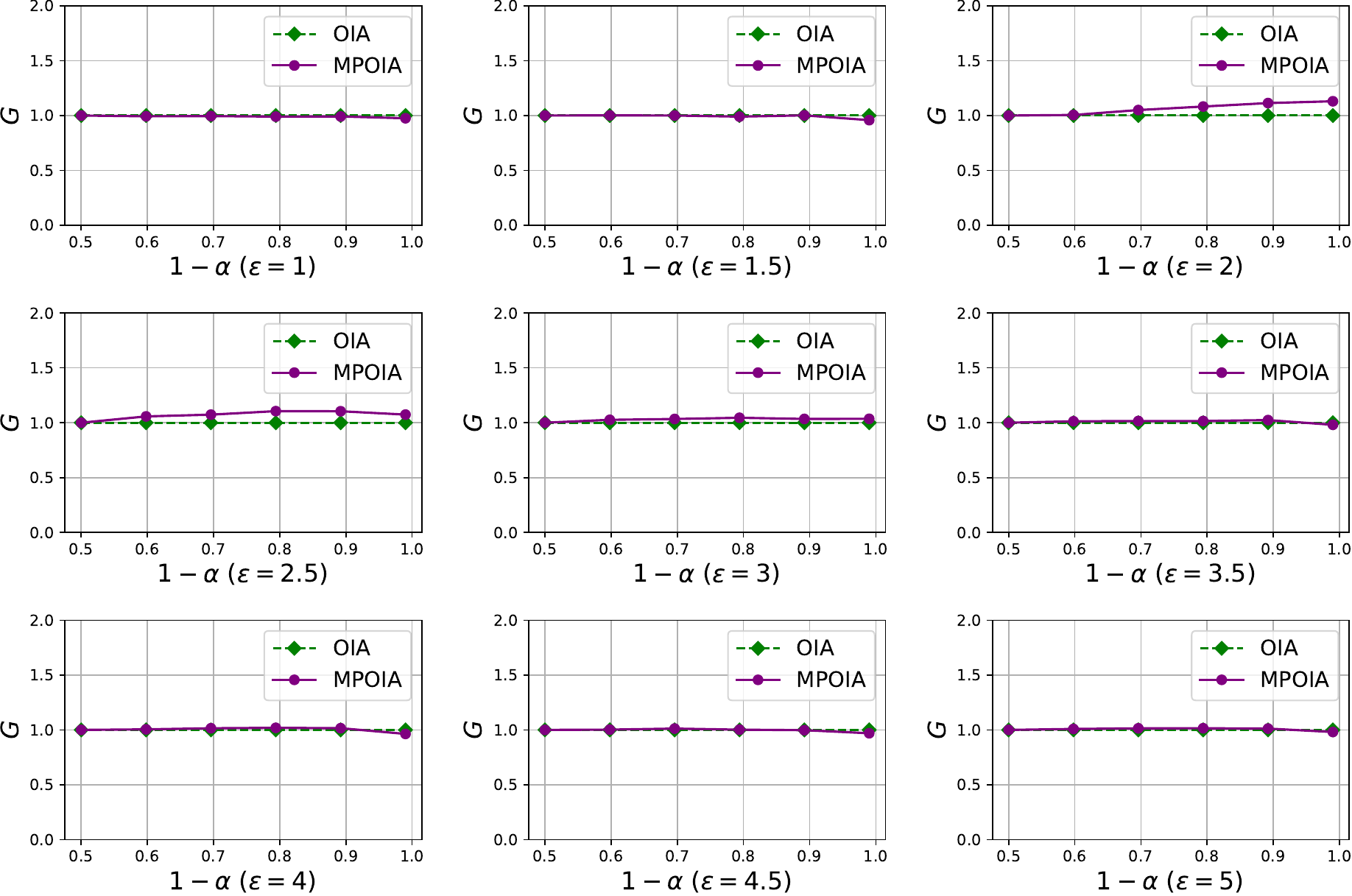}
\end{center}
\caption{\label{fig:MPOIA_OLH_Syn} Impact of MPOIA with different parameters ($\epsilon$, $1-\alpha$) on the overall gains for OLH on Synthetic dataset. }
\end{figure}

\begin{figure}[t]
\begin{center}
\includegraphics[width=\linewidth]{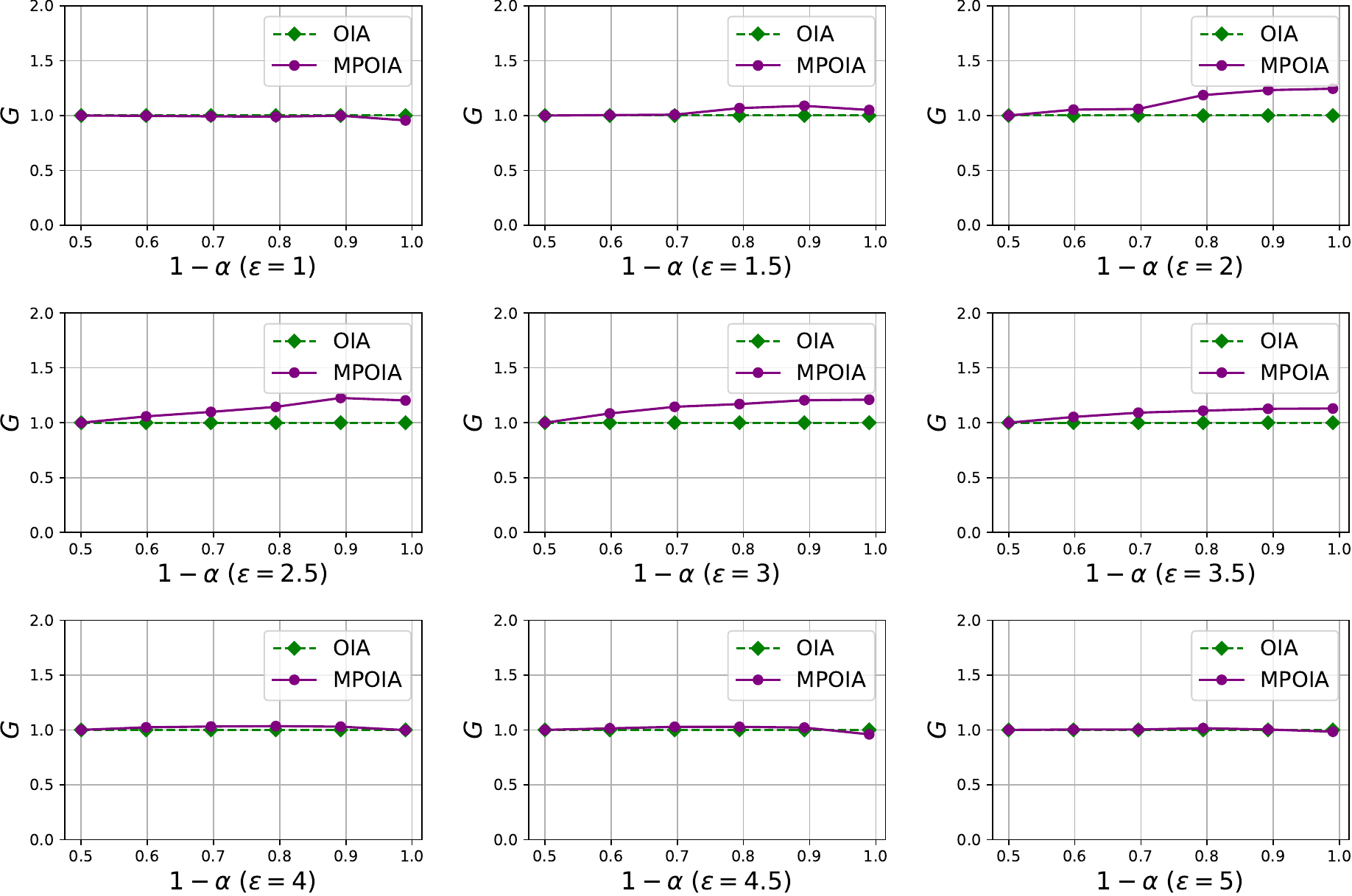}
\end{center}
\caption{\label{fig:MPOIA_OLH_Adult} Impact of MPOIA with different parameters ($\epsilon$, $1-\alpha$) on the overall gains for OLH on Adult dataset. }
\end{figure}

\begin{figure}[t]
\begin{center}
\includegraphics[width=\linewidth]{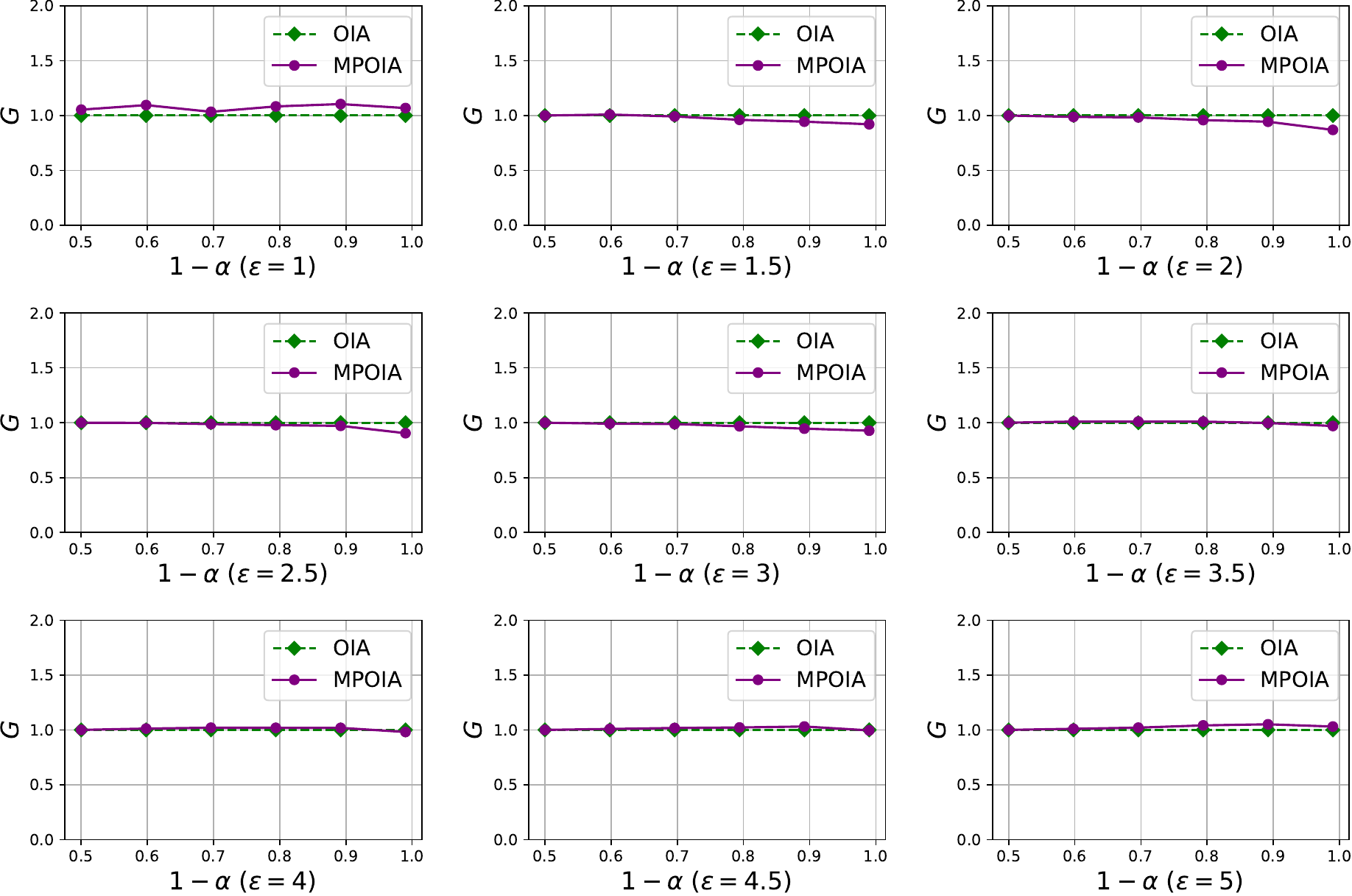}
\end{center}
\caption{\label{fig:MPOIA_OLH_Fire} Impact of MPOIA with different parameters ($\epsilon$, $1-\alpha$) on the overall gains for OLH on Fire dataset. }
\end{figure}

Figures~\ref{fig:MPOIA_kRR_Syn}--\ref{fig:MPOIA_OLH_Fire} illustrate the relative attack effectiveness of MPOIA compared to OIA across three protocols and three datasets, showing how it varies with the confidence level $1-\alpha$ under different privacy budgets $\epsilon$.

\section{Online Resources}
The source code, data, and/or other artifacts have been made avail-
able at https://github.com/LDP-user/LDP-Ranking.git.


\clearpage

\clearpage

\clearpage

\Description{Impact of different parameters on the success rates of the three attacks for PEM, for Synthetic, Adult, and Fire datasets.}

\end{document}